\documentclass[aps,10pt,prd,twocolumn,superscriptaddress,preprintnumbers,%
showkeys,eqsecnum,showpacs]{revtex4-1}
\usepackage{color}
\definecolor{red}{rgb}{0.9, 0, 0}
  
\usepackage{bm} 
\usepackage{graphicx}
\usepackage{hyperref}
\newcommand{\be}{\begin{eqnarray}}
\newcommand{\ee}{\end{eqnarray}}
\newcommand{\beq}{\begin{equation}}
\newcommand{\eeq}{\end{equation}}
\begin{document}
\title{Intrinsic transverse momentum and parton correlations \\
from dynamical chiral symmetry breaking}
\author{P.~Schweitzer}
\affiliation{Department of Physics, University of Connecticut,
Storrs, CT 06269, USA}
\author{M.~Strikman}
\affiliation{Department of Physics, Pennsylvania State University,
University Park, PA 16802, USA}
\author{C.~Weiss}
\affiliation{Theory Center, Jefferson Lab, Newport News, VA 23606, USA}
\begin{abstract}
The dynamical breaking of chiral symmetry in QCD is caused by 
nonperturbative interactions on a distance scale 
$\rho \sim 0.3\, \textrm{fm}$, much smaller than the typical hadronic size 
$R \sim 1 \, \textrm{fm}$. These short--distance interactions influence the 
intrinsic transverse momentum distributions of partons and their correlations
at a low normalization point. We study this phenomenon in an effective 
description of the low--energy dynamics in terms of chiral constituent 
quark degrees of freedom, which refers to the large--$N_c$ limit of QCD. 
The nucleon is obtained as a system of constituent quarks and antiquarks 
moving in a self--consistent classical chiral field (relativistic 
mean--field approximation, or chiral quark--soliton model). 
The calculated transverse momentum distributions of constituent 
quarks and antiquarks are matched with QCD quarks, antiquarks and gluons 
at the chiral symmetry--breaking scale $\rho^{-2}$.
We find that the transverse momentum distribution of valence quarks is 
localized at $p_T^2 \sim R^{-2}$ and roughly of Gaussian shape. 
The distribution of unpolarized sea quarks exhibits a would--be power--like 
tail $\sim 1/p_T^2$ extending up to the chiral symmetry--breaking scale. 
Similar behavior is observed in the flavor--nonsinglet polarized sea. 
The high--momentum tails are the result of short--range correlations
between sea quarks in the nucleon's light--cone wave function, which are
analogous to short--range $NN$ correlations in nuclei. We show that the 
nucleon's light--cone wave function contains correlated pairs of transverse 
size $\rho \ll R$ with scalar--isoscalar ($\Sigma$) and 
pseudoscalar--isovector ($\Pi$) quantum numbers, whose internal wave 
functions have a distinctive spin structure and become identical at 
$p_T^2 \sim \rho^{-2}$ (restoration of chiral symmetry). These features
are model--independent and represent an effect of dynamical 
chiral symmetry breaking on the nucleon's partonic structure.
Our results have numerous implications for the transverse 
momentum distributions of particles produced in hard scattering processes.
Under certain conditions the nonperturbative parton correlations 
predicted here could be observed in particle correlations 
between the current and target fragmentation regions of deep--inelastic
scattering.
\end{abstract}
\preprint{JLAB-THY-12-1641}
\maketitle
\tableofcontents
%
\section{Introduction}
\label{sec:introduction}
The parton model provides the basic script for expressing hadron 
structure as seen by short--distance probes such as local current 
operators or high--momentum transfer processes. Its fundamental assumption 
is that the fast--moving hadron can be regarded as a collection of pointlike
constituents that behave like free particles on the timescale of their
interaction with the external probe. The basic quantities are the
number densities of partons as functions of their longitudinal momentum
fraction $x$. Matrix elements of local current operators measure 
integrals of the parton densities (``sum rules''), while processes
such as deep--inelastic $eN$ scattering (or DIS) or production of high--mass
systems in $NN$ scattering probe them differentially in $x$.

The parton model can be thought of as the limiting case of a
large class of dynamical models of hadron structure, in which the
typical transverse momenta do not grow (or at most logarithmically)
as the longitudinal momentum of the hadron is increased. In this approach 
the parton densities appear as integrals over the transverse momenta 
of the constituents \cite{Bacchetta:2006tn},
\beq
f_1(x| \mu^2) \;\; = \;\; \int_{\mu^2} \!
d^2 p_T \; f_1(x, p_T) ,
\label{fpt_gen}
\eeq
where $\mu^2$ signifies a cutoff (e.g.\ in the parton virtuality, 
or in the invariant mass of configurations in the light--front 
wave functions), which restricts the 
integral over $p_T$ and defines the resolution scale at which 
the picture of pointlike partons is supposed to apply.
The integrand in Eq.~(\ref{fpt_gen}) is referred to as the ``intrinsic'' 
transverse momentum distribution of the partons. It is tempting
to interpret this function as a density of particles and use it 
to study the internal motion of the constituents in the system. 
While this is principally possible, studies of
field--theoretical models show that the transverse motion can 
generally not be separated from the interactions in the system. 
In gauge theories such as QED and QCD the parton picture requires
the light--cone gauge, where the transverse components of the gauge 
potential, $\bm{A}_T$, represent dynamical degrees of freedom.
Gauge invariance implies that the transverse derivatives of the 
fermion fields appear in the combination 
$\nabla_T \equiv \partial_T - i \bm{A}_T$, linking the kinetic
transverse momentum of the fermions to dynamical 
gauge fields in the hadron \cite{Ellis:1982wd}. 
The interpretation of the intrinsic transverse momentum distributions 
is thus generally much more subtle than that of the parton number densities.

A partonic picture of hadron structure is expected to emerge from QCD 
as the result of long--range nonperturbative interactions, 
and is commonly used 
to describe the boundary conditions for perturbative QCD calculations 
of hard processes. In this approach the parton densities can be expressed
as matrix elements of certain quark and gluon light--ray operators of twist 2,
i.e., correlation functions of the fields at light--like separation,
normalized at the scale $\mu^2$ (normalization point) \cite{Collins:1981uw}. 
The dependence on the scale can be calculated perturbatively for 
sufficiently large values; it is governed by the renormalization group 
equations for the composite operators, which coincide with the 
Dokshitzer--Gribov--Lipatov--Altarelli--Parisi (or DGLAP) evolution equations
describing parton decay in the leading logarithmic approximation.
In the region where perturbative evolution is applicable
the intrinsic $p_T$ distribution of partons is well--defined and 
of the form
\beq
f_1(x, p_T) \;\; \sim \;\; \frac{C_{f_1}(x)}{p_T^2} ,
\label{tail_QCD}
\eeq
which reflects the ultraviolet (or UV) divergences of QCD and implies 
a logarithmic dependence of the parton density Eq.~(\ref{fpt_gen}) on the 
scale $\mu^2$. Conversely, the coefficient $C_{f_1}(x)$ can be recovered 
as the logarithmic derivative of the parton density with respect to the scale,
\beq
C_{f_1}(x) \;\; = \;\; \pi^{-1} \; \mu^2 \frac{d}{d\mu^2} \; f_1(x | \mu^2) .
\label{tail_coeff}
\eeq
``Unintegrated'' parton densities defined according to 
Eqs.~(\ref{tail_QCD}) and (\ref{tail_coeff}) have been employed
in phenomenological studies of DIS at small $x$ ($\lesssim 10^{-2}$) 
\cite{Badelek:1996ap}.
A more careful treatment accounts also for the loss of partons
in the course of DGLAP evolution as described by the Sudakov 
survival factor, which becomes more important 
at larger values of $x$ \cite{Dokshitzer:1978hw,Kimber:1999xc}.

The extension of the concept of parton transverse momentum in QCD into 
the nonperturbative domain is inherently not unique, and different
prescriptions have been proposed in the context of studies of various
classes of high--momentum transfer processes. The calculation of power 
corrections to unpolarized DIS structure functions in the 
collinear expansion leads to twist--4 light--ray operators of the 
form $\bar\psi \nabla_{T, i} \ldots \nabla_{T, j} \psi$,
which can be interpreted as measuring the ``average transverse momentum''
of the quarks (keeping in mind that it cannot be separated
from the transverse gauge fields) \cite{Ellis:1982wd,Martinelli:1996pk}. 
Similar operators of twist 3 appear in the study of single--spin 
asymmetries in high--$p_T$ particle production \cite{Qiu:1991pp}. 
An alternative prescription are the 
transverse--momentum--dependent distributions (or TMDs) introduced
in the context of a QCD description of semi--inclusive DIS at
low transverse momenta \cite{Collins:1981uk,Ji:2004wu,Collins:2004nx}. 
They are defined as correlation functions of fields off the light--cone, 
at finite transverse separations, and involve gauge links (or phase factors) 
describing the effect of QCD initial/final--state interactions of the quark 
participating in the hard process. The renormalization properties
of these operators and the proper choice of gauge links are presently 
the subject of intense theoretical 
study \cite{Collins:1984kg,Cherednikov:2007tw,Aybat:2011zv}. 
The formal correspondence between the ``collinear'' and ``TMD'' 
definitions of parton transverse momentum in the perturbative
region of high $p_T$ was studied in Ref.~\cite{Bacchetta:2008xw}.
A natural concept of ``intrinsic'' transverse momentum also appears in
approaches which describe hadron structure in terms of light--cone wave 
functions at a low scale \cite{Lepage:1980fj,Brodsky:1997de}. 

The effective dynamics of strong interactions at the hadronic scale is 
in large measure governed by the spontaneous breaking of chiral symmetry
in the QCD vacuum. It leads to the appearance of a chiral condensate
$\langle \bar\psi\psi \rangle \neq 0$ (order parameter), whose phase
fluctuations give rise to almost massless excitations (Goldstone bosons), 
the pions. Their interactions are summarized by the 
universal chiral Lagrangian and determine the behavior of strong 
interactions over distances of the order $\sim 1/m_\pi$, which can be 
studied using methods of effective field theory. However, the influence 
of chiral symmetry breaking extends down to much shorter distances, 
defined by the range of the nonperturbative QCD interactions that
lead to the appearance of the chiral condensate. Lattice QCD calculations 
and numerous phenomenological observations suggest that the size of the 
nonperturbative field configurations causing the spontaneous 
breaking of chiral symmetry in QCD is much smaller than the typical
hadronic radius $R \sim 1\, \textrm{fm}$. An objective gauge--invariant 
measure of this scale is the average quark virtuality in the chiral 
condensate \cite{foot_m0}
\beq
\frac{m_0^2}{2} \;\; \equiv \;\; 
\frac{\langle\bar\psi \nabla^2 \psi\rangle}{\langle\bar\psi \psi\rangle} .
\label{m_0^2}
\eeq
Lattice simulations give $m_0^2/2 \gtrsim 0.5\, \text{GeV}^2$ at a 
normalization point of $\mu \sim 1\, \text{GeV}$ 
\cite{Kremer:1987ve,Chiu:2003iw}; even larger values were obtained 
in Ref.~\cite{Doi:2002wk}. This is supported by the results of direct 
studies of chirality--flipping topologically charged vacuum fluctuations 
in lattice simulations \cite{Chu:1994vi}; 
see Refs.~\cite{Negele:1998ev,Diakonov:2002fq} for a review.
The same conclusion is obtained from the instanton 
model of the QCD vacuum \cite{Shuryak:1981ff,Diakonov:1985eg} 
(see Refs.\cite{Diakonov:2002fq,Schafer:1996wv} for a review), 
where the typical size of the instantons
is $\rho \sim 0.3 \, \textrm{fm}$ and the average quark virtuality 
in the chiral condensate is found to be $m_0^2/2 = 2 \, \rho^{-2} \approx 
0.7 \, \textrm{GeV}^2$ \cite{Polyakov:1996kh}. Abstracting from these 
findings, one may state that the spontaneous breaking of chiral symmetry 
in QCD is characterized by a dynamical scale much shorter than the 
typical hadronic radius,
\beq
\rho \, \sim \, 0.3 \, \textrm{fm} \;\;\; \ll  \;\;\; 
R \, \sim \, 1 \, \textrm{fm}. 
\label{rho_vs_R}
\eeq
The existence of this
nonperturbative short--distance scale has far--reaching consequences 
for the structure of hadrons and their low--energy interactions
\cite{Diakonov:2002fq,Schafer:1996wv}.

Here we want to ask what dynamical chiral symmetry breaking and the
existence of the short--distance scale $\rho$ imply for the transverse 
momentum distribution of partons in the nucleon at a 
low normalization point. This question is clearly of great importance for
both the general theoretical understanding of partonic structure and the 
phenomenology of hard processes with identified particles,
such as semi--inclusive $ep$ scattering, jets and Drell--Yan pair 
production in $pp$ scattering, and multiparton interactions
in $pp$ collisions. In view of the ambiguities in the very definition of 
intrinsic transverse momentum in QCD we shall not attempt to approach
this problem in a model--independent manner, as the evaluation
of certain a priori defined QCD operators in the nucleon state. 
Instead, we shall study the transverse momentum distributions of partons 
in a model of the effective low--energy dynamics resulting from the
spontaneous breaking of chiral symmetry, which implements
the two dynamical scales of Eq.~(\ref{rho_vs_R}). The dynamical model 
will suggest a natural definition of the intrinsic transverse momentum 
distribution in terms of effective degrees of freedom, including the 
pertinent resolution scale. The matching of the model $p_T$ distributions 
with QCD will then be considered on the basis of their specific form,
and with the help of empirical information on the $p_T$--integrated 
parton densities, at a normalization point determined by the chiral 
symmetry--breaking scale, $\mu^2 \sim \rho^{-2}$. 
In the present state of development such an approach is fully justified 
and provides a useful complement to more abstract studies of transverse 
momentum distributions based on specific QCD operator definitions.

Numerous observations point to the importance of constituent quarks 
and pions as effective degrees of freedom below the chiral 
symmetry--breaking scale. Theoretical arguments suggest that in the 
large--$N_c$ limit of QCD the effective dynamics resulting from the 
spontaneous breaking of chiral symmetry can be approximated by a 
field--theoretical model based on chiral constituent 
quarks \cite{Diakonov:1984tw,Diakonov:2002fq}.
It expresses the fact that the modes of the QCD quark fields with 
virtualities below the chiral--symmetry breaking scale $\rho^{-2}$ 
acquire a dynamical mass. Because of chiral invariance, this is necessarily 
accompanied by a coupling to the Goldstone pion field, which
in the large--$N_c$ limit is itself a composite of constituent
quarks and antiquarks. This effective dynamics is relevant up
to the chiral symmetry--breaking scale, which appears as the
UV cutoff of the model. A crucial point is that the 
chiral symmetry--breaking scale is assumed to be parametrically 
large compared to the dynamical quark mass, such that the massive 
constituent quarks can be regarded as pointlike over a wide range 
of virtualities. This two--scale picture gives a precise meaning 
to the notion of constituent quarks as effective degrees of freedom 
and provides an ordering principle for the calculation of hadron structure.

In the effective chiral model the nucleon is obtained as an extended 
solution in which massive quarks and antiquarks move in the background 
of a self--consistent pion field (relativistic mean--field approximation, 
or chiral quark--soliton model) \cite{Diakonov:1987ty}. Matrix elements
of operators between nucleon states can be computed in a systematic
$1/N_c$ expansion. This picture results
in an essentially parameter--free description of the static nucleon 
observables and form factors \cite{Christov:1995vm}. Because the
description is fully field--theoretical, the model has a partonic limit
and can be used to calculate the nucleon's parton densities at 
a low normalization point \cite{Diakonov:1996sr,Diakonov:1997vc}. 
It provides for a nontrivial antiquark content of the nucleon at 
the starting scale of DGLAP evolution, in agreement with the results 
of global QCD fits of DIS data;
see Ref.~\cite{Gluck:2007ck} for a recent update. 
In particular, it quantitatively reproduces the flavor asymmetry of 
the unpolarized sea quarks, $f_1^{\bar u}(x) - f_1^{\bar d}(x) < 0$, 
observed in DIS \cite{Amaudruz:1991at} and Drell--Yan pair 
production \cite{Baldit:1994jk}. It predicts a large
flavor asymmetry also in the polarized sea, 
$g_1^{\bar u}(x) - g_1^{\bar d}(x) > 0$; there are hints of an
asymmetry of this sign in a recent global QCD fit including 
semi--inclusive data \cite{deFlorian:2008mr}; further clarification 
is expected from $W^\pm$ production
in polarized $pp$ collisions \cite{Dressler:1999zv,Aggarwal:2010vc}. 
These nonsinglet sea quark distributions do not mix with gluons 
under DGLAP evolution and represent clear signals
of the nonperturbative QCD vacuum structure encoded in the model.

The matching of the model parton distributions with QCD quark and gluon 
densities is performed at the chiral symmetry--breaking scale,
$\mu^2 \sim \rho^{-2}$, which represents the UV cutoff of the effective 
dynamics of constituent quarks. Thanks to the field--theoretical 
formulation of the dynamics (completeness of states, local interactions) 
and the relativistic covariance of the mean--field approximation
the chiral quark--soliton model conserves the overall 
light--cone momentum, so that the constituent quarks and antiquarks 
carry the entire light--cone momentum of the nucleon. This provides a 
solid basis for matching the effective degrees of freedom with the 
quarks and gluons of QCD. Physically, the effective degrees of 
freedom are composites of the QCD quark and gluon fields, and the model 
parton distributions should be ``resolved'' into their QCD content
at the scale $\mu^2 \sim \rho^{-2}$.
This process is not governed by intrinsic properties of the effective 
chiral dynamics but requires detailed knowledge of its embedding in QCD, 
which is poorly understood at present. In the simplest approximation 
one assumes that the constituent quarks and antiquarks remain pointlike 
up to the chiral symmetry--breaking scale and matches them with the quarks 
and antiquarks of QCD; the gluon density is zero in this approximation
\cite{Diakonov:1996sr}. This approximation was adopted in most calculations 
of partonic structure in the chiral quark--soliton model so far. 
Its accuracy may be judged from the fact 
that in leading--order fits to the DIS data \cite{Gluck:2007ck} 
at $\mu_{\rm LO}^2 \approx 0.3 \, \textrm{GeV}^2$ about $30\%$ of the 
nucleon's momentum is carried by gluons. This shows that the resolution 
effect is moderately strong in the singlet sector; a substantially 
weaker effect is expected for nonsinglets. More accurate matching 
would be possible either with a microscopic derivation of the effective 
chiral dynamics from QCD (such as the instanton vacuum 
model \cite{Diakonov:2002fq}) or with detailed phenomenological modeling 
based on empirical parton densities.

In this article we explore the role of dynamical chiral symmetry breaking
in the intrinsic transverse momentum distribution of partons, using the 
chiral quark--soliton model as an effective description of the dynamics 
below the chiral symmetry--breaking scale. Our study is
comprehensive and aims to address all relevant aspects of the problem:
the definition of the $p_T$ distributions within the effective model, 
their practical evaluation and numerical study, the implementation of 
the UV cutoff and the matching with QCD, and the implications for 
DIS experiments. We show that the effective 
dynamics suggests a natural definition of the intrinsic transverse momentum 
distributions, as the momentum densities of massive quarks and antiquarks 
in the fast--moving nucleon. We calculate the transverse momentum 
distribution of valence and sea quarks in the model in leading order 
of the $1/N_c$ expansion and study their properties. Our investigation 
leads to several interesting new insights.

First, we find that valence and sea quarks have very different intrinsic
transverse momentum distributions. The distribution of valence quarks 
(quarks minus antiquarks) has a range of the order of the inverse 
nucleon size, $p_T^2 \sim R^{-2}$ and an approximately Gaussian shape. 
The distribution of sea quarks (antiquarks), in contrast,
exhibits a power--like tail $\propto p_T^{-2}$ that extends up 
to the chiral--symmetry breaking scale. Its coefficient is 
determined by low--energy chiral dynamics and quasi model--independent. 
Such behavior is found in the flavor--singlet unpolarized sea quark 
distribution, where it was first observed in the
numerical study of Ref.~\cite{Wakamatsu:2009fn}, and the flavor--nonsinglet 
polarized sea quark distribution, which are the leading combinations in the 
$1/N_c$ expansion. The qualitative difference between valence and sea 
quark transverse momenta represents the imprint of dynamical chiral 
symmetry breaking on the nucleon's partonic structure and has numerous 
potential implications for hard scattering processes.

Second, we show that, under rather general conditions, the sea quark 
transverse momentum distributions do not depend on the details of the
UV cutoff of the effective chiral model. While the chiral 
symmetry--breaking scale represents the generic UV cutoff of 
the effective chiral dynamics, the manner in which it is implemented in 
the model is not dictated by chiral symmetry but must be constrained by 
other physical considerations. Imposing minimal physical conditions on 
the regularization scheme (charge conservation, longitudinal momentum 
conservation, analyticity) we find that the sea quark transverse momentum 
distributions are independent of the regularization scheme up to momenta 
$p_T^2 \sim 1 \, \textrm{GeV}^2$ and represent stable 
predictions of the model.
Since the regularization conserves the overall light--cone momentum, the 
constituent quark and antiquark distributions in the model carry the 
entire light--cone momentum of the nucleon and can consistently be
matched with QCD quarks and gluons at the scale $\mu^2 \sim \rho^{-2}$.

Third, we explore the role of dynamical chiral symmetry breaking 
at a more microscopic level, in terms of the light--cone wave function
of the nucleon in the chiral 
quark--soliton model \cite{Petrov:2002jr,Diakonov:2004as,Lorce:2007as}. 
The large--$N_c$ limit allows us to discuss the nucleon's partonic
structure in terms of the traditional nuclear physics concepts
of the mean field and short--range correlations. We show that the sea 
quarks in the nucleon's light--cone wave functions can exist in 
correlated pairs with a transverse size of the order of the 
chiral--symmetry--breaking scale $\rho$, much smaller than the
nucleon size $R$, which reflects their origin from dynamical chiral 
symmetry breaking. The pairs come in scalar--isoscalar ($\Sigma$) 
and pseudoscalar--isovector ($\Pi$) quantum numbers and have a 
distinctive spin structure; at large transverse momenta $p_T^2 \sim \rho^{-2}$
their internal wave functions become identical, reflecting the ``restoration
of chiral symmetry'' at the cutoff scale. These short--range correlations
represent another imprint of chiral symmetry breaking on the nucleon's 
partonic structure. They provide a natural microscopic explanation of the
high--$p_T$ tails found in the sea quark transverse momentum distributions
and point to an interesting analogy with short--range $NN$ correlations 
in nuclei; see 
Refs.~\cite{Frankfurt:1981mk,Frankfurt:2008zv,Arrington:2011xs}
for a review. Most importantly, it may be possible 
to observe these nonperturbative parton correlations directly in 
measurements of particle correlations between the current and target 
fragmentation regions in deep--inelastic $ep$ scattering or
multiparton processes in $pp$ scattering.

Quantifying the experimental implications of our results is a complex
task, which for the most part we leave to a separate study. Additional 
information on QCD final--state interactions and the fragmentation process 
is needed to relate the intrinsic $p_T$ distribution of partons to the 
observed transverse momentum distributions of hadrons emerging from 
hard scattering processes. Nevertheless, some simple conclusions can be 
drawn already at the present stage, without detailed modeling. 
Semi--inclusive DIS with single
identified hadrons is widely used to measure the flavor decomposition
of the nucleon parton densities. We show that the usual procedure of
combining $\pi^+$ and $\pi^-$ multiplicities to isolate the valence
quark density has to be modified if the intrinsic $p_T$ distributions 
of quarks and antiquarks in the nucleon are not the same and the 
experiment does not cover the full transverse momentum range of the
produced hadrons. A direct test of the nonperturbative parton
short--range correlations predicted here could be performed through
measurements of particle correlations between the current and target 
fragmentation regions in deep--inelastic $ep$ scattering. We show
that there is a kinematic window at moderate $\gamma^\ast N$
center--of--mass energies $W^2 \sim \textrm{few} \times 10 \, \textrm{GeV}^2$
in which the two fragmenting partons could be cleanly separated while 
perturbative QCD radiation does not yet destroy the nonperturbative
correlations. We also comment on the role of nonperturbative correlations
in multiparton processes in high--energy $pp$ scattering. Finally,
the nonperturbative parton correlations predicted here may play an 
important role in exclusive meson production at energies of 
$W \sim \textrm{few GeV}$.

The plan of this paper is as follows. In Sec.~\ref{sec:dynamics} we
summarize the model of the effective dynamics below the chiral 
symmetry--breaking scale and the resulting mean--field description
of the nucleon in the large--$N_c$ limit. In Sec.~\ref{sec:distributions}
we present the definition of the transverse momentum distributions
in the model as momentum densities of constituent quarks and antiquarks
in the fast--moving nucleon and discuss their basic properties.
We evaluate the expressions in terms of quark single--particle wave 
functions and develop their interpretation in the nucleon rest frame. 
We also discuss the coordinate--space correlation function associated
with the transverse momentum distribution in our model, and the 
positivity conditions and inequalities for the polarized 
distributions. In Sec.~\ref{sec:valence} we study the transverse 
momentum distributions of valence quarks 
(quarks minus antiquarks). We calculate the flavor--singlet 
unpolarized and flavor--nonsinglet polarized valence quark
distributions, $f_1^{u + d - \bar u - \bar d}(x, p_T)$ and 
$g_1^{u - d - \bar u + \bar d}(x, p_T)$, which appear in leading order
of the $1/N_c$ expansion, and study the average transverse
momentum $\langle p_T^2 \rangle$. 

In Sec.~\ref{sec:sea} we give
an in--depth treatment of the sea quark transverse momentum distributions 
in our approach. We evaluate them using the gradient expansion of the
quark Green function, an approximation which allows us to analytically
study the behavior of the distributions at large transverse momenta. 
The gradient expansion is formulated in terms of light--cone 
variables, which allows for a simple physical interpretation in terms 
of quark--antiquark pair production by the classical chiral field of
the nucleon. We analytically exhibit the power--like $1/p_T^2$ tail
of the flavor--singlet sea quark distribution $f_1^{\bar u + \bar d}(x, p_T)$ 
and discuss its significance. We then describe the 
physical conditions on the UV cutoff and present two regularization 
schemes that meet them (Pauli--Villars subtraction, and an invariant--mass 
cutoff). We evaluate the distributions numerically and verify that they 
are independent of the form of the UV cutoff. We also compute the 
coordinate--space correlation function in the model; we show that 
at large distances it decays exponentially and is completely governed
by low--energy dynamics. Finally, we also compare the sea quark
with the valence quark distributions at the numerical level and
confirm their qualitative difference. We also compute the flavor--nonsinglet
polarized distribution $g_1^{\bar u - \bar d}(x, p_T)$ and show that 
it exhibits a similar power--like tail at large $p_T$ as the flavor--singlet
unpolarized one.

In Sec.~\ref{sec:correlations} we 
discuss the nucleon's light--cone wave function at large transverse
momenta. We show that it is dominated by configurations in which
a single quark--antiquark pair has momenta of the order of the
chiral symmetry--breaking scale. We compute the internal wave functions
of $\Sigma$-- and $\Pi$--type pairs, study their spin structure, and 
demonstrate that chiral symmetry is effectively restored at large $p_T$.
We then prove that the high--$p_T$ tails in the distribution of sea quarks,
found previously by gradient expansion of the one--body densities, 
is exactly reproduced by 
the momentum density (wave function overlap) of such correlated pairs.
In Sec.~\ref{sec:summary} we summarize the model results for transverse 
momentum distributions and correlations and list problems meriting 
further study. We discuss the matching of the model distributions
with QCD using information about empirical $p_T$--integrated parton 
densities at a low scale, and discuss pertinent open questions. 
Lastly, we develop a qualitative
physical picture how nonperturbative parton correlations emerge from
QCD and discuss its implications. In Sec.~\ref{sec:applications} we 
outline the implications of our results for hard scattering processes. 
We discuss the consequences of different $p_T$ distributions of
valence and sea quarks for quark flavor separation in semi--inclusive
DIS. We also investigate the possibility of probing parton correlations 
by measurements of particle correlations between the current and 
target fragmentation region. Lastly, we comment on the potential role of
nonperturbative parton correlations in multiparton processes in
$pp$ collisions and exclusive meson production in $ep$ scattering.

A numerical study of transverse momentum--de\-pen\-dent quark distributions
in the chiral quark--soliton model was reported in 
Ref.~\cite{Wakamatsu:2009fn} and found a larger average $p_T$ of sea 
compared to valence quarks. Here we reproduce and explain this surprising 
result using analytic approximations, and show that the presence
of a power--like tail of the $p_T$ distribution follows directly from 
the short--range nature of dynamical chiral symmetry breaking in QCD as
encoded in the effective model. We also extend our study to the quark 
helicity distribution, where the tail appears in the flavor nonsinglet 
sector and is less affected by perturbative QCD radiation
in hard processes.

Transverse momentum distributions of quarks were extensively studied 
in diquark spectator models of the nucleon \cite{Jakob:1997wg}, the bag 
model \cite{Avakian:2008dz,Avakian:2010br}, 
light--front quark models \cite{Pasquini:2008ax}, 
and a covariant parton model \cite{Efremov:2009ze}. 
While incorporating some aspects of relativistic kinematics, all of these 
models describe the nucleon as system with a fixed number of particles,
ignoring the essential many--body nature of the parton picture.
Our approach here is field--theoretical and describes the nucleon
as a superposition of configurations with different numbers of particles,
which allows us to uncover the effect of the QCD vacuum on the nucleon's
partonic structure. Also, with the underlying two dynamical scales,
cf.\ Eq.~(\ref{rho_vs_R}), our approach provides a parametric framework 
for defining the transverse momentum distributions at a low 
scale in terms of effective degrees of freedom, which effectively 
include also the original gauge fields of QCD.
\section{Effective dynamics and nucleon structure}
\label{sec:dynamics}
\subsection{Chiral constituent quarks}
\label{subsec:constituent_quarks}
We begin by summarizing the essential elements of the two--scale
model of low--energy dynamics and the resulting description of 
the nucleon as a chiral soliton, following the lines of
Refs.~\cite{Diakonov:1987ty,Diakonov:1996sr,Diakonov:1997vc}.
The effective dynamics resulting from the spontaneous breaking
of chiral symmetry can be approximated as a field theory of massive 
quarks coupled to a Goldstone boson (pion) field in a chirally 
invariant manner. It is described by the Lagrangian density
\beq  
L_{\rm eff} \;\; \equiv \;\; \bar\psi (x) [ i \hat\partial
- M \, U^{\gamma_5}(x) ] \psi(x) ,
\label{L_eff} 
\eeq  
where $\psi (x)$ is the quark field (the sum over light quark 
flavors $u$ and $d$ is implied), 
\beq
\hat\partial \;\; \equiv \;\; \gamma^\alpha\partial_\alpha ,
\label{hat_def}
\eeq
and $M$ is the dynamical quark mass. The pion field is contained 
in the variable
\be
U^{\gamma_5}(x) &\equiv& \exp[ i \gamma_5 \tau^a \pi^a (x) / F_\pi] 
\label{U_gamma_5}
\nonumber \\[1ex]
&=& {\textstyle\frac{1}{2}}(1 + \gamma_5)  U(x)
+ {\textstyle\frac{1}{2}}(1 - \gamma_5)  U^\dagger(x) ,
\label{U_gamma_5_from_U}
\ee
where
\beq
U(x) \;\; \equiv \;\; \exp[i \tau^a \pi^a (x) / F_\pi] 
\label{U_def}
\eeq
is the usual unitary matrix field. Here $\tau^a (a = 1, 2, 3)$ denote 
the isospin Pauli matrices, and $F_\pi = 93\,\textrm{MeV}$ 
is the pion decay constant. Typical values of the dynamical quark mass 
obtained from phenomenological considerations are 
$M \sim \text{0.35--0.4} \, \textrm{GeV}$. In conventional terms, 
the strength of the pion--quark coupling is given by $M/F_\pi \approx 4$, 
as can be seen be expanding 
the exponential in Eq.~(\ref{U_def}) in powers of the pion field. 
The effective theory defined by Eq.~(\ref{L_eff}) is thus strongly coupled 
and has to be solved using nonperturbative methods based on the $1/N_c$ 
expansion (semiclassical or saddle--point approximation)
\cite{foot:pion_cloud}.

The effective dynamics described by Eq.~(\ref{L_eff}) applies to quarks
with virtualities below the chiral--symmetry--breaking scale, which acts 
as an UV cutoff for the model. In practical calculations the cutoff is 
implemented by applying a regularization scheme, and the actual value 
of the cutoff parameter depends on the scheme. In the following we denote 
the generic cutoff parameter by $\Lambda^2$ (not to be confused with
the QCD scale parameter $\Lambda_{\rm QCD}^2$). A crucial point is that 
the dynamical quark mass is assumed to be parametrically small compared 
to the cutoff,
\beq  
M^2 \;\; \ll \;\; \Lambda^2 ,
\label{parametric}
\eeq  
implying that the effective model is applicable in a parametrically 
wide range of quark momenta where the massive quarks behave approximately
as pointlike particles. While the numerical accuracy of this approximation
is limited ($M^2/\Lambda^2 \sim 0.3$ with a typical virtuality cutoff; 
see below), it serves as an ordering principle for the calculation 
of physical quantities and provides a clear mathematical justification for 
the constituent quark picture. 
The choice of regularization scheme for implementing the UV cutoff 
involves physical judgment and is usually motivated by the desire to 
preserve fundamental properties such as analyticity or current conservation.
For the $p_T$--integrated parton densities this question was 
studied in Refs.~\cite{Diakonov:1996sr,Diakonov:1997vc}; for the 
transverse momentum distributions of interest here it will be discussed 
in detail in Sec.~\ref{subsec:cutoff} below.

The effective model Eq.~(\ref{L_eff}) can be motivated by
general considerations about constituent quarks as effective
degrees of freedom, see Ref.~\cite{Diakonov:2002fq} for a review.
In particular, when integrating over the quark fields and expanding
the fermion determinant in gradients of the pion field, one
obtains the chiral Lagrangian of the pion field, with definite
predictions for the coupling constants in terms of the two 
parameters, $M$ and $\Lambda$. In this sense, the effective
theory ``interpolates'' between the chiral Lagrangian and 
a theory of free quarks at large virtualities. Eq.~(\ref{L_eff})
has also been derived from the instanton model of the QCD vacuum,
by large--$N_c$ bosonization of the instanton--induced 'tHooft 
interaction between quarks in the chirally broken 
phase \cite{Diakonov:1986aj,Diakonov:1995qy}.
In this context the UV cutoff is given by the inverse
average instanton size in the vacuum, $\rho^{-2} \approx 
0.4 \, \text{GeV}^2$, and restricts the quarks' Euclidean momenta
(or virtualities); the dynamical quark mass 
is obtained at $M \approx 0.35\, \text{GeV}$. The parametric
smallness of the quark mass compared to the cutoff, 
Eq.~(\ref{parametric}), follows directly from the ``diluteness'' 
of the instanton medium describing the QCD vacuum.

\subsection{Nucleon as chiral soliton}
%
%
\begin{figure}
\includegraphics[width=.46\textwidth]{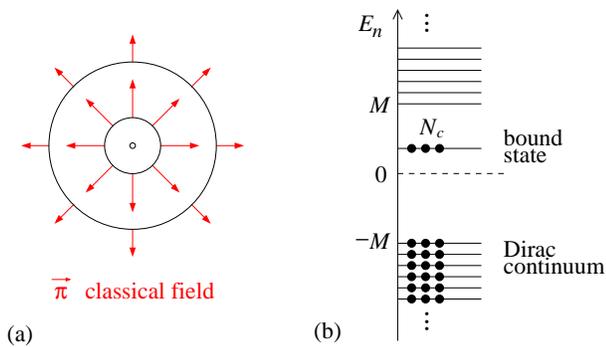} 
\caption{(Color online) Chiral quark--soliton model of the nucleon. 
(a) Classical chiral field in the nucleon rest frame. The isospin 
direction at a given point is defined by the radius vector (``hedgehog'').
(b) Spectrum of quark single--particle levels
in the classical chiral field, Eq.~(\ref{eigen}). It includes a 
discrete bound--state level and the distorted negative and
positive--energy Dirac continua. In the nucleon state (baryon number $+1$) 
the bound--state and negative--energy continuum levels are occupied 
by $N_c$ quarks each.}
\label{fig:chqsm}
\end{figure}
The effective dynamics described by Eq.~(\ref{L_eff}) refers to the
large--$N_c$ limit of QCD, and information on hadron structure is
extracted by calculating hadronic correlation functions in a 
systematic $1/N_c$ expansion. The nucleon is obtained
as an extended system in which quarks move in the background of a 
self--consistent classical chiral field (relativistic mean--field 
approximation, or chiral quark--soliton model) \cite{Diakonov:1987ty}. 
In the nucleon rest frame the chiral field is of 
``hedgehog'' form (see Fig.~\ref{fig:chqsm}a),
\beq
U_{\text{cl}} (\bm{x}) \; = \; 
\exp \left[ \frac{i \tau^a x^a}{r} P(r) \right] ,
\hspace{2em} (r \equiv |\bm{x}|),  
\label{hedgehog}
\eeq
and the profile function $P(r)$ satisfies $P(0) = -\pi$ and 
$P(r) \rightarrow 0$ for $r \rightarrow \infty$ \cite{foot_gamma5}. 
The quarks move in single--particle orbits, whose wave functions are the 
eigenfunctions of the Dirac Hamiltonian in the background chiral field, 
i.e., the solutions of the Dirac equation
\beq
\left( -i \gamma^0 \gamma^i \partial_i + M \gamma^0 U^{\gamma_5}_{\text{cl}} 
\right) \; \Phi_n (\bm{x}) \;\; = \;\; E_n \Phi_n (\bm{x}) .
\label{eigen}
\eeq
The energy spectrum includes a discrete bound--state level and the 
positive and negative Dirac continua, distorted by the chiral field
(see Fig.~\ref{fig:chqsm}b). Each level is occupied with $N_c$ quarks.
The static energy of the system is the sum of energies of the discrete level 
and the energy stored in the negative--energy continuum,
\beq
E[U_{\rm cl}] 
\;\; = \;\;  N_c E_{\rm lev} \; + \; N_c \sum_{E_n<0} (E_n - E_n^{(0)}) ,
\label{E}
\eeq
where $E_n^{(0)}$ denote the energy levels of the vacuum Hamiltonian 
with $U_{\rm cl} = 1$. The profile function $P(r)$ is determined by 
minimizing the static energy Eq.~(\ref{E}).
In Appendix~\ref{app:profile} we show the profile obtained by numerical 
minimization \cite{Weiss:1997rt} with a Pauli--Villars cutoff of the 
Dirac sea contribution to the energy (described in Sec.~\ref{sec:sea}) 
and give a simple analytic parametrization for use in our 
numerical estimates below. The nucleon mass, which is $O(N_c)$
within the $1/N_c$ expansion, is in leading order given by the minimum
value of the classical energy,
\beq
M_N \;\; = \;\; E[U_{\rm cl}]_{\rm min} \; + \; O(N_C^0) .
\label{M_N_as_minimum}
\eeq
Finally, nucleon states of definite spin/isospin and linear momentum 
are constructed by quantizing the $\textrm{(iso--)}$ 
rotational and translational zero modes of the soliton, see 
Ref.~\cite{Diakonov:1987ty} for details. This leads to the appearance
of the $N$ and $\Delta$ as rotational states of the soliton and explains
their mass splitting, $M_\Delta - M_N = O(1/N_c)$, as the difference 
of the rotational energies.

The description of the nucleon obtained 
in this approach is fully field--theoretical and does not involve any 
ingredients ``extraneous'' to the dynamics encoded in Eq.~(\ref{L_eff}).
No further approximations besides the $1/N_c$ expansion are made
in solving the dynamical problem. The description is also fully 
relativistic; it appears noncovariant only because the nucleon in 
the large--$N_c$ limit is heavy, but relativistic corrections 
appear systematically as part of the $1/N_c$ expansion.

Matrix elements of quark one--body operators between 
nucleon states can be calculated in a systematic $1/N_c$ expansion
and are generally expressed as sums of matrix elements between quark
single--particle states in the classical chiral field; see
Ref.~\cite{Christov:1995vm} for a review. The projection
on nucleon spin/isospin and momentum states is done by integrating over 
the rotational/isorotational and translational zero modes of the
classical field with appropriate collective wave functions. The sums 
over single--particle states can then be evaluated numerically by 
constructing the eigenfunctions and eigenvalues (both of the discrete 
bound--state level and the continuous spectrum) through numerical 
diagonalization of the Hamiltonian of Eq.~(\ref{eigen}) in a spherical 
box of finite size \cite{Kahana:1984be}. 

Alternatively, one may express matrix elements of quark one--body operators 
through the quark Green function in the chiral background field, using
the formalism of second quantization. In an arbitrary (generally 
time--dependent) classical chiral field, the Feynman Green function is 
defined as the solution of the inhomogeneous Dirac equation with a delta 
function source,
\beq
[ i \hat{\partial} - M \, U^{\gamma_5}(x) ] G_F (x, y) \;\; = \;\; 
\delta^{(4)}(x - y) ,
\label{inhomogeneous}
\eeq
with causal boundary conditions, corresponding to the advanced solution
for $x^0 < y^0$ and the retarded one for $x^0 > y^0$. In second quantization
this function coincides with the expectation value of the 
time--ordered product of quantized field operators,
\beq
i G_F (x, y) \;\; = \;\; \langle N| \, \textrm{T} \, \psi (x) \,
\bar\psi (y) \, | N \rangle ,
\eeq
where $|N\rangle$ denotes the ground state of the fermionic system
in the background of the (generally time--dependent) classical chiral field,
with the occupation of the single--particle levels as indicated in 
Fig.~\ref{fig:chqsm}b. In the static chiral field in the nucleon rest 
frame, Eq.~(\ref{hedgehog}), the time--ordered product becomes
\be
\lefteqn{i G_F (x, y)} 
\nonumber \\[1ex]
&=& \Theta (x^0 - y^0) \, \sum\limits_{n \; {\rm non-occ}} \!
e^{- i E_n (x^0 - y^0)}
\Phi_n (\bm{x}) \Phi_n^\dagger (\bm{y}) \gamma^0 
\nonumber \\[1ex]
&-& \Theta (y^0 - x^0) \, 
\sum\limits_{n \; {\rm occ}} 
e^{- i E_n (x^0 - y^0)}
\Phi_n (\bm{x}) \Phi_n^\dagger (\bm{y}) \gamma^0
\\[1ex]
\label{green_wf}
&=&
\int_{-\infty}^\infty \frac{d\omega}{2\pi} e^{-i \omega (x^0 - y^0)}
\sum_n \frac{\Phi_n (\bm{x}) \Phi_n^\dagger (\bm{y}) \gamma^0}
{\omega - E_n + i 0 \sigma_n} ,
\label{green_spectral}
\ee
\be
\sigma_n &\equiv & 
\left\{ 
\begin{array}{rr} 
-1 & \text{occupied levels}, \\[0ex]
+1 & \text{non--occupied levels}.
\end{array}
\right.
\label{sigma_n_def}
\ee
Equations~(\ref{green_wf}) and (\ref{green_spectral}) can be used to
convert traces of the Green function into sums over single--particle 
levels and vice versa. There are many practical advantages of working
with the Green function \cite{Diakonov:1996sr,Diakonov:1997vc}. 
It has simple transformation properties under 
Lorentz boosts, which follow directly from Eq.~(\ref{inhomogeneous}) 
(see Sec.~\ref{sec:distributions}). Its analytic properties in
energy allow one to derive sum rules and interchange sums over occupied
and non--occupied states. Finally, the Green function can be evaluated
approximately by expanding it in derivatives of the classical 
chiral field (gradient expansion). In this way one can obtain analytic 
expressions for the leading dependence of nucleon matrix elements in the 
limit of large UV cutoff, $\Lambda \rightarrow \infty$, or other 
limiting cases, such as the parton distributions at large transverse
momenta (see Sec.~\ref{sec:sea}).
\section{Intrinsic transverse momentum distributions}
\label{sec:distributions}
\subsection{Definition and parametric domain}
\label{subsec:definition}
The basic framework for calculating parton densities in the chiral 
quark--soliton model of the nucleon was developed in 
Refs.~\cite{Diakonov:1996sr,Diakonov:1997vc}. Generally, in the 
large--$N_c$ limit one is interested in the parton densities at
momentum fractions of the order 
\beq
x \;\; = \;\; O(N_c^{-1}), 
\eeq
corresponding to average (non--exceptional) configurations in the 
wave function of the fast--moving system \cite{Diakonov:1996sr}. 
The parton densities can be calculated starting from 
their parton model definition as number densities of particles in 
a nucleon moving with momentum $P \rightarrow \infty$ \cite{Diakonov:1997vc}, 
or from their representation as matrix elements of quark light--ray 
operators in the nucleon rest frame \cite{Diakonov:1996sr}.
The equivalence of the two formulations was demonstrated
in Ref.~\cite{Diakonov:1997vc} and is due to the relativistic and 
field--theoretical character of this description of the nucleon.

The relativistic mean--field picture of the nucleon implies a natural 
definition of the intrinsic transverse momentum distribution of partons
at a low scale. Extending the approach of Ref.~\cite{Diakonov:1997vc}, 
we \textit{define} the intrinsic transverse momentum distribution of 
partons in this model as the $p_T$--dependent number densities of 
quarks and antiquarks in the fast--moving nucleon state:
\be
f_1^a(x, p_T) &\equiv& \frac{P}{(2\pi)^{3}} \, \langle N_v | \;
{\textstyle\sum_\sigma} a_{a\sigma}^\dagger (\bm{p})a_{a\sigma}(\bm{p})
\; |N_v \rangle ,
\hspace{1em}
\label{fpt_def}
\\[1ex]
f_1^{\bar a}(x, p_T) &\equiv& \frac{P}{(2\pi)^{3}} 
\, \langle N_v| \;
{\textstyle\sum_\sigma} b_{a\sigma}^\dagger (\bm{p}) b_{a\sigma}(\bm{p})
\; |N_v \rangle ,
\hspace{1em}
\label{fpt_anti_def}
\ee
where $p_T \equiv |\bm{p}_T|$,
\be
\bm{p} &\equiv & (\bm{p}_T, \, x P) ,
\\[1ex]
P &=& \frac{v M_N}{\sqrt{1-v^2}},
\label{P_from_v}
\ee
and the limit
\beq
v \;\; \rightarrow \;\; 1 
\eeq
is understood here and in the following.
Here $a_{a\sigma}, a_{a\sigma}^\dagger$ and 
$b_{a\sigma}, b_{a\sigma}^\dagger$ are the quark and 
antiquark annihilation and creation operators corresponding to the
massive quark fields of the effective model Eq.~(\ref{L_eff}).
They annihilate/create quarks in plane--wave states with
three--momentum $\bm{p}$ and energy $p^0 = \sqrt{|\bm{p}|^2 + M^2}$,
with $a = u, d$ denoting the quark flavor and $\sigma$ the spin 
projection on the $3$--axis.
The notation $\langle N_v|\ldots |N_v \rangle$ indicates
the average in the ground state of the fast--moving many--body 
system, with $\langle N_v |N_v \rangle = 1$.
Equations~(\ref{fpt_def}) and (\ref{fpt_anti_def}) define the
unpolarized distributions (summed over quark/antiquark spin); 
the corresponding expressions for the longitudinally 
polarized distributions are
\be
g_1^a(x, p_T) &\equiv& \frac{P}{(2\pi)^{3}} \, \langle N_v | \;
{\textstyle\Delta_\sigma} a_{a\sigma}^\dagger (\bm{p})a_{a\sigma}(\bm{p})
\; |N_v \rangle ,
\hspace{1em}
\label{gpt_def}
\\[1ex]
g_1^{\bar a}(x, p_T) &\equiv& \frac{P}{(2\pi)^{3}} 
\, \langle N_v| \;
{\textstyle\Delta_\sigma} b_{a\sigma}^\dagger (\bm{p}) b_{a\sigma}(\bm{p})
\; |N_v \rangle ,
\hspace{1em}
\label{gpt_anti_def}
\ee
where
\beq
\Delta_\sigma \;\; \equiv \;\;  (\sigma = +1/2) - (\sigma = -1/2)
\label{gpt_def_Delta}
\eeq
denotes the spin difference. 

An obvious property of the transverse momentum distributions defined
by Eqs.~(\ref{fpt_def}) and (\ref{fpt_anti_def}) is that their integral 
over $p_T$ reproduce the quark/antiquark densities in the model,
\be
\int\limits \!\! d^2 p_T \; f_1^{a, \bar a}(x, p_T)_{\rm reg} 
&=&
f_1^{a, \bar a}(x) ,
\label{f_1_integrated}
\\[1ex]
\int\limits \!\! d^2 p_T \; g_1^{a, \bar a}(x, p_T)_{\rm reg} 
&=&
g_1^{a, \bar a}(x) .
\label{g_1_integrated}
\ee
The integral over transverse momenta would be logarithmically divergent at
large values and is rendered finite by the UV cutoff of the model. In our 
interpretation the cutoff is part of the definition of the transverse 
momentum distribution in the model, as indicated by the label ``reg''
in Eqs.~(\ref{f_1_integrated}) and (\ref{g_1_integrated}), not as an 
operation applied only at the level of the $p_T$ integral. This 
interpretation is more restrictive than the one developed in 
Ref.~\cite{Diakonov:1996sr,Diakonov:1997vc}, where only the integrated
parton densities were regularized, and places stronger demands on the
regularization scheme. A detailed discussion of the implementation of 
the UV cutoff is given in Sec.~\ref{subsec:cutoff} below.

In the sense of the $1/N_c$ expansion the transverse momentum we
consider are of the order
\beq
p_T \;\; = \;\; O(N_c^0) . 
\eeq
The $1/N_c$ expansion of the transverse momentum distributions is 
completely analogous to that of the $p_T$--integrated 
distributions \cite{Diakonov:1996sr}. In the following we consider 
the $p_T$ distributions of those spin--flavor combinations of
parton densities which appear in leading order of the $1/N_c$ expansion,
namely the flavor--singlet unpolarized distributions,
\beq
f_1^{u + d}(x, p_T),  \hspace{1em}
f_1^{\bar u + \bar d}(x, p_T), 
\eeq
and the flavor--nonsinglet polarized distributions,
\beq
g_1^{u - d}(x, p_T),  \hspace{1em}
g_1^{\bar u - \bar d}(x, p_T). 
\eeq

Equations~(\ref{fpt_def})--(\ref{gpt_def_Delta}) define the transverse
momentum distributions of massive constituent quarks and antiquarks ---
effective degrees of freedom which are to be matched with QCD quarks,
antiquarks and gluons at the chiral symmetry--breaking scale. The formulas 
should be regarded as preliminary definitions, whose physical significance 
and UV regularization will be elaborated in the following. The UV cutoff
affects the distribution of valence and sea quarks very differently and
will be discussed separately for the two cases (Secs.~\ref{sec:valence}
and \ref{sec:sea}). We shall see that under rather general conditions 
on the regularization scheme (charge conservation, longitudinal momentum 
conservation, analyticity) the transverse momentum distributions in the 
model are not sensitive to the details of the regularization scheme up to 
momenta $p_T^2 \sim 10 \, M^2$ and represent robust predictions of the model. 
This provides a firm basis for the matching of the model $p_T$ 
distributions with QCD (Sec.~\ref{sec:summary}). In the remainder of
this section we discuss the formal properties of the distributions
defined by Eqs.~(\ref{fpt_def})--(\ref{gpt_def_Delta}) without reference
to the UV cutoff.
\subsection{Evaluation in single--particle states}
\label{subsec:evaluation}
The evaluation of the model transverse momentum distributions defined by
Eqs.~(\ref{fpt_def}) and (\ref{fpt_anti_def}) proceeds along the
same lines as those of the $p_T$--integrated parton 
densities \cite{Diakonov:1997vc}. In the following we summarize the
main steps, emphasizing those aspects that require attention in the
case of transverse momentum dependence.

The one--body momentum density can conveniently be calculated in terms 
of the Feynman Green function of the system, using the formalism of
second quantization. The quark field operator is defined as
\be
\psi(x) &=& \int\!\frac{d^3p}{(2\pi)^3\sqrt{2 p^0}} \sum_\sigma 
\left[ a_\sigma (\bm{p}) \; u(\bm{p}, \sigma) \; e^{-i px} \right.
\nonumber \\
&+& \left. b^+_\sigma (\bm{p}) \; v(\bm{p}, \sigma) \; e^{ipx} 
\right] ,
\label{field}
\ee
where $u(\bm{p}, \sigma)$ and $v(\bm{p}, \sigma)$ are the spinor 
wave function of the free quarks and antiquarks, normalized as
$\bar{u}u = -\bar{v}v = 2M$, with $\bar u \equiv u^\dagger \gamma^0$
and $\bar v \equiv v^\dagger \gamma^0$. The quark and antiquark 
number operators can be expressed as equal--time products of the 
field operators, 
\be
&& \int\! d^3 x_1 \int \! d^3 x_2 \; e^{-i{\bm p}(\bm{x}_1 - \bm{x}_2)} 
\; \bar\psi_j (\bm{x}_2, t) \; \psi_i (\bm{x}_1, t) 
\nonumber \\
&=& 
\sum_{\sigma_1 \sigma_2} 
a^+_{\sigma_2}(\bm{p}) \, a_{\sigma_1}(\bm{p}) \; 
\frac{\bar u_j (\bm{p}, {\sigma_1}) u_i (\bm{p}, {\sigma_2})}{2p_0}
\; + \; \ldots , \hspace{2.5em}
\label{equal_time_a}
\\[1ex]
&& \int\! d^3 x_1 \int \! d^3 x_2 \; e^{-i{\bm p}(\bm{x}_1 - \bm{x}_2)} 
\; \psi_j (\bm{x}_2, t) \; \bar\psi_i (\bm{x}_1, t) 
\nonumber \\
&=& 
\sum_{\sigma_1 \sigma_2} 
b^+_{\sigma_2}(\bm{p}) \, b_{\sigma_1}(\bm{p}) \; 
\frac{v_j (\bm{p}, {\sigma_1}) \bar v_i (\bm{p}, {\sigma_2})}{2p_0}
\; + \; \ldots , \hspace{2em}
\label{equal_time_b}
\ee
where we exhibit the bispinor indices for clarity. The ellipsis here 
represent terms corresponding to the creation or annihilation of particles 
moving in the ``opposite'' direction (negative momentum in the 
$3$--direction), which disappear when evaluated in the many--body state 
with momentum $P \rightarrow \infty$. It is important to note that 
this happens irrespectively of whether one integrates over the transverse 
momentum $\bm{p}_T$ or keeps it fixed, and that the transverse momentum 
dependence of the structures in Eqs.~(\ref{equal_time_a}) and 
(\ref{equal_time_b}) surviving in the
$P \rightarrow \infty$ limit is unambiguously defined in our model
(in conventional terminology these are the leading--twist distribution
functions).

The density operators required in Eqs.~(\ref{fpt_def}) and 
(\ref{fpt_anti_def}) can be obtained by applying appropriate
spin projectors to Eqs.~(\ref{equal_time_a}) and (\ref{equal_time_b}).
Using the standard expressions for the spin density matrix of
a pure state ($\sigma_1 = \sigma_2 \equiv \sigma$) one finds that
in the limit $p^3 \rightarrow \infty$ and for fixed $\bm{p}_T$
\be
\frac{u (\bm{p}, \sigma) \bar u (\bm{p}, \sigma)}{2p_0}
&\rightarrow & \gamma^0 \frac{1 \pm \gamma_5}{2}
\hspace{1em} (\sigma = \pm 1/2) . \;\;
\ee
Here $\sigma$ is defined as the quark/antiquark spin projection 
along the $3$--axis. Thus the unpolarized density is obtained
by contracting both sides of Eqs.~(\ref{equal_time_a}) and 
(\ref{equal_time_b}) with $\gamma^0$,
\be
\textrm{tr}[\gamma^0 \bar \psi \psi]
&\rightarrow& 
\sum_\sigma a_{a\sigma}^\dagger (\bm{p})a_{a\sigma}(\bm{p}) ,
\\[1ex]
\textrm{tr}[\gamma^0 \psi \bar\psi]
&\rightarrow& 
\sum_\sigma b_{a\sigma}^\dagger (\bm{p})b_{a\sigma}(\bm{p}) ;
\ee
the longitudinally polarized density is obtained by contracting 
with $\gamma^0 \gamma_5$.

The equal--time products in Eqs.~(\ref{equal_time_a}) and 
(\ref{equal_time_b}) can be represented as appropriate limits of the 
time--ordered product of field operators, which is described by
the Feynman Green function in the fast--moving nucleon,
\beq
\langle N_v| \, \textrm{T} \, \psi({\bf x}_1,t_1) \,
\bar{\psi}({\bf x}_2,t_2) \, |N_v \rangle \;\; = \;\; i\, G_F^v (x_1, x_2) .
\label{time_ordered}
\eeq
Without loss of generality we choose
the equal time moment at $t = 0$ and obtain
\be
\left. \begin{array}{cc} 
f_1^{u + d}(x, p_T) \\[1ex]
f_1^{\bar u + \bar d}(x, p_T)
\end{array} \right\} 
&=& \frac{N_c P}{(2\pi)^3}
\int\! d^3 x_1 \int \! d^3 x_2 \; 
e^{-i{\bm p}(\bm{x}_1 - \bm{x}_2)} 
\nonumber \\
&\times & 
(\mp i) \; \textrm{tr} [ G_F^v (0, \bm{x}_1; \pm 0, \bm{x}_2)\, \gamma^0 ].
\hspace{3em}
\label{feynman_fast}
\ee
The Feynman Green function is given by the solution of the 
inhomogeneous Dirac equation Eq.~(\ref{inhomogeneous}) in the classical 
chiral field corresponding to the fast--moving nucleon, which is the static
field of the rest frame boosted along the $3$--direction with
velocity $v$,
\be
U^{\gamma_5, v}_{\rm cl} (\bm{x}, t) 
&\equiv& U^{\gamma_5}_{\rm cl} (\bm{x}')
\nonumber
\ee
\beq
\left[ \bm{x}' \;\; \equiv \;\; 
\left(\bm{x}_T, \frac{x^3 - vt}{\sqrt{1-v^2}} \right) \right] .
\eeq
Here and in the following we use primed variables to denote coordinates 
in the rest frame. Thanks to the Lorentz covariance of the Dirac
equation Eq.~(\ref{inhomogeneous}) and the source term on its right--hand 
side it is possible to express this Green function in terms of the
solutions of the rest--frame equation. Indeed, the Feynman Green function
in the fast--moving nucleon in the limits of Eq.~(\ref{feynman_fast})
can be obtained as the Lorentz boost of the occupied and non--occupied
level parts of the rest--frame Green function,
\be
\lefteqn{
G_F^v (0, \bm{x}_1; \pm 0, \bm{x}_2) \;\;  =  \;\;
(\pm i ) \left\{
\begin{array}{c}
\sum\limits_{n \; {\rm occ}}
\\[4ex]
\sum\limits_{n \; {\rm non-occ}}
\end{array}\right\} } &&
\nonumber 
\\[2ex]
&\times&
S(v) \; \Phi_n(\bm{x}^\prime_1) \Phi_n^\dagger (\bm{x}^\prime_2) \gamma^0
S^{-1}(v) \; e^{-iE_n(t^\prime_1-t^\prime_2)} . 
\hspace{2em}
\label{feynman_boosted}
\ee
Here $(t_{1, 2}', \bm{x}_{1, 2}')$ denote the space--time coordinates that
transform into $(0, \bm{x}_{1, 2})$ under a Lorentz boost in the 
3--direction with velocity $v$,
\be
(t_{1, 2}', \bm{x}_{1, 2}') &\equiv& 
\left( \frac{-v x_{1, 2}^3}{\sqrt{1 - v^2}}, \; \bm{x}_T, \; 
\frac{x_{1, 2}^3}{\sqrt{1 - v^2}} \right) .
\ee
The $\Phi_n (\bm{x}_{1, 2}')$ are the time--independent single--particle 
wave functions in the nucleon rest frame, given by the solution of the 
Dirac equation Eq.~(\ref{eigen}), and $S(v)$ is the transformation
matrix corresponding to the Lorentz boost,
\be
S(v) &\equiv& \exp \left( \frac{\eta}{2} \gamma^0 \gamma^3 \right)
= \cosh \frac{\eta}{2} + \gamma^0 \gamma^3 \sinh \frac{\eta}{2}  
\hspace{2em}
\\[1ex]
&& (v = \tanh \eta) . 
\nonumber
\ee
The formal proof that Eq.~(\ref{feynman_boosted}) represents the 
discontinuity of the Feynman Green function in the fast--moving mean 
field was given in Ref.~\cite{Diakonov:1997vc} and relies essentially
on the completeness of the set of single--particle wave functions.
Substituting Eq.~(\ref{feynman_boosted}) into Eq.~(\ref{feynman_fast})
one can straightforwardly evaluate the transverse momentum distributions 
in terms of the rest--frame single--particle wave functions. In the limit 
$v \rightarrow 1$ one has
\be
S^{-1}(v) \, \gamma^0 \, S(v) 
&=&
\frac{\gamma^0 + v \gamma^3}{\sqrt{1 - v^2}}
\;\; \sim \;\; 
\frac{\gamma^+}{\sqrt{1 - v^2}} ,
\hspace{2em}
\\[1ex]
\gamma^+ &\equiv& \gamma^0 + \gamma^3 .
\ee
Changing the integration variables in Eq.~(\ref{feynman_fast}) from 
$\bm{x}_{1, 2}$ to the rest frame coordinates $\bm{x}_{1, 2}'$
and substituting the momentum representation of the rest--frame 
single--particle wave functions,
\be
\Phi_n (\bm{x}_{1, 2}') &=& 
\int d^3p_{1, 2}' \; e^{i\bm{p}_{1, 2}'\bm{x}_{1, 2}'} \; 
\Phi_n (\bm{p}_{1, 2}')
\label{phi_momentum_def} ,
\ee 
one finds that they are effectively evaluated at rest--frame momenta
\beq
\bm{p}_1' \;  =  \; \bm{p}_2' \; = \; (\bm{p_T}, \, \pm x M_N - E_n) . 
\label{p_in_sum}
\eeq
This assignment is obtained here from the constraints inherent in the
fast--moving nucleon expression Eq.~(\ref{feynman_fast}) in the 
limit $v \rightarrow 1$; an alternative physical interpretation 
directly in the rest frame is provided in Sec.~\ref{subsec:restframe} below.
Finally, substituting the nucleon momentum $P$ as defined by 
Eq.~(\ref{P_from_v}), and taking the limit $v \rightarrow 1$, one obtains
\be
\left. \begin{array}{cc} 
f_1^{u + d}(x, p_T) \\[4ex]
f_1^{\bar u + \bar d}(x, p_T)
\end{array} \right\} 
&=& \frac{N_c M_N}{(2\pi)^3} \;
\left\{ \!
\begin{array}{c}
\sum\limits_{n \; {\rm occ}}
\\[3ex]
\sum\limits_{n \; {\rm non-occ}}
\end{array}\right\}
\nonumber 
\\[2ex]
&\times& 
\Phi_n (\bm{p})^\dagger \, \gamma^0 \gamma^+ 
\Phi_n(\bm{p}) \hspace{2em}
\nonumber
\ee
\beq
[\bm{p} \equiv (\bm{p_T}, \, \pm x M_N - E_n)] .
\label{fpt_sum}
\eeq
[For ease of notation we have dropped the prime on the rest frame momenta; 
all momenta in the following refer to the rest frame unless 
specified otherwise (in Sec.~\ref{sec:correlations}).] Formally one still
needs to average Eq.~(\ref{fpt_sum}) over the (iso--) rotational zero modes
and project on states with definite spin--isposin quantum numbers
\cite{Diakonov:1996sr}; however, in the case of the flavor--singlet 
unpolarized distributions here this operation is trivial and does not 
change the form of the expression.

In the quark distribution in Eq.~(\ref{fpt_sum}) the summation extends over 
all occupied quark single--particle states, including the discrete level 
and the negative--energy Dirac continuum (see Fig.~\ref{fig:chqsm}b).
The antiquark distribution is given by the corresponding sum over 
non--occupied states in the spectrum, with the sign of the $x M_N$ term
reversed in the 3--component of the rest frame momentum Eq.~(\ref{p_in_sum}). 
Equivalent representations of the quark distribution as a sum over 
non--occupied states, and of the antiquark distribution as a sum over 
occupied states, can be derived by making use of the completeness of 
the single--particle states (see below) \cite{Diakonov:1997vc},
\be
\left. \begin{array}{cc} 
f_1^{u + d}(x, p_T) \\[4ex]
f_1^{\bar u + \bar d}(x, p_T)
\end{array} \right\} 
&=& -\frac{N_c M_N}{(2\pi)^3} \;
\left\{ \!
\begin{array}{c}
\sum\limits_{n \; {\rm non-occ}}
\\[3ex]
\sum\limits_{n \; {\rm occ}}
\end{array}\right\}
\nonumber 
\\[2ex]
&\times& 
\Phi_n (\bm{p})^\dagger \, \gamma^0 \gamma^+ 
\Phi_n(\bm{p}) \hspace{2em}
\label{fpt_sum_alt}
\ee
\be
&& \left[ \bm{p} \equiv (\bm{p_T}, \, \pm x M_N - E_n) \right] . 
\nonumber
\ee

In the case of the polarized distributions, the flavor--nonsinglet
(isovector) combination appears in leading order of the $1/N_c$ expansion. 
An analogous calculation as in the case of the unpolarized distribution 
leads to the expressions (cf.~Ref.\cite{Diakonov:1997vc})
\be
\left. \begin{array}{cc} 
g_1^{u - d}(x, p_T) \\[4ex]
g_1^{\bar u - \bar d}(x, p_T)
\end{array} \right\} 
&=& \mp \frac{N_c M_N}{3(2\pi)^3} \;
\left\{ \!
\begin{array}{c}
\sum\limits_{n \; {\rm occ}}
\\[3ex]
\sum\limits_{n \; {\rm non-occ}}
\end{array}\right\}
\nonumber 
\\[2ex]
&\times& 
\Phi_n (\bm{p})^\dagger \, \tau^3 \, \gamma^0 \gamma^+ \gamma_5
\Phi_n(\bm{p}) 
\hspace{2em} 
\label{gpt_sum}
\ee
\be
&& \left[ \bm{p} \equiv (\bm{p_T}, \, \pm x M_N - E_n) \right] .
\nonumber
\ee
Here $\tau^3$ denotes the Pauli matrix in quark flavor indices.
The factor $-1/3$ arises from the integration over the $\textrm{(iso--)}$ 
rotational zero modes and projection on the nucleon state with spin 
$S_3 = 1/2$ and isospin $T_3 = 1/2$ \cite{Diakonov:1996sr}. An alternative
representation, analogous to Eq.~(\ref{fpt_sum_alt}), is
\be
\left. \begin{array}{cc} 
g_1^{u - d}(x, p_T) \\[4ex]
g_1^{\bar u - \bar d}(x, p_T)
\end{array} \right\} 
&=& \pm \frac{N_c M_N}{3(2\pi)^3} \;
\left\{ \!
\begin{array}{c}
\sum\limits_{n \; {\rm non-occ}}
\\[3ex]
\sum\limits_{n \; {\rm occ}}
\end{array}\right\}
\nonumber 
\\[2ex]
&\times& 
\Phi_n (\bm{p})^\dagger \, \tau^3 \, \gamma^0 \gamma^+ \gamma_5
\Phi_n(\bm{p}) 
\hspace{2em} 
\label{gpt_sum_alt}
\ee
\be
&& \left[ \bm{p} \equiv (\bm{p_T}, \, \pm x M_N - E_n) \right] . 
\nonumber
\ee

The transverse momentum distributions can also be expressed in terms of the
Feynman Green function of quarks in the nucleon rest frame. It is convenient 
to introduce the energy--momentum representation of the Feynman Green 
function as
\be
G_F (x, y) &=&  
\int\!\frac{d^4 p_1}{(2\pi )^4} 
\int\!\frac{d^4 p_1}{(2\pi )^4}
\; e^{-i p_1 x + i p_2 y} 
\nonumber \\
&\times & 2\pi \delta (p_1^0 - p_2^0) \; S_F (p_1^0; \bm{p}_1, \bm{p}_2),  
\label{S_F_def}
\ee
whose spectral representation in the rest frame is
\beq
S_F (p_1^0; \bm{p}_1, \bm{p}_2) \;\; = \;\;
\sum_n \frac{\Phi_n (\bm{p}_1) \Phi_n (\bm{p}_2)^\dagger \gamma^0}
{p_1^0 - E_n + i 0 \sigma_n} ,
\label{S_F_spectral}
\eeq
where $\sigma_n$ is defined in Eq.~(\ref{sigma_n_def}). The sums over 
occupied and non--occupied single--particle levels in 
Eq.~(\ref{fpt_sum}) can be then expressed as contour
integrals in the complex energy variable encircling the negative and 
positive--energy poles. These integrals can in turn be converted
into integrals over the real energy axis by deforming the contours
at infinity; see Ref.~\cite{Diakonov:1996sr} for details. In this it 
is important that the UV cutoff respect the analytic
properties of the unregularized theory and does not introduce
spurious singularities. For the quark distribution one obtains
\be
\left. \begin{array}{cc} 
f_1^{u + d}(x, p_T) \\[4ex]
f_1^{\bar u + \bar d}(x, p_T)
\end{array} \right\} 
&=& \pm \textrm{Im} \, \frac{N_c M_N}{(2\pi)^3}
\; \int_{-\infty}^\infty \!\frac{dp^0}{2\pi} 
\nonumber \\
&\times & 
\textrm{tr}[ S_F (p^0; \bm{p}, \bm{p}) \, \gamma^+ ] 
\label{fpt_green}
\ee
\be
&& \left[ \bm{p} \equiv (\bm{p}_T, \pm x M_N - p^0) \right] .
\nonumber
\ee
The antiquark distribution is given by the negative of the expression
for the quark distribution with $x$ replaced by $-x$. The corresponding
expressions for the flavor--nonsinglet polarized distribution are
\be
\left. \begin{array}{cc} 
g_1^{u - d}(x, p_T) \\[4ex]
g_1^{\bar u - \bar d}(x, p_T)
\end{array} \right\} 
&=& -\textrm{Im} \, \frac{N_c M_N}{3(2\pi)^3}
\; \int_{-\infty}^\infty \!\frac{dp^0}{2\pi} 
\nonumber \\
&\times & 
\textrm{tr}[ S_F (p^0; \bm{p}, \bm{p}) \, \tau^3 \, \gamma^+ \gamma_5] 
\hspace{2em}
\label{gkt_green}
\ee
\be
&& \left[ \bm{p} \equiv (\bm{p}_T, \pm x M_N - p^0) \right].
\nonumber
\ee
In the polarized case the antiquark distribution is given by the 
expression for the quark
distribution with $x \rightarrow -x$ without change of the overall sign.
Equations~(\ref{fpt_green}) and (\ref{gkt_green}) are the starting 
point for calculating the $p_T$ distribution by gradient expansion in 
the classical chiral field. They can also be used to derive equivalent
representations of the quark distribution as a sum over non--occupied states,
or of the antiquark distribution as a sum over occupied states, by deforming
the integration contour in the $p^0$ integral.
\subsection{Rest frame interpretation}
\label{subsec:restframe}
The expression Eq.~(\ref{fpt_sum}) for the quark and antiquark distributions 
permit a simple interpretation directly in the nucleon rest frame, which 
explains the form of the rest--frame three--momentum Eq.~(\ref{p_in_sum}) 
at which the single--particle wave functions are evaluated. It also reveals 
the role of the large--$N_c$ limit in the derivation and emphasizes the 
analogy with the conventional mean--field approximation of nuclear physics.

\begin{figure}
\includegraphics[width=.45\textwidth]{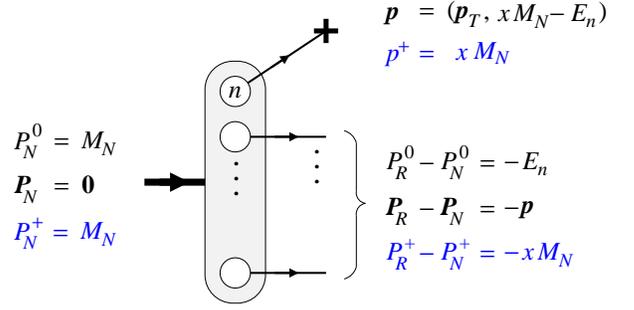} 
\caption{(Color online) Interpretation of the rest frame expression 
of the transverse momentum--dependent quark density in the chiral 
quark--soliton model, Eq.~(\ref{fpt_sum}). 
The operator removes a quark with 3--momentum $\bm{p}$
from the single--particle level $n$ in the nucleon wave function
(in the case of the antiquark density it would place a quark in an
unoccupied level). The remnant system recoils
with momentum $-\bm{p}$ and energy loss $-E_n$. The plus momentum
loss calculated as the difference of the plus momenta of the 
remnant system and the initial nucleon state exactly compensates 
the plus momentum of the active quark, $x M_N$. This shows that the 
relativistic mean--field approximation conserves plus momentum 
at the order $O(N_c^0)$ (details see text).}
\label{fig:restframe}
\end{figure}

The parton densities defined by Eqs.~(\ref{fpt_def}) and 
(\ref{fpt_anti_def}) can equivalently be regarded as the densities of 
quarks and antiquarks carrying a fraction $x$ of the nucleon's 
light--cone ``plus'' momentum $P_N^+\equiv P_N^0 + P_N^3$.
As such they may be evaluated in any reference frame. In the rest 
frame $P_N^3 = 0$ and $P_N^0 = M_N$, and thus
\beq
P_N^+ \;\; = \;\; M_N
\eeq
(see Fig.~\ref{fig:restframe}). In this frame the operator measuring 
the parton density counts quarks and antiquarks with plus momentum
\beq
p^+ \;\; = \;\; x M_N .
\eeq
In the mean--field approximation this operator removes a quark from an 
occupied single--particle state $n$ with energy $E_n$ and three--momentum 
$\bm{p}$, which has both longitudinal and transverse components (in the 
case of the antiquark density, the operator puts a quark with the same 
momentum in an unoccupied state $n$). As the quark removed from the 
interacting system is not on mass--shell, the relation between its energy 
and three--momentum, and hence the relation between these variables and 
the external parameter $x$, are a priori not obvious and require discussion. 
The assignment of the quark three--momentum of Eq.~(\ref{fpt_sum})
obtained from the derivation in Sec.~\ref{subsec:evaluation} 
\cite{Diakonov:1997vc} corresponds to
\beq
p^+ \;\; = \;\; E_n + p^3 .
\label{p_plus_removed}
\eeq
We now show that this assignment can be justified by inspection of the 
remnant system in the rest frame, \textit{i.e.}, the many--body state 
from which the quark or antiquark was removed. This system is on mass 
shell, and its kinematic variables can unambiguously be related to the 
external parameter $x$.
%
%

Because three--momentum is conserved in equal--time quantization, 
the three--momentum imparted to the remnant system in the rest frame is
\beq
\bm{P}_R - \bm{P}_N \;\; = \;\; \bm{P}_R \;\; = \;\; -\bm{p} 
\;\; = \;\; O(N_c^0)
\label{p_remnant}
\eeq
(see Fig.~\ref{fig:restframe}). At the same time, in the mean--field
approximation the energy difference between the remnant system and the 
initial nucleon state is determined by the energy of the removed quark,
\beq
P_R^0 - P_N^0 \;\; = \;\; -E_n \;\; = \;\; O(N_c^0).
\label{E_remnant}
\eeq
Note that the individual energies $P_R^0$ and $P_N^0$ are $O(N_c)$,
but their difference is $O(N_c^0)$ and can be discussed quantitatively 
at this level. An important point is also that the kinetic energy 
associated with the recoil of the remnant system is of the order
\beq
E_{\rm recoil} \;\; \sim \;\; \frac{\bm{p}^2}{2 M_N} \;\; = \;\; O(N_c^{-1})
\eeq
and can be neglected relative to Eq.~(\ref{E_remnant}). 
From Eqs.~(\ref{p_remnant}) and (\ref{E_remnant}) we can calculate
the plus momentum difference between the remnant system and the
initial nucleon,
\beq
P_R^+ - P_N^+ \;\; = \;\; -E_n - p^3 \;\; = \;\; O(N_c^0) .
\eeq
This is precisely the negative of the plus momentum of the
removed quark, Eq.~(\ref{p_plus_removed}), obtained from the
earlier derivation in Sec.~\ref{subsec:evaluation}.
Thus we see that this assignment implies the conservation of plus momentum 
of the interacting system in leading order of the $1/N_c$ expansion, 
i.e., the individual changes in plus momenta of $O(N_c^0)$ add up to zero 
at that order. The same applies of course to the transverse momenta. 
We emphasize that this argument does not neglect the interactions in the 
system but only relies on the mean--field approximation to the nucleon
wave function. The conservation of plus momentum represents a nontrivial 
consequence of the consistency of the approximations made in this model.
\subsection{Positivity and inequalities}
\label{subsec:positivity}
The unpolarized transverse momentum distributions of quarks and 
antiquarks in the chiral quark--soliton model defined by 
Eqs.~(\ref{fpt_def}) and (\ref{fpt_anti_def}) are particle number 
densities in longitudinal and transverse momentum and should 
therefore be positive. It is easy to verify that their expressions
in terms of single--particle wave functions, Eq.~(\ref{fpt_sum}),
are indeed manifestly positive. Noting
that the hermitean matrix $\gamma^0\gamma^+/2$ is a projector,
\beq
\left(\frac{\gamma^0\gamma^+}{2}\right)^2
\;\; = \;\; \frac{\gamma^0\gamma^+}{2}
\eeq
we can rewrite Eq.~(\ref{fpt_sum}) as
\be
\left. \begin{array}{cc} 
f_1^{u + d}(x, p_T) \\[4ex]
f_1^{\bar u + \bar d}(x, p_T)
\end{array} \right\} 
&=& \frac{2 N_c M_N}{(2\pi)^3} \;
\left\{ \!
\begin{array}{c}
\sum\limits_{n \; {\rm occ}}
\\[3ex]
\sum\limits_{n \; {\rm non-occ}}
\end{array}\right\} 
\nonumber 
\\[2ex]
&\times& 
\left| \frac{\gamma^0\gamma^+}{2}\, \Phi_n(\bm{p}) \right|^2 \hspace{2em}
\ee
\be
&& \left[ \bm{p} \equiv (\bm{p_T}, \, \pm x M_N - E_n) \right] . 
\label{positivity}
\ee
Here the quark and antiquark distribution are expressed as a sum of
explicitly positive terms. This property essentially relies on the
completeness of quark single--particle states in the model.
We note that the UV cutoff implicit in the effective chiral 
dynamics may in principle violate the positivity of the unregularized 
expression. We shall show in Sec.~\ref{subsec:cutoff}
that the physical regularization schemes proposed there naturally
preserve the positivity of the transverse momentum distributions.

The completeness of quark single--particle states also implies the 
existence of inequalities between the polarized and unpolarized
transverse momentum distributions. An obvious consequence of the
probabilistic nature of the densities defined by Eqs.~(\ref{fpt_def}) 
and (\ref{fpt_anti_def}), and Eqs.~(\ref{gpt_def}) and 
(\ref{gpt_anti_def}), is that
\beq
f_1^{a, \bar a}(x,p_T) \;\; \ge \;\; |g_1^{a, \bar a}(x,p_T)|
\eeq
for any given quark or antiquark flavor $a$ \cite{Bacchetta:1999kz}.
Actually, in the large--$N_c$ limit 
a stronger inequality was proved for the $p_T$--integrated densities
\cite{Pobylitsa:2000tt}, namely $f_1^a(x) \ge |3g_1^a(x)|$. The
generalization of this proof to the case of the $p_T$ distributions
is straightforward, and one has
\beq
f_1^{a, \bar a}(x,p_T) \;\; \ge \;\; |3 \, g_1^{a, \bar a}(x,p_T)| .
\eeq
In particular, for the flavor combinations appearing in leading
order of the $1/N_c$ expansion we obtain
\be
f_1^{u + d}(x,p_T) &\ge& |3 \, g_1^{u - d}(x,p_T)| ,
\label{ineq}
\\[1ex]
f_1^{\bar u + \bar d}(x,p_T) &\ge& |3 \, g_1^{\bar u - \bar d}(x,p_T)| .
\label{ineq_anti}
\ee
In Secs.~\ref{sec:valence} and \ref{sec:sea} we show that these 
inequalities are satisfied by the model distributions obtained here, 
including the effects of the UV cutoff. This again
illustrates the consistency of the scheme of approximations proposed here.
The probabilistic character of the model $p_T$ distributions 
expressed in the positivity condition and the inequalities also facilitates
their matching with QCD quark and gluon distributions at the chiral 
symmetry--breaking scale; cf.\ the discussion in Sec.~\ref{subsec:matching}.
\subsection{Coordinate--space correlation function}
\label{subsec:coordinate}
\begin{figure}
\includegraphics[width=.26\textwidth]{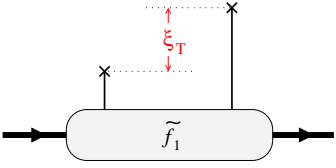} 
\caption{(Color online) Coordinate--space correlation function
$\tilde f_1^{u + d} (x, \xi_T)$ and 
$\tilde f_1^{\bar u + \bar d} (x, \xi_T)$, Eq.~(\ref{fkt_corr})
representing the Fourier transform of the transverse momentum 
distributions in the chiral quark--soliton model.}
\label{fig:coordinate}
\end{figure}

The total ($p_T$--integrated) parton densities in the chiral--quark soliton 
model can be represented as correlation functions of the massive quark 
fields of the model at light--like space--time 
separations \cite{Diakonov:1996sr}. Denoting the space--time separation 
four--vector by $\xi$, with $\xi^2 = 0$, one usually chooses
a frame where $\xi^+ = 0, \, \bm{\xi}_T = 0,$ and $\xi^- \neq 0$ and performs
the Fourier transform of the correlation function with respect to $\xi^-$. 
The equivalence of this representation of the parton density with the 
number density of quarks and antiquarks in the infinite--momentum frame was 
demonstrated in the chiral quark--soliton model in 
Ref.~\cite{Diakonov:1997vc}. In a similar manner, the transverse momentum 
distributions \textit{in the model} can now be represented in terms of 
correlation functions of the massive quark fields at a finite transverse 
separation $\bm{\xi}_T \neq 0$, corresponding to a space--like separation of 
the fields, $\xi^2 < 0$ (see Fig.~\ref{fig:coordinate}). 
It is straightforward to show that Eq.~(\ref{fpt_sum}) 
can formally be represented as
\be
\left. \begin{array}{cc} 
f_1^{u + d}(x, p_T) \\[4ex]
f_1^{\bar u + \bar d}(x, p_T)
\end{array} \right\} 
&=&
\int\! \frac{d^2 \xi_T}{(2\pi )^2} \; e^{i (\bm{p}_T \bm{\xi}_T)} \; 
\left\{ \begin{array}{cc} 
\tilde f_1^{u + d}(x, \xi_T) \\[4ex]
\tilde f_1^{\bar u + \bar d}(x, \xi_T) ,
\end{array} \right.
\nonumber \\[2ex]
\label{fpt_fourier}
\\[2ex]
\left. \begin{array}{cc} 
\tilde f_1^{u + d}(x, \xi_T) \\[4ex]
\tilde f_1^{\bar u + \bar d}(x, \xi_T)
\end{array} \right\}
&\equiv& \pm \frac{1}{8 \pi} \,
\int_{-\infty}^\infty\! d\xi^- \, e^{\pm ix P_N^+ \xi^-/2} 
\nonumber \\
&\times& \langle P_N| 
\bar\psi (0) \gamma^+ \psi (\xi) \; | P_N \rangle_{\xi^+ = 0} .
\nonumber
\\[2ex]
\label{fkt_corr}
\ee
Here $P_N$ is the nucleon four--momentum, with $P_N^+$ arbitrary, 
$\bm{P}_{N, T} = 0$, and $P_N^- = M_N^2/P_N^+$. The sum over quark flavors
$(u, d)$ is implied in Eq.~(\ref{fkt_corr}). Conversely, the 
coordinate--space correlation function can be obtained as the 
two--dimensional Fourier transform of the $p_T$ distribution,
\be
\tilde f_1^{u + d}
(x, \xi_T) &\equiv& \int d^2 p_T \; e^{-i \bm{p}_T \bm{\xi}_T}
\; f_1^{u + d}(x, p_T) 
\nonumber
\\[1ex]
&=& 2\pi \int_0^\infty \!\!\! dp_T \, p_T \, J_0 (p_T \xi_T) \,
f_1^{u + d}(x, p_T) ,
\nonumber \\
\label{f_corr_def}
\ee
and similarly for the antiquark distribution. Here $J_0$ is the 
cylindrical Bessel function. In particular, the value at $\xi_T = 0$ 
corresponds to the total ($p_T$--integrated) parton density,
\be
\tilde f^{u + d} (x, \xi_T = 0)  
&=& f^{u + d} (x) ,
\nonumber
\\[0ex]
\tilde f^{\bar u + \bar d} (x, \xi_T = 0)  
&=& f^{\bar u + \bar d} (x) .
\ee
Here as before in Sec.~\ref{subsec:definition}, it is assumed that the
integral over transverse momenta in the model converge; the role of
the UV cutoff in the coordinate--space correlation function
will be discussed in Sec.~\ref{subsec:correlator}.
%
%

We emphasize that we regard Eq.~(\ref{fkt_corr})
strictly as a definition within the effective model and make no attempt to 
relate the bilinear quark operator in massive quark fields to any QCD 
operators with finite transverse separation. As discussed in 
Sec.~\ref{sec:introduction}, the proper definition of such operators 
in QCD faces considerable challenges (gauge invariance and path dependence, 
rapidity divergences, renormalization properties) and is the subject of
on--going research. In our dynamical model the operator with finite 
transverse separation of the fields is well--defined, and its nucleon
matrix element can be calculated within the $1/N_c$ expansion.

Finally, the corresponding coordinate--space correlation functions for 
the flavor--nonsinglet polarized distribution are
\be
\left. \begin{array}{cc} 
\tilde g_1^{u + d}(x, \xi_T) \\[4ex]
\tilde g_1^{\bar u + \bar d}(x, \xi_T)
\end{array} \right\}
&\equiv& \frac{1}{8\pi} \,
\int_{-\infty}^\infty\! d\xi^- \, e^{ix P_N^+ \xi^-/2}
\nonumber \\
&\times& \langle P_N| 
\bar\psi (0) \tau^3 \gamma^+ \gamma_5 \psi (\xi) \; | P_N \rangle_{\xi^+ = 0} .
\nonumber
\\[2ex]
\ee
Their relation to the $p_T$ distributions is analogous to the unpolarized
case, cf.\ Eq.~(\ref{fpt_fourier}).
\section{Valence quark transverse momentum}
\label{sec:valence}
\subsection{Transverse momentum distribution}
Using the expressions derived in Sec.~\ref{subsec:evaluation} we now 
want to calculate the transverse momentum distributions of quarks and 
antiquarks in the chiral quark--soliton model and study their behavior.
It will be convenient to discuss separately the ``valence'' 
(charge--nonsinglet, or quark minus antiquark) and the ``sea'' (antiquark) 
distributions, as the two receive contributions from different classes of
configurations in the nucleon wave function and consequently show 
different behavior, especially at large values of $p_T$. In this
section we study the valence quark transverse momentum distributions;
the sea quark distributions will be considered in Sec.~\ref{sec:sea}.

\begin{figure}
\includegraphics[width=.48\textwidth]{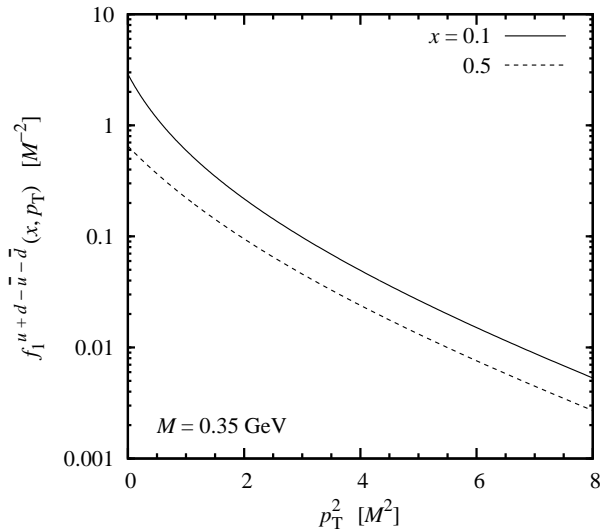}
\caption{Transverse momentum distribution of flavor--singlet valence quarks,
$f_{1, {\rm val}}^{u + d} (x, p_T)$, Eq.~(\ref{fkt_val}), as a function 
of $p_T^2$, at $x = 0.1$ (solid line) and $0.5$ (dotted line). A Gaussian 
distribution would correspond to a straight line in the coordinates of 
this plot. [Self--consistent soliton profile Eq.~(\ref{P_self}) with 
$M = 0.35\, \textrm{GeV}$ and $M_N = 3.26 \, M$.]}
\label{fig:f1_val}
\end{figure}

The flavor--singlet unpolarized valence quark density
\beq
f^{u + d - \bar u - \bar d}_{1}(x) \;\; \equiv \;\;
f^{u + d}_1(x) - f^{\bar u + \bar d}_1(x)
\label{fkt_val}
\eeq
in the chiral quark--soliton model is dominated by the contribution
of the discrete bound--state level in the quark single--particle spectrum. 
The contribution of the Dirac continuum to the valence quark density was 
calculated in Ref.~\cite{Diakonov:1996sr} and found to be $\lesssim 5\%$ of
the total at non--exceptional values of $x$; its integral over $x$ is 
zero, as the discrete level occupied by $N_c$ quarks already accounts 
for the entire baryon number of the nucleon. The transverse momentum 
distribution of valence quarks is therefore also dominated by the 
bound--state level, and we can 
neglect the Dirac continuum contribution in our numerical study.
The explicit form of the bound state wave function of the 
discrete level is given in Appendix~\ref{app:level} (see also 
Appendix~B of Ref.~\cite{Diakonov:1996sr}). 
Evaluating the discrete level term in the quark distribution given by
Eq.~(\ref{fpt_sum}), and the antiquark
distribution given by Eq.~(\ref{fpt_sum_alt}), 
we obtain
\be
\left. \begin{array}{cc} 
f_{1, {\rm lev}}^{u + d}(x, p_T) \\[4ex]
f_{1, {\rm lev}}^{\bar u + \bar d}(x, p_T)
\end{array} \right\} 
&=& \pm \frac{N_c M_N}{4 \pi p^2}
      \left[ h^2(p) + j^2(p) \phantom{\frac{0^0}{0}} \right. \nonumber \\
&-& \left. \frac{2p^3}{p} \, h(p) \, j(p) \right] 
\label{level_f1}
\ee
\be
&& \left[ p^3 \equiv \pm xM_N - E_{\rm lev} , \;
p \equiv |\bm{p}| = \sqrt{\bm{p}_T^2+(p^3)^2} \right] .
\nonumber 
\ee
Here $h(p)$ and $j(p)$ are the radial wave functions of the bound--state
level in momentum representation as defined in Eq.~(\ref{level_h_j_momentum}).
The valence quark $p_T$ distribution in our approximation is then given 
by [cf.\ Eq.~(\ref{fkt_val})]
\beq
f^{u + d - \bar u - \bar d}_{1}(x, p_T) \;\; = \;\;
f^{u + d}_{1, {\rm lev}}(x, p_T) - f^{\bar u + \bar d}_{1, {\rm lev}}(x, p_T) .
\label{approx_val}
\eeq
We evaluate the $p_T$ distributions with $M = 0.35\, \textrm{GeV}$, 
the self--consistent soliton profile described in Appendix~\ref{app:profile}, 
Eqs.~(\ref{P_self})--(\ref{P_self_end}), $E_{\rm lev} = 0.466\, M$ as obtained
from Eq.~(\ref{level_equation}), and $M_N = 3.26 \, M$. As always, we put 
$N_c = 3$ when using large--$N_c$ expressions such as Eq.~(\ref{level_f1}) 
as a numerical approximation. The resulting $p_T$ distributions are 
shown in Fig.~\ref{fig:f1_val} for two values of $x$. One sees that the 
valence distributions are concentrated at momenta of the order 
$p_T^2 \sim M^2$, which corresponds to the inverse spatial size 
of the mean field; cf.\ Appendix~\ref{app:profile} and Fig.~\ref{fig:prof}.
The distributions are roughly of Gaussian shape. Such behavior is 
obtained in a variety of bound--state models with fixed particle number 
(see Ref.~\cite{Schweitzer:2010tt} for a recent discussion) and seems
to be a generic consequence of the spatial localization of the 
bound--state wave function. 
%
%

The transverse momentum integral of the valence quark distribution
is UV--finite and does not require a cutoff. Because the $p_T$ 
distribution is numerically concentrated at low values 
$p_T^2 \sim \textrm{few} \, M^2$ we do not consider
the effect of possible UV regularizations on the valence quark $p_T$ 
distribution. Indeed, there are physical reasons why the $p_T$--integrated 
valence quark density should not be regularized. Leaving it
unregularized ensures that the baryon sum rule is satisfied in the
model, as the baryon number resulting from the bound--state level 
contribution is unity and that from the Dirac continuum integrates 
to zero \cite{Diakonov:1996sr,Diakonov:1997vc}. It also ensures
that the momentum sum rule is satisfied in the model
\cite{Diakonov:1996sr,Weiss:1997rt}, i.e., that the
valence and sea quark distributions together carry the entire
light--cone momentum of the nucleon --- a property of central importance 
for the interpretation of the model parton distributions and their matching
with QCD (cf.\ the discussion in Sec.~\ref{subsec:matching}).

Using the fact that $|p^3| \leq p$ it is easily seen that the expression 
in brackets in Eq.~(\ref{level_f1}) is positive, irrespectively of the 
sign of the radial wave functions. The discrete level contribution to the 
unpolarized quark and antiquark $p_T$ distributions, Eq.~(\ref{level_f1}), 
thus satisfies
\beq
f_{1, {\rm lev}}^{u + d}(x, p_T) \; > \; 0, \hspace{2em}
f_{1, {\rm lev}}^{\bar u + \bar d}(x, p_T) \; < \; 0 ,
\eeq
for all $x$ and $p_T$. For the quark distribution this in accordance
with the positivity condition Eq.~(\ref{positivity}). The antiquark
$p_T$ distribution in the model includes the positive Dirac continuum 
contribution in addition to the discrete level contribution
Eq.~(\ref{level_f1}) and is also positive (see Sec.~\ref{sec:sea}).
This also guarantees positivity of the total ($p_T$--integrated) 
antiquark density, as discussed in Ref.~\cite{Diakonov:1996sr}.
\subsection{Average transverse momentum}
%
%
\begin{figure}
\includegraphics[width=.47\textwidth]{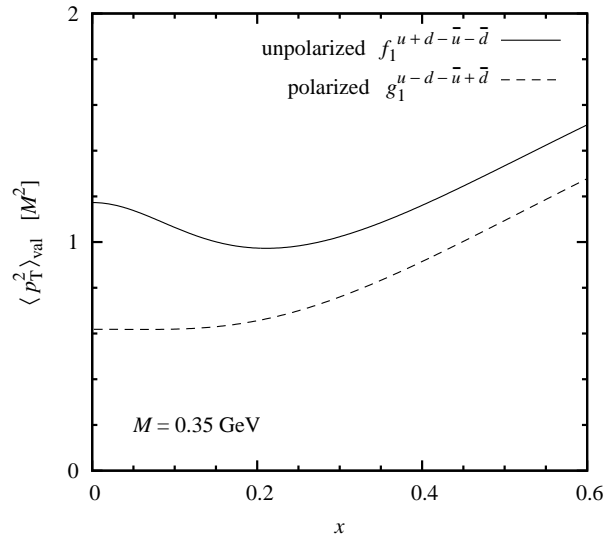}
\caption{Average transverse momentum squared of the valence quark
distribution, $\langle p_T^2 \rangle_{\rm val}(x)$, 
Eq.~(\ref{Eq:av-pT2-sea}), as a function of $x$. 
Solid line: Flavor--singlet unpolarized distribution
$f_{1}^{u + d - \bar u - \bar d}$. Dashed line: Flavor--nonsinglet 
polarized distribution, $g_{1}^{u - d - \bar u + \bar d}$.}
\label{fig:ptav}
\end{figure}
Because the distribution of valence quarks decreases rapidly
at large transverse momenta, one may consider the average 
of $p_T^2$ at fixed $x$ as a measure of the overall width of 
the distribution,
\beq
\langle p_T^2 \rangle_{\rm val} (x)
\;\; \equiv \;\; 
\frac{\displaystyle \int d^2p_T\;p_T^2 \; 
f_{1}^{u+d - \bar u - \bar d}(x,p_T)}
     {\displaystyle \int d^2p_T\;      f_{1}^{u+d - \bar u - \bar d}(x,p_T)} .
\label{Eq:av-pT2-sea}
\eeq
Figure~\ref{fig:ptav} (solid line) shows this quantity as a function 
of $x$, as obtained within our approximation. One sees that the typical
transverse momenta in the valence quark distribution are of the order
$p_T^2 \sim M^2$, corresponding to the inverse spatial size of the 
mean field [cf.\ Eqs.~(\ref{P_self})--(\ref{P_self_end})], and that
the average varies only weakly with $x$ over a wide range, as
expected from phenomenological models \cite{Schweitzer:2010tt}.
The increase toward larger values of $x$ indicates that the spatial 
size of configurations contributing to the parton density decreases
at larger values of $x$, which is a general feature of a bound state 
with fixed particle number. Note, however, that the mean--field approximation 
employed here is justified only for non--exceptional momentum fractions 
$x \sim 1/N_c$ and cannot be used to study the large--$x$ limit of 
parton densities.
\subsection{Polarized distribution}
\label{subsec:valence_polarized}
\begin{figure}
\includegraphics[width=.48\textwidth]{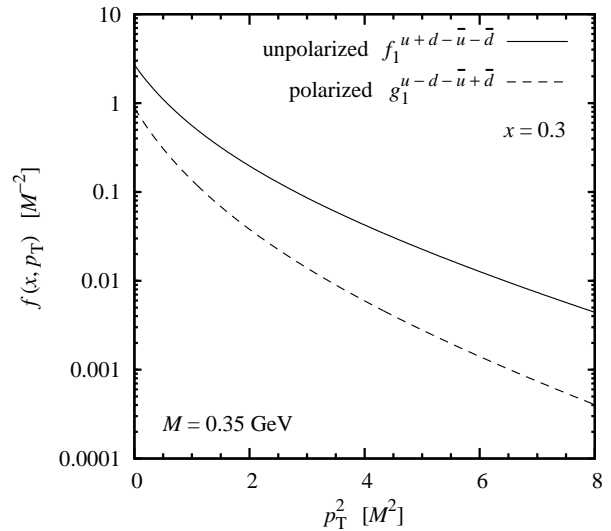}
\caption{The transverse momentum distribution of fla\-vor--nonsinglet 
polarized valence quarks, $g_{1}^{u - d - \bar u + \bar d} (x, p_T)$, 
Eq.~(\ref{gkt_val}), as a function 
of $p_T^2$, at $x = 0.3$.
Also shown for comparison is the flavor--singlet unpolarized 
valence quark distribution $f_{1}^{u + d - \bar u - \bar d} (x, p_T)$ 
at the same value of $x$, cf.\ Eq.~(\ref{fkt_val}) and Fig.~\ref{fig:f1_val}.}
\label{fig:g1_val}
\end{figure}

In a similar manner, the flavor--nonsinglet polarized valence 
quark density
\beq
g^{u - d - \bar u + \bar d}_{1}(x) \;\; \equiv \;\;
g^{u - d}_1(x) - g^{\bar u - \bar d}_1(x) ,
\label{gkt_val}
\eeq
in the chiral quark--soliton model is dominated by the contribution 
of the discrete bound state level. The level contributions to the
$p_T$ distribution of polarized quarks defined by Eq.~(\ref{gpt_sum}), 
and to the antiquark distribution defined
by Eq.~(\ref{gpt_sum_alt}), evaluate to [cf.\ Eq.~(\ref{level_f1})]
\be
\left. \begin{array}{cc} 
g_{1, {\rm lev}}^{u - d}(x, p_T) \\[4ex]
g_{1, {\rm lev}}^{\bar u - \bar d}(x, p_T)
\end{array} \right\} 
&=& \frac{N_c M_N}{3 (4\pi p^2)}
\left\{ h^2(p) + \left[ \frac{2 (p^3)^2}{p^2} - 1 \right]  
j^2(p) \right.
\nonumber \\
&-& \left. \frac{2 p^3}{p} \,h(p)\,j(p) \right\} 
\label{gpt_val_res}
\ee
\be
&& \left[ p^3 \equiv \pm xM_N - E_{\rm lev} , \;
p \equiv |\bm{p}| = \sqrt{\bm{p}_T^2+(p^3)^2} \right] .
\nonumber 
\ee
The resulting distribution is positive and similar in shape
to the unpolarized distribution (see Fig.~\ref{fig:g1_val}, dashed line). 
The somewhat faster decrease of the polarized distribution at
large $p_T$ is due to the lower component of the bound--state
level wave function [for $j(p) \equiv 0$ the shape would be the 
same as that of the unpolarized distribution]
and attests to the relativistic nature of the mean--field
picture of the nucleon at large $N_c$. The average $\langle p_T^2 \rangle$ 
of the polarized distribution, defined in analogy with 
Eq.~(\ref{Eq:av-pT2-sea}), is shown in Fig.~\ref{fig:ptav} (dashed line) 
as a function of $x$. One sees that it is systematically smaller than
that of the flavor--singlet unpolarized distribution, which again
is a relativistic effect \cite{foot:width}. Similar behavior is found
in relativistic bound state models of the valence quark distributions
with fixed particle number; see 
e.g.\ Refs.~\cite{Avakian:2010br,Pasquini:2008ax}.
%
%

It is interesting to note that the unpolarized and polarized quark 
distributions resulting from the bound--state level, 
Eqs.~(\ref{level_f1}) and (\ref{gpt_val_res}), 
satisfy the general large--$N_c$ inequality for the transverse 
momentum distribution, Eq.~(\ref{ineq}). Because numerically 
$g_{1,\rm lev}^{u-d}(x,p_T) > 0$, and also 
$f_{1,\rm lev}^{u+d}(x,p_T) > 0$, we may replace the absolute values 
of the distributions by the distributions themselves when testing the
inequality. If we then form the difference between the left-- and 
right--hand sides of Eq.~(\ref{ineq}) with the expressions 
Eqs.~(\ref{level_f1}) and (\ref{gpt_val_res}), we obtain 
\be
&& f_{1,\rm lev}^{u+d}(x,p_T) \; - \; 3 g_{1,\rm lev}^{u-d}(x,p_T) 
\nonumber \\[1ex]
&=& \frac{2 N_c M_N}{(4\pi p^2)}
\left[ 1 - \frac{(p^3)^2}{p^2} \right] \, j^2 (p) ,
\ee
which is manifestly positive because $|p^3| \leq p$.
The inequality for the corresponding sea quark distributions,
which are dominated by the contribution of the Dirac continuum
of single--particle quark states, is discussed in 
Sec.~\ref{subsec:polarized_sea}.
\section{Sea quark transverse momentum}
\label{sec:sea}
\subsection{Gradient expansion}
\label{subsec:gradient}
We now turn to the transverse momentum distributions of sea quarks in
the chiral quark--soliton model. The sea quark distributions 
receive contributions from a broad range of quark single--particle states
extending up to the cutoff scale. Our first concerns are 
to study how the distributions of sea quarks behave at large $p_T$, 
how they are affected by the UV cutoff, and how they can be 
regularized in a way that satisfies basic physical requirements 
(longitudinal momentum conservation, charge conservation, analyticity) 
and provides distributions with a clear physical interpretation. 
We can then compute the sea quark $p_T$ distributions numerically
and compare them to those of the valence quarks. For simplicity we consider 
first the distribution of unpolarized flavor--singlet sea quarks; 
the flavor--nonsinglet polarized sea will be treated summarily 
in Sec.~\ref{subsec:polarized_sea}.

A powerful analytic method for evaluating the sea quark densities
in the chiral quark--soliton model is the gradient expansion,
in which one expands the quark Green function in powers of derivatives 
of the classical chiral field \cite{Diakonov:1996sr}. Here we adapt
this method to study the $p_T$ distributions. The leading--order 
gradient expansion contains the (exact) leading power behavior of 
the $p_T$ distributions at large $p_T$, which reveals the role of 
dynamical chiral symmetry breaking and allows us to study the
effect of the UV cutoff . The leading--order expression also provides 
us with an accurate ``interpolating'' approximation to the sea quark 
distributions at all values of $p_T$, which we use for a 
numerical estimate of the distributions in Sec.~\ref{subsec:numerical}.

To derive the gradient expansion, we start from the expression of the 
quark and antiquark $p_T$ distributions in terms of the Feynman Green 
function in the classical chiral field in the nucleon rest frame, 
Eq.~(\ref{fpt_green}). The Green function is defined
as the solution of the inhomogeneous Dirac equation 
Eq.~(\ref{inhomogeneous}). Following Ref.~\cite{Schwinger:1951nm}, 
we can regard it as the matrix element of a formal operator 
between 4--dimensional ``position eigenstates'' 
$|x) \equiv |x^0, \bm{x})$,
\be
G_F (x, y) &=& (x| \, 
[ i\hat\partial - M U^{\gamma_5}]^{-1} \, |y) ,
\ee
where $i\partial$ is the 4--dimensional ``momentum operator'' and 
$\hat\partial \equiv \gamma^\alpha\partial_\alpha$, cf.\ Eq.~(\ref{hat_def}).
Equivalently, the en\-ergy--momentum representation of the Green function
Eq.~(\ref{S_F_def}) can be expressed as the matrix element of the same 
operator between conjugate momentum eigenstates $|p) \equiv |p^0, \bm{p})$, 
with $(x|p) = e^{-ipx}$,
\be
\lefteqn{2\pi \delta (p_1^0 - p_2^0) S_F (p_1^0; \bm{p}_1, \bm{p}_2)} &&
\nonumber \\[1ex]
&=&  (p_1 | \, [ i\hat\partial - M U^{\gamma_5}]^{-1} \, | p_2) .
\ee
The gradient expansion is now performed by formally expanding the
inverse Dirac operator in gradients of the classical chiral field:
\be
\lefteqn{[i\hat\partial - M U^{\gamma_5}]^{-1}} && \nonumber \\
&=& [ D^{-1} - M (i\hat\partial U^{\gamma_5})]^{-1} \; 
(i\hat\partial + M U^{-\gamma_5}) 
\nonumber \\
&=& [ D \, + \, M D (i\hat\partial U^{\gamma_5}) D
\nonumber \\
&+& M^2 \, D (i\hat\partial U^{\gamma_5}) D (i\hat\partial U^{\gamma_5}) D
+ \ldots ] 
\nonumber \\
&\times & (i\hat\partial + M U^{-\gamma_5}), 
\label{expansion}
\ee
where
\be
D &\equiv& (-\partial^2 - M^2 + i0)^{-1}
\ee
is the free massive scalar Feynman Green function. The leading--order 
contribution to the quark and antiquark $p_T$ distributions 
Eq.~(\ref{fpt_green}) is contained in the terms of 
order $M^2$ collected after multiplying out the factors in 
Eq.~(\ref{expansion}). Their matrix elements between 4--dimensional
momentum eigenstates are calculated by inserting complete sets of 
momentum eigenstates between the operators. The basic matrix elements 
are
\be
(k_2 | U^{\pm\gamma_5}|k_1) &=& 2\pi \delta (k_1^0 - k_2^0)
\; \widetilde U_{\rm cl}^{\pm\gamma_5} (\bm{k}_2 - \bm{k}_1) ,
\hspace{2em}
\\[1ex]
\widetilde U_{\rm cl}^{\pm \gamma_5}(\bm{k}) 
&\equiv& {\textstyle\frac{1}{2}}(1 \pm \gamma_5) 
\widetilde U_{\rm cl} (\bm{k}) 
\nonumber \\
&+& {\textstyle\frac{1}{2}}(1 \mp \gamma_5) \widetilde U_{\rm cl} 
(-\bm{k})^\dagger , \;\;
\label{U_gamma_5_k}
\ee
where
\be
\widetilde{U}_{\text{cl}} (\bm{k})
&\equiv& \int d^3 x\; e^{-i\bm{x}\bm{k}} \;
[U_{\text{cl}} (\bm{x}) - 1] 
\label{U_fourier_def}
\ee
is the Fourier transform of the static classical chiral field 
in the rest frame [the Fourier transform of $U^\dagger (\bm{x})$ 
is given by $\widetilde U(-\bm{k})^\dagger$], and
\be
(p_2 | D |p_1) &=& (2\pi )^4 \, \delta^{(4)} (p_2 - p_1) \, D(p_1) ,
\\[1ex]
D(p) &\equiv & \frac{1}{p^2 - M^2 + i0} .
\label{D_def}
\ee
The relevant part of the Green function thus becomes
\be
\lefteqn{S_F(p^0; \bm{p}, \bm{p})} 
\nonumber \\
&=& M^2 \int\!\frac{d^3 k}{(2\pi )^3} \;
[ D(p) D(p - k) \hat k 
\widetilde U_{\rm cl}^{\gamma_5} (\bm{k}) 
\widetilde U_{\rm cl}^{-\gamma_5} (-\bm{k})
\nonumber \\
&+&  D^2(p) D(p - k) \hat k 
\widetilde U_{\rm cl}^{\gamma_5} (\bm{k}) (-\hat k) 
\widetilde U_{\rm cl}^{\gamma_5} (-\bm{k}) \hat p ]
\\[1ex]
&& \left[ p \equiv (p^0, \bm{p}), \; k \equiv (0, \bm{k}) \right] ,
\nonumber
\ee
where $\hat k \equiv k^\alpha \gamma_\alpha$ as in Eq.~(\ref{hat_def}).
Substituting this expression in 
Eq.~(\ref{fpt_green}), evaluating
the trace over spinor indices, and using the symmetry of the
bilinear forms under $\bm{k} \rightarrow -\bm{k}$, we finally obtain
\be
\lefteqn{
\left. \begin{array}{cc} 
f_{1, {\rm grad}}^{u + d}(x, p_T) \\[4ex]
f_{1, {\rm grad}}^{\bar u + \bar d}(x, p_T)
\end{array} \right\}
} 
\nonumber \\[2ex]
&=& \pm \frac{4 N_c M_N M^2}{(2\pi )^3}
\int\!\frac{d^3 k}{(2\pi )^3} \; 
\, \textrm{tr}_{\rm fl} \,
[ \widetilde U_{\rm cl}(\bm{k}) \widetilde U_{\rm cl}(\bm{k})^\dagger ]
\nonumber \\
&\times& \textrm{Im} \, \int \frac{dp^0}{2\pi} \;
[ D(p - k) D(p) k^+
\nonumber \\[1ex]
&-& D(p - k) D^2(p) \, k^2 \, p^+ ]
\label{f_grad_energy}
\\[2ex]
&& \left[ p \equiv (p^0, \bm{p}_T, \pm x M_N - p^0) \right].
\nonumber
\ee
Because of the symmetry of the combined momentum integrals the 
expression on the right--hand side is actually the same in both 
cases, and one has 
\beq
f_{1, {\rm grad}}^{\bar u + \bar d}(x, p_T) \;\; = \;\; 
f_{1, {\rm grad}}^{u + d}(x, p_T)
\eeq
in leading--order gradient expansion.
In the following, for simplicity, we use the upper expression also for
the antiquark distribution. 

Equation~(\ref{f_grad_energy}) expresses
the sea quark transverse momentum distribution in terms of the Fourier 
transform of the classical chiral field in the rest frame 
and an explicitly calculable
loop integral over the free massive quark propagators, and are our
starting point for the discussion of their physical properties.
When integrating over the transverse momentum $\bm{p}_T$, 
Eq.~(\ref{f_grad_energy}) reproduces 
the gradient expansion for the parton densities of 
Refs.~\cite{Diakonov:1996sr,Diakonov:1997vc}. The latter were shown 
to be equivalent to the gradient expansion of moments of local 
twist--2 operators in the effective chiral model.

The gradient expansion Eq.~(\ref{f_grad_energy}) 
contains the leading power--like asymptotic behavior of the transverse 
momentum distribution at large momenta $p_T^2 \gg M^2$, as shown in detail 
in Sec.~\ref{subsec:pair} below.
This important property follows from chiral invariance, which dictates 
that coefficients with higher mass dimension in an expansion in $1/p_T^2$ 
necessarily involve higher derivatives of the classical chiral field
\cite{foot:uv}. The gradient expansion therefore represents an ideal 
tool to evaluate the $p_T$ distributions at large $p_T$ and study the 
role of the UV regularization.

In the context of the exact representation of the quark/antiquark
distributions as sums over quark single--particle levels, 
Eqs.~(\ref{fpt_sum}) and (\ref{fpt_sum_alt}), the gradient
expansion approximates the contribution from the Dirac continuum, 
i.e., quark scattering states distorted by the classical chiral field. 
The contribution of the discrete bound--state level with its normalizable 
wave function is not contained in the expanded Green function.
In the sea quark $p_T$ distribution the Dirac continuum contribution
is numerically dominant, especially at large momenta 
$p_T^2 \gg M^2$, and can reliably be computed using the 
gradient expansion. At lower momenta $p_T^2 \sim M^2$, as well as
in the integral over $p_T$, the contribution of the discrete level 
becomes numerically relevant. An ``interpolating'' approximation 
which includes the contribution from the bound--state level will 
be discussed in Sec.~\ref{subsec:sea_val}. In the following studies
of the large--$p_T$ behavior of the distributions and the role of
the UV cutoff we can safely neglect the level contribution and take 
the sea quark distribution as defined by the gradient expansion.
\subsection{Representation in light--cone variables}
\label{subsec:lightcone}
%
%
\begin{figure}
\includegraphics[width=.36\textwidth]{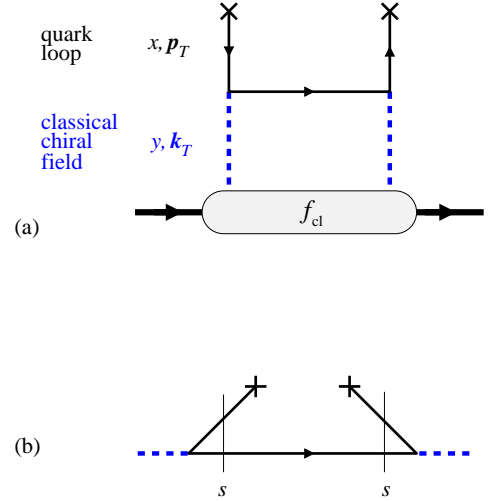}
\caption[]{(Color online) 
(a) Structure of the leading--order gradient expansion of
the sea quark distribution Eq.~(\ref{f_grad_lc}). The classical 
chiral field carries light--cone momentum fraction $y$ and transverse
momentum $\bm{k}_T$. It couples to a quark loop in which the
the antiquark carries light--cone momentum fraction $x < y$
and transverse momentum $\bm{p}_T$. (b)~Time--ordered interpretation
of the quark loop integral. The classical chiral field creates a
quark--antiquark pair with invariant mass $s$, Eq.~(\ref{s_def}). 
The UV cutoff of the chiral model can be implemented in the form of 
a cutoff in $s$, Eq.~(\ref{f_qqbar_inv}).}
\label{fig:seagrad}
\end{figure}
An interesting interpretation of the sea quark distribution is obtained 
by expressing the result of the leading--order gradient expansion 
Eq.~(\ref{f_grad_energy}) in terms of 
light--cone variables. The integral over the quark rest--frame energy 
$p^0$ in Eq.~(\ref{f_grad_energy}) can be rewritten as
\be
&&
\int\!\frac{dp^0}{2\pi} \; \{ \ldots \}_{p = (p^0, \bm{p}_T, x M_N - p^0)}
\nonumber \\
&=& 
\int\!\frac{dp^0}{2\pi} \int \frac{dp^3}{2\pi} \; 2\pi 
\delta (p^0 + p^3 - x M_N)
\; \{\ldots \}
\nonumber \\
&=& 
\frac{1}{2} \int\!\frac{dp^+}{2\pi} \int \frac{dp^-}{2\pi} \; 
2\pi \delta (p^+ - x M_N) \; \{\ldots \}
\nonumber \\
&=& 
\frac{1}{2} \int \frac{dp^-}{2\pi} \; \{\ldots \}_{p^+ = x M_N} .
\ee
We introduce the light--cone fraction $y$ of the momentum of the
static chiral field ($k^0 \equiv 0$) as
\beq
y \;\; \equiv \;\; k^3 / M_N .
\eeq
The light--cone components of the chiral field's momentum in the 
rest frame are then given by
\beq
k^+ \;\; = \;\; -k^- \;\; = \;\; k^3 \;\; = y M_N.
\eeq
Equation~(\ref{f_grad_energy}) can then be expressed as an integral 
over light--cone momentum fractions as
\be
f_{1, {\rm grad}}^{\bar u + \bar d}(x, p_T) &=& \int\!\frac{dy}{y} 
\; \int\! d^2 k_T \; f_{\rm cl} (y, \bm{k}_T) 
\nonumber
\\[1ex]
&\times& f_{q\bar q}(x, y; \bm{p}_T, \bm{k}_T) .
\label{f_grad_lc}
\ee
Here $f_{\rm cl}$ denotes the light--cone momentum distribution of the
classical chiral field of the soliton, defined as
\be
f_{\rm cl} (y, \bm{k}_T) &\equiv& \frac{F_\pi^2 M_N^2 y}{(2\pi)^3} \; 
\textrm{tr}_{\rm fl} \, [ \widetilde U_{\rm cl} (\bm{k}) 
\widetilde U_{\rm cl}(\bm{k})^\dagger ] ,
\label{f_pi_def}
\\[1ex]
&& \left[ \bm{k} = (\bm{k}_T,  y M_N)\right] .
\nonumber 
\ee
The function $f_{q\bar q}$ describes the light--cone momentum distribution
of a quark--antiquark pair,
\be
f_{q\bar q}(x, y; \bm{p}_T, \bm{k}_T) 
&\equiv& \frac{2 N_c M^2}{(2\pi)^3 F_\pi^2} 
\; \textrm{Im} \, \int\!\frac{dp^-}{2\pi} D(p - k)
\nonumber \\
&\times & 
\left[ D(p) k^+ - D^2(p) \, k^2 \, p^+ \right] \hspace{1em}
\label{f_qqbar_def}
\\[1ex]
&& (p^+ = x M_N, k^+ = - k^- = y M_N) .
\nonumber
\ee
Equation~(\ref{f_grad_lc}) permits a simple interpretation
of the nucleon's sea quark distribution in gradient expansion
(see Fig.~\ref{fig:seagrad}a). The classical chiral with 
light--cone momentum fraction $y$ and transverse
momentum $\bm{k}_T$ ``creates'' a quark--antiquark pair, of
which either the quark or the antiquark is observed with light--cone 
momentum fraction $x$ and transverse momentum $\bm{p}_T$. 
Explicit calculation below shows that $f_{q\bar q}$ is nonzero 
only if $x < y$, as required by light--front kinematics.

The gradient expansion Eq.~(\ref{f_grad_lc}) contains the leading asymptotic 
behavior of the transverse momentum distribution at large $p_T$. For our 
subsequent discussion we would like to exhibit this behavior in a simple 
form. It turns out that in the region $p_T^2 \gg M^2$ Eq.~(\ref{f_grad_lc})
and its ingredients permit further simplification. The components of the 
3--momentum of the classical chiral field in the rest frame are of the 
order of the inverse radius of the chiral field,
\be
|\bm{k}_T| &\sim& R^{-1} ,
\label{neglect_kt}
\\
k^3 \; \equiv \; y M_N &\sim& R^{-1} .
\label{neglect_k3}
\ee
The typical radius, as determined by the minimum of the classical energy 
of the soliton, is of the order $R \sim M^{-1}$
[cf.\ Appendix~\ref{app:profile}]. For transverse momenta in the region 
$p_T^2 \gg M^2$ we can therefore neglect the dependence of the quark loop 
integral in Eq.~(\ref{f_grad_lc}) on $\bm{k}_T$ and $k^-$. We suppose that 
the integral over $y$ is dominated by non--exceptional values $y \gtrsim x$, 
which will be borne out by explicit calculation (see below). Furthermore, 
we can neglect the second term in Eq.~(\ref{f_grad_energy}), as it carries 
an extra factor of $-k^2 \sim R^{-2}$. With these simplifications 
Eq.~(\ref{f_grad_lc}) takes the form
\be
f_{1, {\rm grad}}^{\bar u + \bar d}(x, p_T) &=& \int_x^\infty \!\frac{dy}{y} 
\; f_{\rm cl} (y) 
\; f_{q\bar q}(x/y, p_T) , 
\hspace{1em}
\label{f_grad_lc_simple} 
\ee
where $f_{\rm cl} (y)$ is the light--cone momentum distribution 
of the classical chiral field integrated over transverse momenta,
\be
f_{\rm cl} (y) &\equiv& \int d^2 \! k_T \; f_{\rm cl} (y, \bm{k}_T)
\nonumber \\
&=& \frac{F_\pi^2 M_N^2 y}{(2\pi)^3} \int \! d^2 k_T \;
\textrm{tr}_{\rm fl} \,
[ \widetilde U_{\rm cl}(\bm{k}) 
\widetilde U_{\rm cl}(\bm{k})^\dagger ] \hspace{2em}
\label{f_pi}
\\[1ex]
&& (k^3 = y M_N) .
\nonumber 
\ee
The function $f_{q\bar q}(z, p_T)$ describes the light--cone and 
transverse momentum distribution of the quark--an\-ti\-quark 
pair created by the classical chiral field in the collinear approximation, 
$\bm{k}_T = 0$ and $k^2 = 0$,
\be
f_{q\bar q}(z, p_T) &\equiv& \frac{2 N_c M^2}{(2\pi)^3 F_\pi^2} 
\; \textrm{Im} \, \int\!\frac{dp^-}{2\pi} D(p - k) 
\nonumber 
\\[1ex]
&\times& D(p) k^+ 
\label{f_qqbar_simple}
\\[1ex]
&& (p^+ = z k^+, \; k^- = 0, \; \bm{k}_T = 0 ) .
\nonumber
\ee
Here $z$ denotes the fraction of the chiral field's light--cone 
momentum carried by the quark in the pair,
\beq
z \;\; \equiv \;\; x/y ;
\label{z_def}
\eeq
the fraction carried by the antiquark is
\beq
\bar z \;\; \equiv \;\; 1 - z .
\eeq
We retain the quark mass $M$ in the free propagators
Eq.~(\ref{D_def}) in Eq.~(\ref{f_qqbar_simple}). Thus, formally, 
the approximation Eq.~(\ref{f_grad_lc_simple}) corresponds to 
the limit of large radius of the chiral field, $R \gg M^{-1}$, at
fixed quark mass $M$. We refer to Eq.~(\ref{f_grad_lc_simple}) as the 
collinear approximation. It has the same structure as the DGLAP evolution 
equations describing parton splitting in perturbative QCD in the 
collinear approximation. Note that the simplifications made here
--- integration over the chiral field's $k_T$ independently of $p_T$,
and neglect of $k^2$ --- are parametrically justified in the domain
$p_T^2 \gg M^2$, and that Eq.~(\ref{f_grad_lc_simple}) captures
the exact leading behavior of the model $p_T$ distribution in this region.

%
%
\begin{figure}
\includegraphics[width=.48\textwidth]{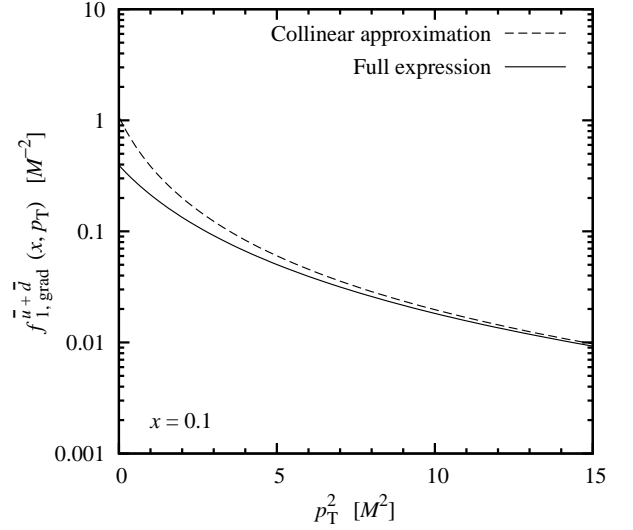}
\caption[]{Gradient expansion of the sea quark transverse momentum
distribution at $x = 0.1$. Solid line: Full expression, 
Eq.~(\ref{f_grad_lc}). Dashed line: Collinear approximation
for $p_T^2 \gg M^2$, Eq.~(\ref{f_qqbar_simple}).
[Self--consistent soliton profile Eq.~(\ref{P_self}), 
$M = 0.35\, \textrm{GeV}, M_N = 3.26 \, M$.]}
\label{fig:fpt_app}
\end{figure}
In Fig.~\ref{fig:fpt_app} we compare the $p_T$ distributions
obtained from the full expression Eq.~(\ref{f_grad_lc}) and the
collinear approximation Eq.~(\ref{f_qqbar_simple}) at a typical 
values $x = 0.1$ (the distributions shown here were regularized 
with a Pauli--Villars cutoff, described in Sec.~\ref{subsec:cutoff}
below). One sees that the collinear approximation accurately reproduces 
the full gradient expansion with better than $30\%$ accuracy above
$p_T^2 \sim 5 \, M^2$. In the following investigations of the 
structure of the distributions at large $p_T$ and the role of the
UV cutoff we can therefore use the simpler collinear expression 
Eq.~(\ref{f_grad_lc}); for numerical estimates at finite $p_T$ 
we use shall use the full expression Eq.~(\ref{f_grad_lc}).
\subsection{Momentum distribution of classical field}
\label{subsec:momentum_classical}
\begin{figure}
\includegraphics[width=.47\textwidth]{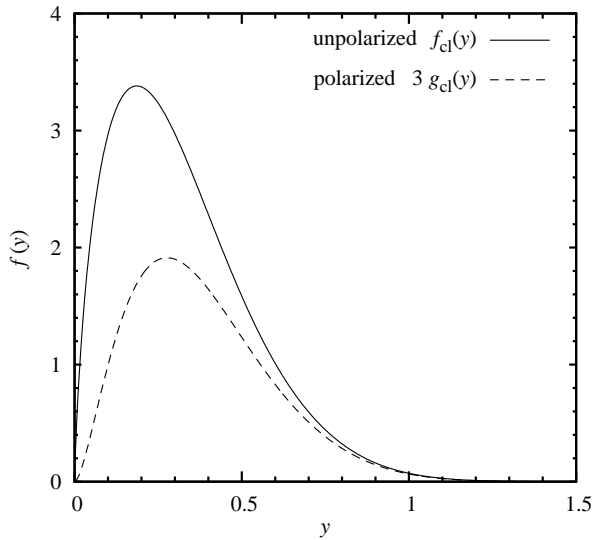}
\caption[]{Light--cone momentum distributions of the classical chiral
field in the large--$N_c$ nucleon. Solid line: Flavor--singlet unpolarized 
distribution $f_{\rm cl} (y)$, Eq.~(\ref{f_pi}). 
as appears in the convolution integral Eq.~(\ref{f_grad_lc_simple}).
The distribution shown here corresponds to the self--consistent soliton 
profile Eq.~(\ref{P_self}), with $M = 0.35 \, \textrm{GeV}$ and 
$M_N = 3.26 \, M$. Dashed line: Flavor--nonsinglet polarized 
distribution $g_{\rm cl} (y)$, Eq.~(\ref{g_cl}), as appears in 
Eq.~(\ref{g_grad_lc_simple}).
The plot shows $3 \, g_{\rm cl} (y)$, which is constrained by the
positivity condition Eq.~(\ref{positivity_classical}).}
\label{fig:fy}
\end{figure}

We first consider the light--cone momentum distribution of the classical 
chiral field, defined by Eq.~(\ref{f_pi}). It can be evaluated using 
the explicit expression given in Eq.~(\ref{utilde_trace}) in 
Appendix~\ref{app:fourier} and the soliton profile parametrization 
Eqs.~(\ref{P_self})--(\ref{P_self_end}) in Appendix~\ref{app:profile}. 
The numerical distribution is shown in Fig.~\ref{fig:fy} (solid line). 
One sees that the light--cone momentum distribution of the chiral
field extends over a broad range of momentum fractions 
$y \sim 0.1-0.5$. It is \textit{not} limited to values $y \ll 1$, 
as would be the case for single pions emitted by valence quarks, and in 
this sense reflects the complex interactions in the nucleon in the 
mean--field approximation (cf.\ our discussion of parton correlations in 
Sec.~\ref{subsec:matching} below). Note that in this classical approach 
based on large--$N_c$ limit the chiral field's light--cone momentum 
fraction $y$ is not limited to values $y < 1$; however, the distributions 
become exponentially small in the limit 
$y \rightarrow \infty$ \cite{Diakonov:1996sr}.
Comparison of the results obtained with the self--consistent
soliton profile Eq.~(\ref{P_self}) and the variational profile 
Eq.~(\ref{P_var}) shows that the bulk distributions are not sensitive 
to the details of the profile and represent stable features of the model 
\cite{Strikman:2003gz}.
%
%
\subsection{Structure of quark--antiquark pair}
\label{subsec:pair}
We now turn to the light--cone momentum distribution of the quark--antiquark 
pair in the gradient expansion of the sea quark distribution, defined by 
Eq.~(\ref{f_qqbar_def}). We first study it without 
the UV cutoff; the implementation of the cutoff will be considered 
in the following subsection. The integral over $p^-$ can
be calculated straightforwardly by closing the integration contour in
the complex plane. The poles in $p^-$ are located at
\be
p^- - k^- &=& \frac{(\bm{p}_T - \bm{k}_T)^2 + M^2 - i0}{p^+ - k^+} ,
\label{pole_pminusk}
\\
p^- &=& \frac{\bm{p}_T^2 + M^2 - i0}{p^+} .
\ee
The integral is nonzero only if they lie on opposite sides of the real 
axis, which requires $k^+ > |p^+|$. In terms of the variable
\beq
z \;\; = \;\; p^+ / k^+ ,
\eeq
which measures the fraction of the pair's light--cone momentum 
carried by the quark, cf.\ Eq.~(\ref{z_def}), this condition implies
$|z| < 1$. Because $x > 0$ we can limit ourselves to $z > 0$, and thus
$0 < z < 1$. We close the contour around the pole Eq.~(\ref{pole_pminusk}) 
in the upper half plane. At the pole, the virtuality of the active quark is
\be
p^2 - M^2  &\equiv& t - M^2
\nonumber \\
&=& -\frac{1}{\bar z} [(\bm{p}_T - z \bm{k}_T)^2 + M^2 - z\bar z k^2] ,
\hspace{2em}
\label{t_pole}
\ee
where we have used that $k^+k^- = k^2 + \bm{k}_T^2$. The integral 
in Eq.~(\ref{f_qqbar_def}) then becomes
\be
f_{q\bar q}(x, y; \bm{p}_T, \bm{k}_T) 
&=& \frac{2 N_c M^2}{(2\pi)^3 F_\pi^2} \;
\frac{\Theta (z) \Theta (\bar z)}{\bar z}
\nonumber \\
&\times& \left[ \frac{1}{M^2 - t} + \frac{z k^2}{(M^2 - t)^2} \right] .
\hspace{2em}
\label{f_qqbar_full}
\ee

We can equivalently express the quark--antiquark momentum distribution 
in terms of the variables of light--front time--ordered perturbation 
theory (see Fig.~\ref{fig:seagrad}b). In this approach light--front
3--momenta (i.e., the plus and transverse components of the 4--momentum) 
are conserved and intermediate particles are on mass shell, but the 
light--front energy (i.e., the minus component of the 4--momentum) of 
the intermediate state is generally different from that of the initial 
state, resulting in nonconservation of 4--momentum. With the 
$(+, -, \perp)$ components of the 4--momenta of the quark and antiquark
given by
\be
p_1 &=& \left( z k^+, \; \frac{\bm{p}_T^2 + M^2}{z k^+}, \; \phantom{-}
\bm{p}_T \right) ,
\\
p_2 &=& \left( \bar z k^+, \; 
\frac{(\bm{p}_T - \bm{k}_T)^2 + M^2}{\bar z k^+}, \; 
\bm{k}_T - \bm{p}_T \right) ,
\hspace{2em}
\ee
the invariant mass of the quark--antiquark pair is
\be
s &\equiv& (p_1 + p_2)^2 
\nonumber \\[1ex]
&=& \frac{(\bm{p}_T - z \bm{k}_T)^2 + M^2}{z\bar z} .
\label{s_def}
\ee
This corresponds to the invariant mass of a pair with zero 
overall transverse momentum, subjected to a transverse boost by $\bm{k}_T$.
Thus, we find the following simple relation between the active quark 
virtuality in the invariant approach and the invariant mass difference
in the time--ordered approach:
\beq
t - M^2 \;\; = \;\; -z (s - k^2) .
\eeq
The quark--antiquark distribution Eq.~(\ref{f_qqbar_full})
then takes the form
\be
f_{q\bar q}(x, y; \bm{p}_T, \bm{k}_T) 
&=& \frac{2 N_c M^2}{(2\pi)^3 F_\pi^2} \;
\frac{\Theta (z) \Theta (\bar z)}{z \bar z}
\nonumber
\\
&\times& \left[ \frac{1}{s - k^2} + \frac{k^2}{(s - k^2)^2} \right]
\nonumber \\
&=& \frac{2 N_c M^2}{(2\pi)^3 F_\pi^2} \;
\frac{\Theta (z) \Theta (\bar z)}{z \bar z} \; \frac{s}{(s - k^2)^2} .
\hspace{3em}
\ee

In the collinear approximation Eq.~(\ref{f_qqbar_simple}), 
if we neglect the overall transverse momentum $\bm{k}_T$ and the 
virtuality $k^2$ according to Eqs.~(\ref{neglect_kt})--(\ref{neglect_k3}), 
the quark--antiquark momentum distribution becomes
\be
f_{q\bar q}(z, p_T) &=& \frac{2 N_c M^2}{(2\pi)^3 F_\pi^2} \;
\frac{\Theta (z) \Theta (\bar z)}{p_T^2 + M^2} .
\label{f_qqbar_unreg}
\ee
This result has several interesting features. First, the transverse 
momentum distribution in the pair exhibits a power--like ``tail'' at 
large values,
\beq
f_{q\bar q}(z, p_T) \;\; \sim \;\;  \frac{2 N_c M^2}{(2\pi)^3 F_\pi^2} \;
\frac{\Theta (z) \Theta (\bar z)}{p_T^2} \hspace{3em} (p_T^2 \gg M^2) .
\label{tail_pair}
\eeq
Through Eq.~(\ref{f_grad_lc_simple}) it produces a similar tail in 
the nucleon's sea quark transverse momentum distribution; this feature
will be discussed in detail in Sec.~\ref{subsec:tail} below.
Second, the $\sim 1/p_T^2$ tail results in a would--be logarithmic 
divergence of the quark--antiquark density in the pair when integrated 
over $p_T$ up to the cutoff scale. The coefficient of this logarithmic
divergence is the same as that of the logarithmic divergence of
$F_\pi^2$ in the effective theory,
\be
F_\pi^2 \;\; \sim \;\; \frac{N_c M^2}{(2\pi)^2} \log \frac{\Lambda^2}{M^2}
\hspace{2em} (\Lambda^2 \gg M^2).
\ee
As a consequence, in the limit $\Lambda \rightarrow \infty$ the 
quark--antiquark distribution in the pair is normalized such that
\be
\int_{\Lambda^2} \! 
d^2 p_T \; f_{q\bar q}(z, p_T) \;\; = \;\; 1
\hspace{2em} (\Lambda^2 \rightarrow \infty).
\ee
We shall require that this condition be satisfied also for finite values 
of the cutoff when we introduce the UV regularization in 
Sec.~\ref{subsec:cutoff}.

We note that in Eq.~(\ref{f_qqbar_unreg}) the quark--antiquark 
momentum distribution in the pair is actually independent of the 
momentum fraction $z$. This is because at this level of approximation
there is no restriction on the constituent quarks' virtuality or 
invariant mass. A nontrivial $z$--dependence of the distribution 
will appear when a cutoff is introduced.

\subsection{Power--like tail at large momenta}
\label{subsec:tail}
Through the convolution formula of Eq.~(\ref{f_grad_lc_simple}) the
$1/p_T^2$ tail in the momentum distribution of the quark--antiquark
pair, Eq.~(\ref{tail_pair}), causes a similar behavior in 
the nucleon's sea quark transverse momentum distribution:
\be
f_{1, {\rm grad}}^{\bar u + \bar d}(x, p_T) 
&\sim& \frac{C_{f_1}^{\bar u + \bar d}(x)}{p_T^2} \hspace{2em} 
(p_T^2 \gg M^2) ,
\label{tail_nucleon_general}
\\[1.5ex]
C_{f_1}^{\bar u + \bar d}(x) 
&=& \frac{2 N_c M^2}{(2\pi)^3 F_\pi^2} \; \int_x^\infty \frac{dy}{y}
\, f_{\rm cl}(y) .
\label{tail_nucleon_coeff}
\ee
The $p_T$ distribution of sea quarks thus has a power--like behavior at
momenta $p_T^2 \gg M^2$. This fact is of central importance and implies
that the distribution of sea quarks is qualitatively different
from that of valence quarks (see Sec.~\ref{sec:valence}).
It is a direct consequence of dynamical chiral symmetry
breaking and shows the imprint of the QCD vacuum on the nucleon's
partonic structure. Note that the coefficient of the tail is determined 
by the effective chiral dynamics \textit{at the scale} $\sim M$ and can be
computed without explicit reference to the UV cutoff
of the effective theory. In fact, the ratio $M/F_\pi$ appearing
in Eqs.~(\ref{tail_pair}) and (\ref{tail_nucleon_coeff})
is just the coupling constant of the massive constituent quarks to 
the chiral field, as it follows from expanding the effective 
interaction Eq.~(\ref{L_eff}), cf.\ Sec.~\ref{subsec:constituent_quarks}. 
A corresponding interpretation of the tail as due to quark--antiquark 
pairs created by the classical chiral field will be developed in 
Sec.~\ref{sec:correlations}. 

Parametrically, the power--like tail in the sea quark $p_T$ distribution
extends up to the UV cutoff scale $\Lambda^2$. It is interesting 
to note that the coefficient of the tail related to the would--be 
logarithmic divergence of the $p_T$--integrated parton density and can be 
recovered as the derivative of the $p_T$--integrated parton density with 
respect to the upper limit of the $p_T$ integral, in formal analogy to the 
relation for the unintegrated parton density in perturbative QCD, 
Eqs.~(\ref{tail_QCD}) and (\ref{tail_coeff}). In fact, the coefficient
Eq.~(\ref{tail_nucleon_coeff}) is nothing but the $p_T$ integrated
parton density in gradient expansion,
\beq
C_{f_1}^{\bar u + \bar d}(x) 
\;\; = \;\; f_{1, {\rm grad}}^{\bar u + \bar d}(x) .
\label{tail_nucleon_gradient}
\eeq
However, we caution that 
Eqs.~(\ref{tail_nucleon_general})--(\ref{tail_nucleon_gradient})
apply with logarithmic accuracy only, and that the numerical values 
of the $p_T$ distribution for finite cutoff are strongly affected
by the cutoff. In Sec.~\ref{subsec:cutoff} we formulate the physical
conditions under which the UV regularization should be implemented.
and show that the high--momentum tail of the sea quark distribution 
--- albeit in numerically modified form --- is indeed a robust feature 
of the model. 
\subsection{Implementation of ultraviolet cutoff}
\label{subsec:cutoff}
To proceed further with our study of sea quark transverse momentum
distributions we now have to discuss the implementation of the
UV cutoff of the model and its effect on the distributions. This 
will allow us not only to evaluate the distributions quantitatively, 
but also to integrate them over $p_T$ and relate them to the 
total parton densities. To study the effects of the UV cutoff
we use the gradient expansion in the collinear approximation,
Eq.~(\ref{f_grad_lc_simple}), which captures the leading behavior
of the unregularized distributions at large $p_T$ and allows
us to illustrate the essential points in analytic form.

As explained in Sec.~\ref{sec:introduction}, the manner in which the 
cutoff is implemented \textit{defines} the effective degrees of freedom 
of the model and must be based on physical considerations going
beyond the intrinsic structure of the effective chiral theory. 
Here we \textit{require} that the regularization procedure satisfy 
the following conditions:
\begin{itemize}
\item[I)] The regularized distribution should preserve the symmetry of 
quarks and antiquarks in the pair:
\beq
f_{q\bar q} (z, p_T)_{\rm reg} \;\; = \;\; 
f_{q\bar q} (\bar z, p_T)_{\rm reg} .
\eeq
Exchange of quark and antiquark amounts to exchanging $z \rightarrow 
\bar z = 1 - z$ and $\bm{p}_T \rightarrow -\bm{p}_T$; because the 
unpolarized distribution is a function only of $p_T \equiv |\bm{p}_T|$ 
the latter change is trivial.
\item[II)] The regularized distribution should be normalized such that the
total number of quarks and antiquarks in the pair is unity, and that they 
carry the entire longitudinal momentum of the chiral field.
This implies that the $p_T$--integrated distribution
\beq
f_{q\bar q} (z)_{\rm reg} \;\; \equiv \;\; 
\int\! d^2 p_T \; f_{q\bar q} (z, p_T)_{\rm reg}
\label{f_z_def}
\eeq
satisfy
\be
\int_0^1 \! dz \; f_{q\bar q} (z)_{\rm reg} &=& 1,
\label{cond_norm}
\\
\int_0^1 \! dz \; z\; f_{q\bar q} (z)_{\rm reg} &=& \frac{1}{2}.
\label{cond_z}
\ee
Because of the symmetry with respect to $z \rightarrow 1 - z$ the two
requirements are actually identical. Namely,
\be
\int_0^1 \! dz \; z\; f_{q\bar q} (z)_{\rm reg} &=&
\int_0^1 \! dz \; \bar z\; f_{q\bar q} (z)_{\rm reg}
\nonumber \\
&=& 
\frac{1}{2}
\int_0^1 \! dz \; (z + \bar z) \; f_{q\bar q} (z)_{\rm reg} 
\nonumber \\
&=& 
\frac{1}{2}
\int_0^1 \! dz \; f_{q\bar q} (z)_{\rm reg} .
\ee
Physically, these conditions imply that the massive quarks and 
antiquarks are the only constituents of the nucleon's light--front
wave function up to the cutoff scale, and that there is no momentum
``hidden'' in other degrees of freedom.
\item[III)] The regularization should not change the large--distance 
behavior of the quark field correlator in coordinate space. 
This requirement will be discussed in detail in 
Sec.~\ref{subsec:correlator}, and implies
that the cutoff should not modify the analytic properties of the
$p_T$ distribution in the vicinity of the leading singularity
in $p_T$ at complex values of the order $M$, which governs 
the fall-off at large distances. In the collinear approximation
this is the pole at $p_T^2 = -M^2$ in Eq.~(\ref{f_qqbar_unreg}).
\end{itemize}
The above represents a minimal set of physical requirements based
on our present understanding; they may be amended by further conditions 
if more insights into the nature of the effective degrees of freedom 
were gained in the future. In the following we explore to what extent
these minimal requirements determine the sea quark $p_T$ distributions
quantitatively. We first present two regularization schemes that meet 
these requirements.

\textit{Pauli--Villars subtraction.}
In the Pauli--Villars (PV) regularization 
scheme \cite{Diakonov:1996sr,Diakonov:1997vc} one subtracts from the 
original distribution of pointlike quarks with mass $M$ a certain multiple 
of the analogous distribution of quarks with a regulator 
mass $M_{\rm PV} > M$, 
\be
f_{q\bar q} (z, p_T)_{\rm PV} &\equiv&
f_{q\bar q} (z, p_T | M) \nonumber \\
&-& \frac{M^2}{M_{\rm PV}^2} f_{q\bar q} (z, p_T | M_{\rm PV}) .
\ee
The coefficient is chosen such that the subtraction cancels the
logarithmic divergence associated with the integral over $p_T$.
Applying this subtraction to Eq.~(\ref{f_qqbar_unreg}) we get
\be
f_{q\bar q} (z, p_T)_{\rm PV} &=&
\frac{2 N_c M^2}{(2\pi)^3 F_\pi^2} \; \Theta (z)  \Theta (\bar z)  
\nonumber \\
&\times& \frac{M_{\rm PV}^2 - M^2}
{(p_T^2 + M^2)(p_T^2 + M_{\rm PV}^2)} .
\label{f_qqbar_pv}
\ee
The regularized distribution drops as $\sim 1/p_T^4$ at 
$p_T \rightarrow \infty$ and is integrable. At the same time,
we replace $F_\pi^2$ in the normalization factor by the
result obtained with PV regularization,
\beq
F_\pi^2 \;\; = \;\; \frac{N_c M^2}{(2\pi)^2} \log \frac{M_{PV}^2}{M^2} .
\label{fpi_pv}
\eeq
One can easily verify that with this normalization the conditions
Eqs.~(\ref{cond_norm}) and (\ref{cond_z}) are satisfied, and that
\beq
f_{q\bar q} (z)_{\rm PV} \;\; \equiv \;\; \int \! d^2 p_T \;
f_{q\bar q} (z, p_T)_{\rm PV} \;\; = \;\; 1.
\eeq
Note also that the subtraction does not change the residue of the
pole at $p_T^2 = -M^2$; i.e., Eq.~(\ref{f_qqbar_pv}) has the 
same behavior near $p_T^2 \rightarrow -M^2$ as the unregularized
expression Eq.~(\ref{f_qqbar_unreg}).
The numerical value of the regulator mass is fixed by requiring that
Eq.~(\ref{fpi_pv}) reproduce the physical value of the pion decay
constant, $F_\pi = 93\, \textrm{MeV}$. One obtains
\beq
M_{\rm PV}^2/M^2 \; = \; 2.52 
\hspace{2em}
(M = 0.35 \, \textrm{GeV}).
\label{mpv}
\eeq
%
%
\begin{figure}
\begin{tabular}{l}
\includegraphics[width=.45\textwidth]{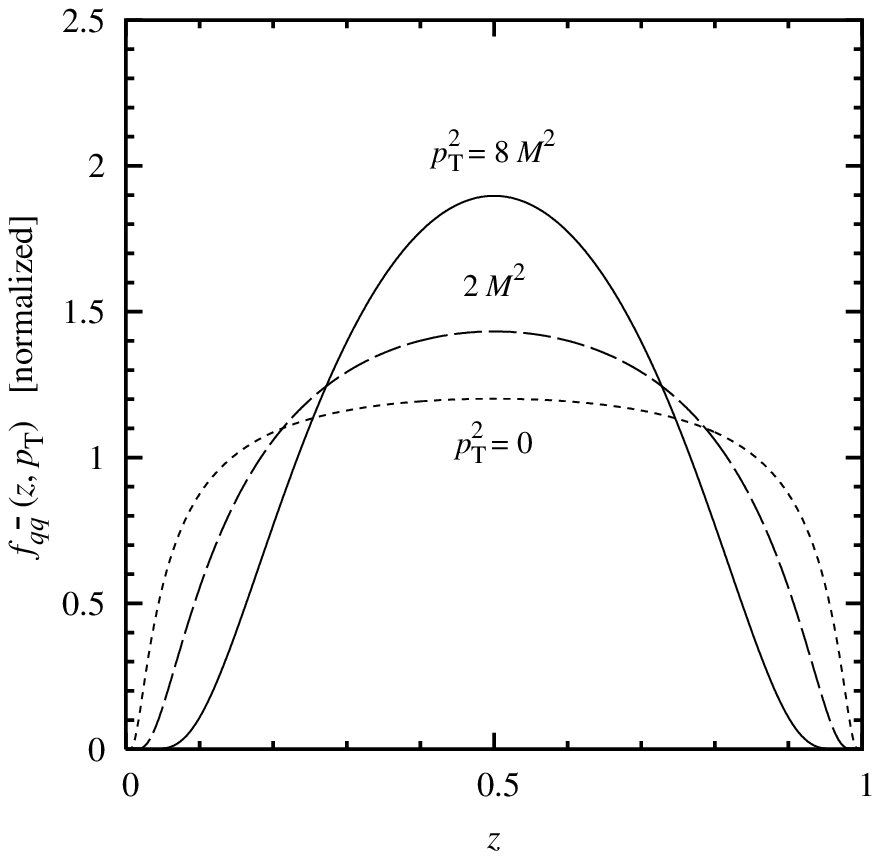}
\\[-4ex]
(a)
\\
\includegraphics[width=.45\textwidth]{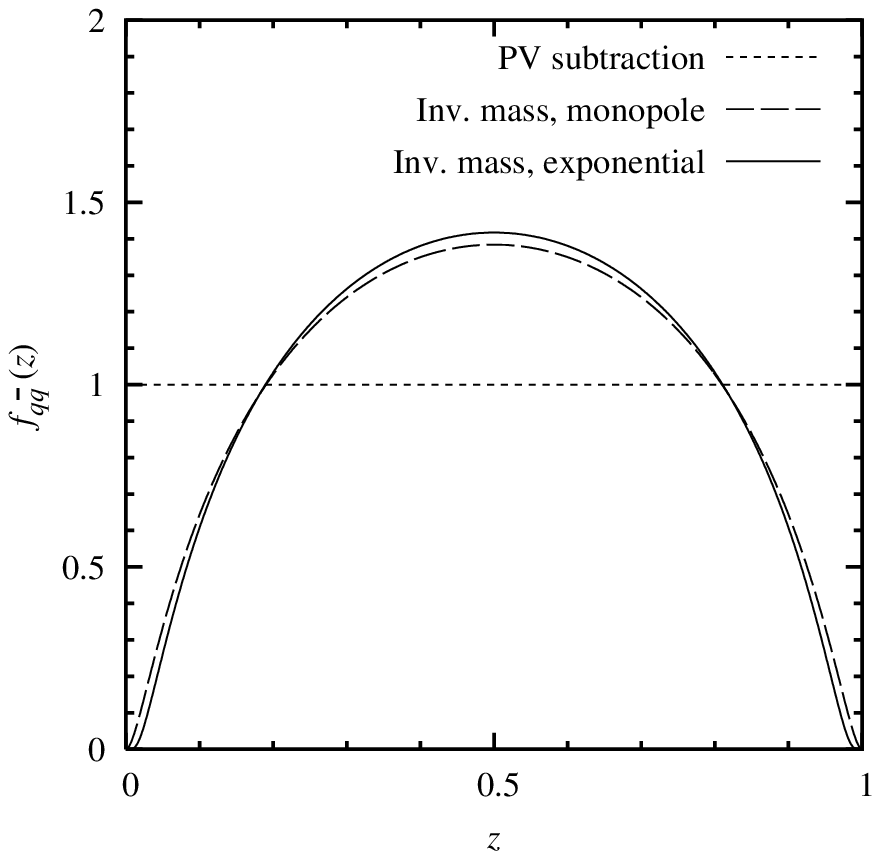}
\\[-4ex]
(b)
\end{tabular}
\caption[]{(a) Quark--antiquark light--cone  momentum distribution
$f_{q\bar q} (z, p_T)$ as function of $z$ for fixed $p_T$
[exponential invariant mass cutoff, Eq.~(\ref{ff_exp})].
The values of $p_T^2$ are indicated above/below the curves.
For the sake of comparison the distributions at all values of $p_T^2$ 
were normalized to unit integral over $z$. 
(b) The $p_T$--integrated quark--antiquark light--cone 
momentum distribution $f_{q\bar q} (z)$, Eq.~(\ref{f_z_def}), 
with different UV cutoffs.
Dotted line: PV subtraction. Dashed line: Invariant mass cutoff
by monopole form factor, Eq.~(\ref{ff_mon}). Solid line: 
Invariant mass cutoff by exponential form factor, Eq.~(\ref{ff_exp}).}
\label{fig:fz}
\end{figure}
\textit{Invariant mass cutoff.}
Another way of implementing the cutoff is to limit the invariant mass
of the quark--antiquark pair in the time--ordered approach, Eq.~(\ref{s_def})
(see Fig.~\ref{fig:seagrad}b). In the collinear approximation
the invariant mass Eq.~(\ref{s_def}) is given by
\beq
s \;\; = \;\; \frac{p_T^2 + M^2}{z\bar z} \;\; > \; 0.
\eeq
The cutoff is implemented by multiplying the vertices of the 
quark--antiquark pairs in the initial and final state with a form factor 
$F(s)$ that suppresses invariant masses of the order $s \sim \Lambda^2$. 
We consider a monopole and an exponential form factor,
\be
F(s)_{\rm mon} &=& \frac{1}{1 +  s/\Lambda^2},
\label{ff_mon}
\\[1ex]
F(s)_{\rm exp} &=& \exp (- s/\Lambda^2) .
\label{ff_exp}
\ee
The form factor is normalized to unity at the unphysical point
$ s = 0$, which corresponds to $p_T^2 = -M^2$; the significance
of this choice will be explained in Sec.~\ref{subsec:correlator}.
The regularized quark/antiquark distribution is then given by
\be
f_{q\bar q} (z, p_T)_{\rm inv} &\equiv & 
f_{q\bar q} (z, p_T) \; F^2(s) 
\nonumber
\\[1ex]
&=& 
\frac{2 N_c M^2}{(2\pi)^3 F_\pi^2} \;
\frac{\Theta (z) \Theta (\bar z) \; F^2( s)}{p_T^2 + M^2} .
\label{f_qqbar_inv}
\ee
The normalization condition now takes the form
\beq
\frac{2 N_c M^2}{(2\pi)^3 F_\pi^2} \;
\int_0^1 \! dz \; \int \! d^2 p_T \;
\frac{F^2(s)}{p_T^2 + M^2} \;\; = \;\; 1.
\label{f_qqbar_inv_norm}
\eeq
The value of $\Lambda^2$ is fixed such that Eq.~(\ref{f_qqbar_inv_norm})
is satisfied with the physical value of $F_\pi^2$. For 
$M = 0.35 \, \textrm{GeV}$ this gives
\be
\Lambda^2_{\rm mon} &=& 31.1 \, M^2 \;\; = \;\; 7.78 \times 4 M^2 , \\
\Lambda^2_{\rm exp} &=& 44.8 \, M^2 \;\; = \;\; 11.2 \times 4 M^2 .
\ee
The latter values are given as multiples of the minimum value of 
$s$ in the physical region, $4 M^2$. Figure~\ref{fig:fz}a
shows the effective $z$--distribution obtained with an invariant
mass cutoff (exponential form factor) for several values of $p_T$;
for the sake of comparison all distributions were normalized to 
unit integral over $z$. One sees that the $z$--distribution is
rather flat for $p_T = 0$ and becomes progressively more concentrated
around $z = 1/2$ as $p_T$ increases.

The $p_T$--integrated distributions in the quark--an\-ti\-quark pair,
Eq.(\ref{f_z_def}), obtained with the two regularization schemes, 
are shown in Fig.~\ref{fig:fz}b. Several features are worth noting:
(a) The light--cone momentum distribution obtained with the PV 
cutoff is independent of $z$ and given by $f_{q\bar q}(z) = 1$.
In the context of the convolution integral, Eq.~(\ref{f_grad_lc_simple}),
this turns out to be a reasonable approximation for $x$ that are 
not parametrically small (see below). (b) The light--cone momentum 
distribution obtained with the invariant mass cutoff vanishes at the 
end points $z \rightarrow 0, 1$. In the convolution integral
Eq.~(\ref{f_grad_lc_simple}), because $f_{\rm cl} (y)/y$ is finite in
the limit $y \rightarrow 0$, the vanishing of $f_{q\bar q}(z)$
at $z \rightarrow 0$ ensures that $f(x, p_T) \rightarrow 0$
for $x \rightarrow 0$, i.e., the sea quark density in the nucleon 
vanishes at $x \rightarrow 0$. (c) With both cutoff schemes the 
convolution integral for the sea quark transverse momentum 
distribution Eq.~(\ref{f_grad_lc_simple}) involves a broad distribution 
of quark momentum fractions centered around $z \sim 1/2$,
at least for non--exceptional values of $x$.

%
%
\begin{figure}
\includegraphics[width=.47\textwidth]{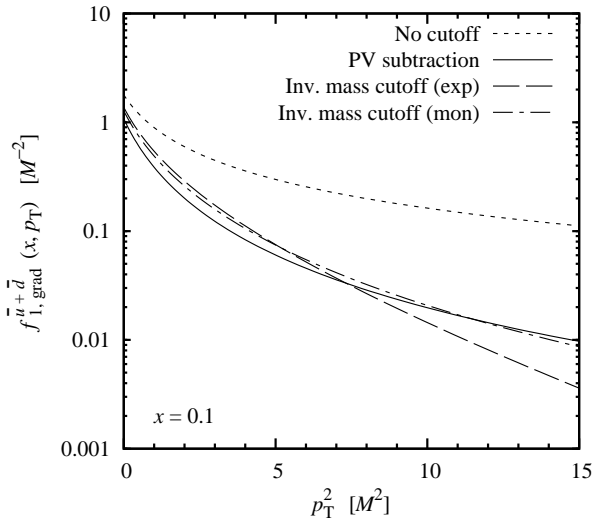}
\caption[]{Sea quark transverse momentum distribution 
$f_{1, {\rm grad}}^{\bar u + \bar d}(x, p_T)$ calculated using 
the gradient expansion in collinear approximation, 
Eq.~(\ref{f_grad_lc_simple}), as a function of $p_T^2$, at $x = 0.1$.
Shown are the distributions obtained with various UV 
cutoff schemes (see text). 
Both $p_T$ and $f_{1, {\rm grad}}^{\bar u + \bar d}$ are given in 
units of the constituent quark mass $M$. Dotted line: Unregularized 
distribution (no UV cutoff).
Solid line: PV subtraction. Dashed line: Invariant mass cutoff,
exponential form. Dash--dotted line: Invariant mass cutoff, monopole form.
[Self--consistent soliton profile Eq.~(\ref{P_self}) with 
$M = 0.35\, \textrm{GeV}$ and $M_N = 3.26 \, M$.]}
\label{fig:fpt}
\end{figure}
With all the ingredients in place, we can now calculate the sea quark
transverse momentum distributions for a finite cutoff and study their
dependence on the UV regularization. We use the gradient expansion
in the collinear approximation, Eq.~(\ref{f_grad_lc_simple}),
with the light--cone momentum distribution of the chiral field given 
by Eq.~(\ref{f_pi}) and evaluated with the soliton profile of
Eqs.~(\ref{P_self})--(\ref{P_self_end}) [cf.\ Fig.~\ref{fig:fy}].
The $p_T$ distribution in the quark--antiquark pair is now given by 
the regularized expressions Eq.~(\ref{f_qqbar_pv}) or Eq.~(\ref{f_qqbar_inv}).
The numerical results for the $p_T$ distributions at a representative
value of $x = 0.1$ are summarized in Fig.~\ref{fig:fpt}.
The $p_T$--distribution without cutoff (or, what is the same, in the 
limit of infinite cutoff), as obtained from the unregularized 
quark--antiquark distribution Eq.~(\ref{f_qqbar_unreg}), is shown by 
the short--dashed line; this distribution is not integrable over $p_T$ 
and shown for comparison only. The distributions obtained with the
PV cutoff and the invariant mass cutoff (exponential and monopole
form factor) are shown by the solid, long--dashed, and dash--dotted
lines. The results show several notable features. First, the distributions 
with any UV cutoff differ from the one without cutoff already at 
low values $p_T^2 \sim \textrm{few} \; M^2$. It shows that the 
cutoff plays an essential role in the numerical value of the $p_T$ 
distribution already at low $p_T$. This fact is not obvious from parametric 
considerations based on the hierarchy $\Lambda \gg M$, as the 
distributions at fixed $p_T$ are UV finite and thus do
not ``require'' regularization.

Second, the distributions obtained with PV subtraction and the invariant
mass cutoff (exponential and monopole) are very close up to values
$p_T^2 \approx 10 \, M^2$. In this region of $p_T$ they are determined by
generic features of the cutoff as are fixed by our general conditions.
This finding is very important, as it means that the $p_T$ distributions 
in this range are robust predictions of the model and can be discussed at 
the same level as other low--energy characteristics of the nucleon.

Third, at $p_T^2 \gtrsim 10 \, M^2$ the distributions obtained with the
different cutoffs show large differences, as expected. One important 
implication of this is that the averages $\langle p_T^2 \rangle$ differ 
substantially and do not serve as reliable measures of the width of the
bulk of the $p_T$ distribution. These averages assign disproportionate 
weight to the high--$p_T$ region where the distributions are not
constrained by our requirements on the cutoff.
\subsection{Quark field correlator in coordinate space}
\label{subsec:correlator}
Further insight into the role of the UV cutoff in the sea quark transverse 
momentum distributions can be gained by studying the corresponding 
coordinate--space correlation functions (cf.\ Sec.~\ref{subsec:coordinate}). 
The essential points can again be illustrated using the gradient expansion 
in the collinear approximation, Eq.~(\ref{f_grad_lc_simple}).
In this approximation the coordinate--space correlation function of sea
quarks, Eq.~(\ref{fkt_corr}), is given by the transverse 
Fourier transform of the convolution formula, Eq.~(\ref{f_grad_lc_simple}), 
\be
\tilde f_1^{\bar u + \bar d}(x, \xi_T) 
&=& \int_x^\infty \!\frac{dy}{y} \; f_{\rm cl} (y) 
\; \tilde f_{q\bar q}(x/y, \xi_T) , 
\hspace{2em}
\label{f_coord_grad} 
\ee
where $f_{q\bar q}(z, \xi_T) \, (z \equiv x/y)$ is the transverse Fourier 
transform of the $p_T$ distribution in the quark--antiquark pair,
\be
\tilde f_{q\bar q}(z, \xi_T) 
&\equiv& \int d^2 p_T \; e^{-i \bm{p}_T \bm{\xi}_T}
\; \tilde f_{q\bar q}(z, \xi_T) 
\nonumber
\\[1ex]
&=& 2\pi \int_0^\infty \!\!\! dp_T \, p_T \, J_0 (p_T \xi_T) \,
\tilde f_{q\bar q} (z, p_T). \hspace{2.5em}
\ee
This function describes the spatial structure of the quark--antiquark 
pairs created by the chiral field. It is interesting to see how
it behaves at small and large distances, and how its behavior is
modified by the UV cutoff of the model. In particular, the behavior
at large distances sheds new light on the regularization conditions
put forward in Sec.~\ref{subsec:cutoff}.

Without a cutoff, i.e., with the distribution Eq.~(\ref{f_qqbar_unreg}), 
the Fourier transform of the $p_T$ distribution in the $q\bar q$ pair
is given by
\be
\tilde f_{q\bar q}(z, \xi_T) &=& 2\pi \; K_0(M \xi_T).
\label{fzt_free}
\ee
At $\xi_T \rightarrow 0$ this function diverges logarithmically,
\be
\tilde f_{q\bar q}(z, \xi_T) &\sim& 2\pi \; \log \frac{1}{M \xi_T} 
\hspace{2em} (\xi_T \rightarrow 0),
\ee
which reflects the logarithmic divergence of the total ($p_T$--integrated)
parton density in the model without cutoff. With a cutoff this divergence 
is regularized. For example, with the PV subtraction the distribution becomes
\be
\tilde f_{q\bar q}(z, \xi_T)_{\rm PV} 
&=& 2\pi \; [ K_0(M \xi_T) - K_0(M_{\rm PV} \xi_T)]
\hspace{2em}
\nonumber \\
&\sim & 2\pi \; \log \frac{M_{\rm PV}}{M} \hspace{2em} (\xi_T \rightarrow 0) .
\hspace{2em}
\ee
A similar result is obtained with the invariant mass cutoff.
Thus we see that at small distances the behavior of the correlation
function is qualitatively changed by the cutoff, in accordance with the
fact that the function at $\xi_T = 0$ coincides with the total parton 
density

%
%
\begin{figure}
\includegraphics[width=.48\textwidth]{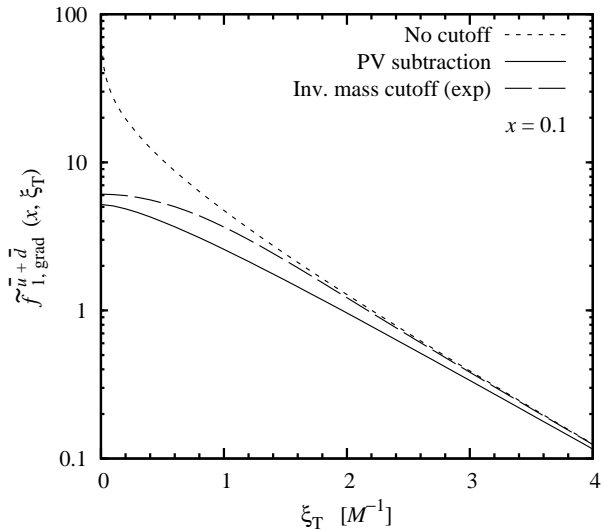}
\caption[]{Coordinate--space correlator
$\tilde f_{1, {\rm grad}}^{\bar u + \bar d}(x, \xi_T)$, corresponding
to the gradient expansion of the sea quark distribution in 
collinear approximation, Eq.~(\ref{f_coord_grad}), as a function of 
$\xi_T$, at $x = 0.1$.
Dotted line: No cutoff. Dashed line: PV subtraction. 
Solid line: Invariant mass cutoff
(exponential).
[Self--consistent soliton profile Eq.~(\ref{P_self}) with $M_\pi = 0$, 
$M = 0.35\, \textrm{GeV}, M_N = 3.26 \, M$.]}
\label{fig:fzt}
\end{figure}
At large distances the $\xi_T$ correlation in the $q\bar q$ pair 
decays exponentially, 
\be
\tilde f_{q\bar q}(z, \xi_T) &\sim& 2\pi \; [\pi / (2 M \xi_T)]^{1/2} \;
\nonumber \\
&\times & 
\exp (-M \xi_T) \hspace{2em} (\xi_T \rightarrow \infty) ,
\hspace{2em}
\label{exp_large_z}
\ee
with a range determined by the constituent quark mass $M$. This appears
natural, as the constituent quark mass represents the ``mass gap'' of
the effective chiral model and controls the long--distance behavior
of quark correlations. Since it reflects a low--energy property of 
our model we should require that the UV cutoff do not modify 
this behavior.
The exponential fall--off at large distances is related to the singularity 
of the unregularized $p_T$ distribution at $p_T^2 = - M^2$, corresponding 
to complex values of $p_T$. Our requirement therefore implies that the
cutoff should not modify this free--field singularity. It is easy to see 
that the schemes discussed in Sec.~\ref{subsec:cutoff} satisfy this 
requirement. PV subtraction leaves the residue of the pole at 
$p_T^2 = - M^2$ unchanged, cf. Eq.~(\ref{f_qqbar_pv}).
Likewise, with the invariant mass cutoff the residue remains unchanged
because the invariant mass vanishes at the pole, $s \rightarrow 0$, 
and the form factors are normalized such that $F(s = 0) = 1$, 
cf.\ Eq.~(\ref{f_qqbar_inv}). This explains the physical basis of 
regularization condition III presented in Sec.~\ref{subsec:cutoff}.

The quark field correlator $\tilde f_1^{\bar u + \bar d}(x, \xi_T)$ 
in the nucleon obtained from Eq.~(\ref{f_coord_grad}) is presented in 
Fig.~\ref{fig:fzt}. 
The plot shows the $\xi_T$ dependence of the correlation function 
at a representative value of $x = 0.1$. 
The following features are worth noting:
(a) The regularized distributions rapidly approach exponential behavior 
at $\xi_T \sim \textrm{few} \, M^{-1}$, as implied by Eq.~(\ref{exp_large_z}). 
(b) The distributions obtained with the PV and invariant mass cutoffs
are overall rather close at all distances. This explains the 
approximate cutoff--independence of the $p_T$--distributions at 
low $p_T$ observed in Sec.~\ref{subsec:cutoff} (see Fig.~\ref{fig:fpt}). 
We note that the results would be even closer if we required that 
the different regularizations reproduce the same total parton density, 
e.g. by adjusting the cutoff values, or by performing a second PV 
subtraction. (c) The curvature of the $\xi_T$ distributions at $\xi_T = 0$, 
which is proportional to $\langle p_T^2 \rangle$, is not effectively
constrained and can thus vary considerably between different 
regularization schemes. This reaffirms our earlier conclusion regarding 
the model dependence of $\langle p_T^2 \rangle$.

In sum, we find that the combined requirements of matching the
parton density at $\xi_T = 0$ and decaying exponentially at 
$\xi_T \rightarrow \infty$ effectively constrain the correlator 
at all distances. The Fourier transform 
of these correlators results in stable, cutoff--independent 
$p_T$--distribution at low $p_T$ as seen in Fig.~\ref{fig:fpt}.
The study of the coordinate--space correlation function of
sea quarks thus reaffirms our conclusion that the physical
regularization conditions presented in Sec.~\ref{subsec:cutoff}
result in robust transverse momentum distributions at low $p_T$.

Our aim in this section was to investigate the influence of the UV 
cutoff on the coordinate--space distribution, which could be done in a 
simple way with the gradient expansion in collinear approximation, 
Eq.~(\ref{f_grad_lc_simple}). In this approximation the information 
about the finite spatial size of the classical chiral field is lost,
and the exponential decay of the correlator at large $\xi_T$ is due
entirely to the intrinsic size of the quark--antiquark pair.
To study the ``true'' large--distance behavior of the sea quark
correlator in the nucleon one should use the full gradient expansion 
result, Eq.~(\ref{f_grad_lc}), in which the finite size of the chiral 
field is encoded in the $\bm{k}_T$ dependence. Numerical studies show
that also in the full correlation function the large--$\xi_T$ behavior 
is independent of the UV cutoff; i.e., our conclusions are general 
and do not depend on the additional simplifications made in 
Eq.~(\ref{f_grad_lc_simple}).
\subsection{Numerical evaluation}
\label{subsec:numerical}
Having established the behavior of the sea quark transverse momentum 
distribution at large momenta $p_T^2 \gg M^2$ and the role of the 
UV cutoff, we now want to make a numerical estimate of the distributions
also at lower momenta $p_T^2 \sim \textrm{few} \, M^2$. This will allow 
us to compare the sea quark transverse momentum distribution
with those of the valence quarks at a quantitative level
(see Sec.~\ref{subsec:sea_val}). 

In its representation as a sum over occupied quark single--particle
levels, Eq.~(\ref{fpt_sum_alt}), the sea quark
distribution receives contributions both from the negative--energy 
Dirac continuum and the discrete bound--state level. The gradient
expansion approximates the contribution from the Dirac
continuum, i.e., quark scattering states distorted by the classical 
chiral field, which dominates at momenta $p_T^2 \gg M^2$.
The contribution from the discrete bound--state level with its normalizable 
wave function is not contained in the expanded Green function.
At lower momenta $p_T^2 \sim \textrm{few} \, M^2$, as well as
in the integral over $p_T$, the contribution of the discrete level 
becomes numerically relevant. A more accurate approximation in this region
for numerical purposes is obtained by adding to the leading--order gradient 
expansion Eq.~(\ref{f_grad_lc_simple}) the contribution from the discrete 
bound--state level given by Eq.~(\ref{level_f1}),
\beq
f_{1}^{\bar u + \bar d}(x, p_T) \;\; \approx \;\; 
f_{1, {\rm grad}}^{\bar u + \bar d}(x, p_T) \; + \; 
f_{1, {\rm lev}}^{\bar u + \bar d}(x, p_T) .
\label{interpolation}
\eeq
This approximation is known as the ``interpolation formula,'' as it 
becomes exact both in the limit of large soliton size, where the
gradient expansion is parametrically justified and the discrete level 
energy becomes negative, and in the limit of small soliton size, 
where the level contribution dominates \cite{Diakonov:1996sr}.
Numerical studies show that Eq.~(\ref{interpolation}) reproduces
the exact numerical result for the $p_T$--integrated sea quark
distribution \cite{Weiss:1997rt} with an accuracy of far better than 
$20\%$ for $x = 0.1 - 0.5$ when evaluated with the self--consistent 
soliton profile Eqs.~(\ref{P_self})--(\ref{P_self_end}). We  
therefore expect it to provide a reasonable approximation also for the
$p_T$ distributions at $p_T^2 \sim \textrm{few}\, M^2$. 

%
%
\begin{figure}
\includegraphics[width=.48\textwidth]{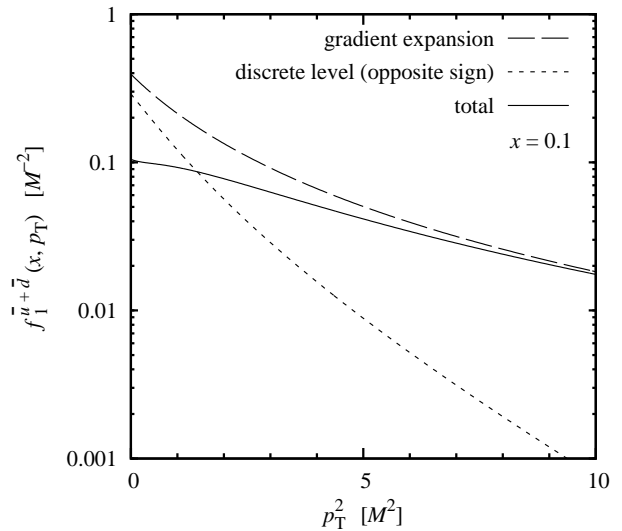}
\caption[]{Numerical approximation for the sea quark transverse momentum
distribution $f_1^{\bar u + \bar d} (x, p_T)$, Eq.~(\ref{interpolation}) 
(``interpolation formula''). Dashed line: Gradient expansion 
$f_{1, {\rm grad}}^{\bar u + \bar d}(x, p_T)$, Eq.~(\ref{f_grad_lc_simple}), 
approximating the Dirac continuum contribution in 
Eq.~(\ref{fpt_sum_alt}) (PV regularization).
Dotted line: Contribution from the discrete bound--state level, 
$f_{1, {\rm lev}}^{\bar u + \bar d}(x, p_T)$
Eq.~(\ref{level_f1}), shown with opposite sign.
Solid line: Total distribution according to Eq.~(\ref{interpolation}).
[Self--consistent soliton profile 
Eq.~(\ref{P_self}) with $M = 0.35\, \textrm{GeV}, M_N = 3.26 \, M$.]}
\label{fig:interpolation}
\end{figure}
Figure~\ref{fig:interpolation} shows the different contributions
to the sea quark distribution in Eq.~(\ref{interpolation}) at a 
representative value of $x = 0.1$
(note that the level contribution is negative and shown with
opposite sign in the figure). One sees that the gradient expansion 
contribution approximating the Dirac continuum clearly dominates 
at large $p_T$; the level contribution to the total
$f_1^{\bar u + \bar d}$ is $<20\%$ above $p_T^2 = 5 \, M^2$ and decreases
rapidly at larger $p_T$. This justifies our earlier use of the gradient 
expansion to study the large--$p_T$ behavior. At low $p_T$ there are 
very significant cancellations between the gradient expansion and 
the discrete level contribution, causing the sum to be 3--4 times
smaller than the individual terms. Since the gradient expansion provides
only a rough approximation to the Dirac continuum at low $p_T$
(cf.\ Fig.~\ref{fig:fpt_app}, which compares different variants),
and is subject to some uncertainty resulting from the UV cutoff,
we cannot use Eq.~(\ref{interpolation}) to predict the total
$f_1^{\bar u + \bar d}$ with any meaningful \textit{relative} accuracy
at low $p_T$, but can conclude only that it is substantially 
smaller than the gradient expansion result alone. More quantitatively,
if we require that the level contribution be $<50\%$ of the gradient
expansion we conclude that we can safely use Eq.~(\ref{interpolation})
for a numerical estimate at $p_T^2 > 2\, M^2$. At lower values of
$p_T$ one should use methods based on exact summation over levels
to calculate the sea quark transverse momentum distribution 
\cite{Wakamatsu:2009fn}. Note, however, that the contribution of
this region to the $p_T$--integral determining the total sea quark 
density is rather small (see below), so that the question of the 
exact behavior of the sea quark $p_T$ distribution in this region 
is somewhat academic.
\subsection{Sea vs. valence quark distribution}
\label{subsec:sea_val}
%
%
\begin{figure}
\begin{tabular}{ll}
\includegraphics[width=.47\textwidth]{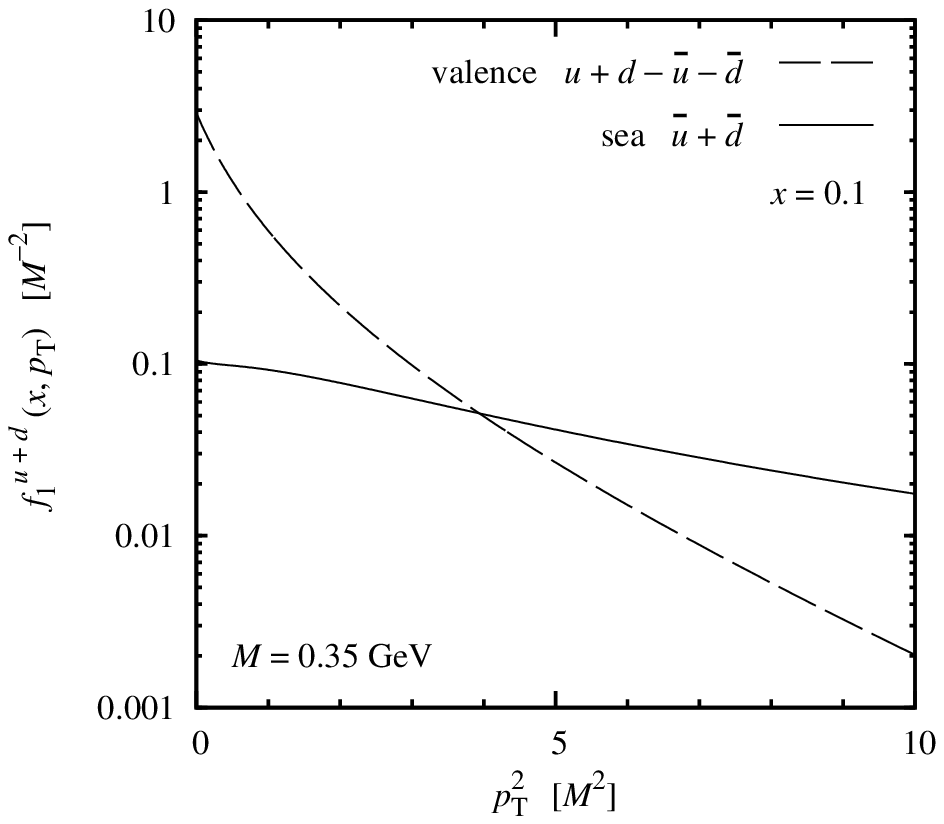}
\\[-4ex]
(a)
\\
\includegraphics[width=.47\textwidth]{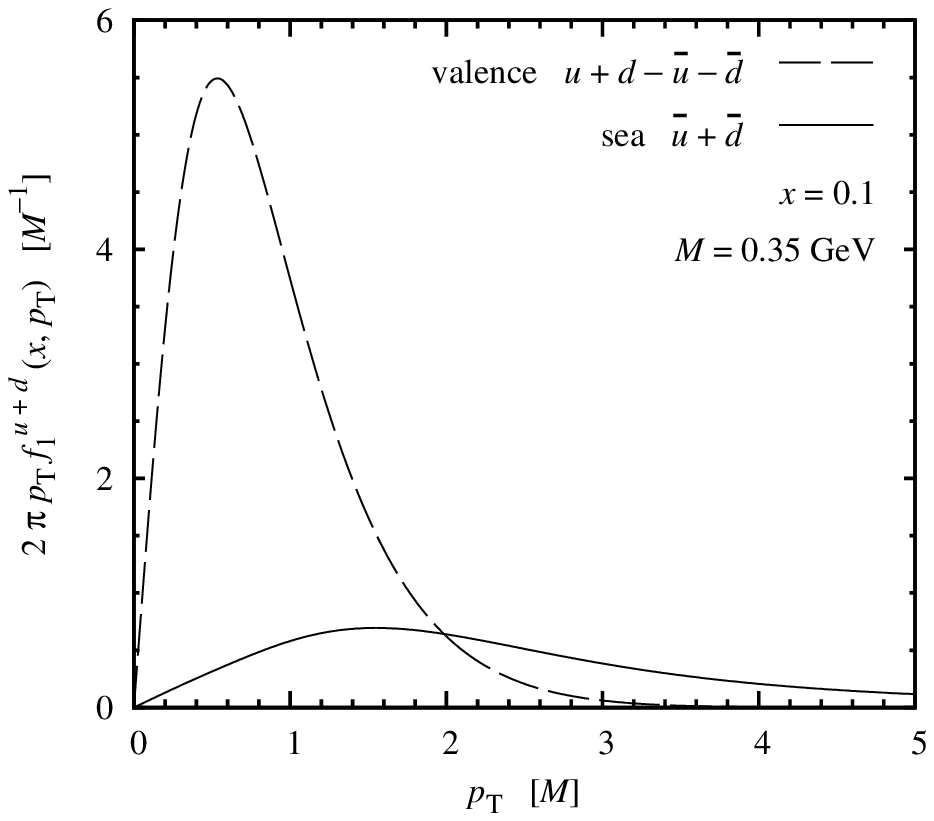}
\\[-4ex]
(b)
\end{tabular}
\caption[]{Transverse momentum distributions of flavor--singlet 
unpolarized valence and sea quarks at $x = 0.1$. Panel (a) shows 
$f_1^{u + d - \bar u - \bar d}$ and $f_1^{\bar u + \bar d}$
as functions of $p_T^2$ on a logarithmic scale; panel (b)
shows the radial distribution $2\pi p_T f_1^{u + d - \bar u - \bar d}$
and $2\pi p_T f_1^{\bar u + \bar d}$ on a linear scale, such that the area 
under the curves corresponds to the integral over $p_T$.
Dashed lines: Valence quark distribution $f_{1}^{u + d - \bar u - \bar d}$
(see Fig.~\ref{fig:g1_val}). Solid lines: Sea quark distribution 
$f_1^{\bar u + \bar d}$ (PV regularization). [Self--consistent soliton 
profile Eq.~(\ref{P_self}) with $M = 0.35\, \textrm{GeV}, 
M_N = 3.26 \, M$.]}
\label{fig:f1_val_sea}
\end{figure}
Using the numerical approximation of Sec.~\ref{subsec:numerical}
we now want to compare our results for the sea quark transverse momentum
distribution with those of the valence quarks calculated in 
Sec.~\ref{sec:valence}. Figure~\ref{fig:f1_val_sea} summarizes the 
numerical results for the valence distribution 
$f_1^{u + d - \bar u - \bar d} (x, p_T)$
and the sea quark distribution $f_1^{\bar u + \bar d} (x, p_T)$ 
at a representative value of $x = 0.1$. 
Panel (a) shows the distributions themselves on a logarithmic scale; 
panel (b) the radial distributions on a linear scale, 
such that the area under the curves corresponds directly to their 
integral over $p_T$. Similar results are obtained at other values 
of $x$: the shape of the individual $p_T$ distribution changes
little with $x$ (cf.\ Fig.~\ref{fig:f1_val} for the valence distribution);
only their normalization changes in proportion to the total valence
and sea quark density.

The numerical estimates clearly show very different shapes 
of the valence and sea quark transverse momentum distributions, 
especially at large values of $p_T$, as first observed in the 
calculation of Ref.~\cite{Wakamatsu:2009fn}. Based on our theoretical
analysis we can now explain this striking behavior as the effect of
dynamical chiral symmetry breaking in the QCD vacuum on the intrinsic
transverse momentum distribution of the sea quarks. Even with the 
strong modification of the would--be $1/p_T^2$ tail by the UV cutoff, 
the sea quark transverse momentum distribution in the chiral 
quark--soliton model is qualitatively different from that of the 
valence quarks. While the precise numerical values depend on the model 
implementation (see e.g.\ Fig.~\ref{fig:fpt}), the fact as such is 
rooted in the basic structure of the effective dynamics chiral and 
should be model--independent. 

When interpreting the results of Figure~\ref{fig:f1_val_sea} one should
keep in mind that the accuracy of the approximation Eq.~(\ref{interpolation})
used in our numerical estimate of $f_1^{\bar u + \bar d}(x, p_T)$ is not
sufficient to predict the values at $p_T^2 \lesssim 2\, M^2$ with
meaningful relative accuracy (cf.\ the discussion in 
Sec.~\ref{subsec:numerical}).
In this sense the plot of the radial distribution, in which the low--$p_T$
region is suppressed, conveys a more realistic picture. This
uncertainty, however, in no way influences our conclusions regarding 
the qualitatively different behavior of valence and sea quark 
distributions at large $p_T$.

The qualitative difference between the $p_T$ distribution of valence and
sea quarks is the most important practical result of our study. Its
numerous implications for deep--inelastic processes are explored in 
Sec.~\ref{sec:applications}.
\subsection{Polarized sea quark distribution}
\label{subsec:polarized_sea}
To complete our study of the sea quark transverse momentum distribution 
we want to investigate also the flavor--nonsinglet polarized sea quark 
distribution. The gradient expansion of this distribution can be
carried out in complete analogy to the flavor--singlet unpolarized
case starting from Eq.~(\ref{gkt_green}), cf.\ Secs.~\ref{subsec:gradient} 
and \ref{subsec:lightcone}; we do not present the intermediate steps here. 
The result can again be represented as a convolution integral over the 
momentum of the classical chiral field, analogous to Eq.~(\ref{f_grad_lc}),
\be
g_{1, {\rm grad}}^{\bar u - \bar d}(x, p_T) &=& \int\!\frac{dy}{y} 
\; \int\! d^2 k_T \; g_{\rm cl} (y, \bm{k}_T) 
\nonumber
\\[1ex]
&\times& g_{q\bar q}(x, y; \bm{p}_T, \bm{k}_T) .
\label{g_grad_lc}
\ee
The relevant momentum distribution of the classical field is now
\be
g_{\rm cl} (y, \bm{k}_T) &\equiv& \frac{F_\pi^2 M_N^2 y}{3 (2\pi)^3} \; 
\textrm{tr}_{\rm fl} \, [ \tau^3 \widetilde U_{\rm cl} (\bm{k}) 
\widetilde U_{\rm cl}(\bm{k})^\dagger ] \hspace{2em}
\label{g_pi_def}
\\[1ex]
&& \left[ \bm{k} = (\bm{k}_T,  y M_N) \right] .
\nonumber 
\ee
The momentum distribution resulting from the quark loop integral
turns out to be the same as in the flavor--singlet unpolarized 
case
\beq
g_{q\bar q}(x, y; \bm{p}_T, \bm{k}_T) \;\; = \;\; 
f_{q\bar q}(x, y; \bm{p}_T, \bm{k}_T) 
\label{g_qqbar_equals_f_qqbar}
\eeq
(for the $p_T$--integrated distributions this was already noted in 
Ref.~\cite{Diakonov:1996sr}). This remarkable fact can be understood as 
an instance of ``restoration of chiral symmetry.'' When expanding in 
gradients of the chiral fields the numerator of the quark loop integral 
becomes independent of the dynamical quark mass $M$, and the same 
coefficient is obtained for the axial vector--type operator in the polarized 
distribution (Dirac matrix $\gamma^+ \gamma_5$) as for the vector--type 
operator in the unpolarized distribution ($\gamma^+$).
A more microscopic interpretation of this result in terms of pair correlations
in the nucleon wave function will be discussed below. 
Equation~(\ref{g_qqbar_equals_f_qqbar}) has important consequences for 
the behavior of the flavor--nonsinglet polarized $p_T$ distribution at 
large transverse momenta. 

%
%
\begin{figure}
\begin{tabular}{ll}
\includegraphics[width=.47\textwidth]{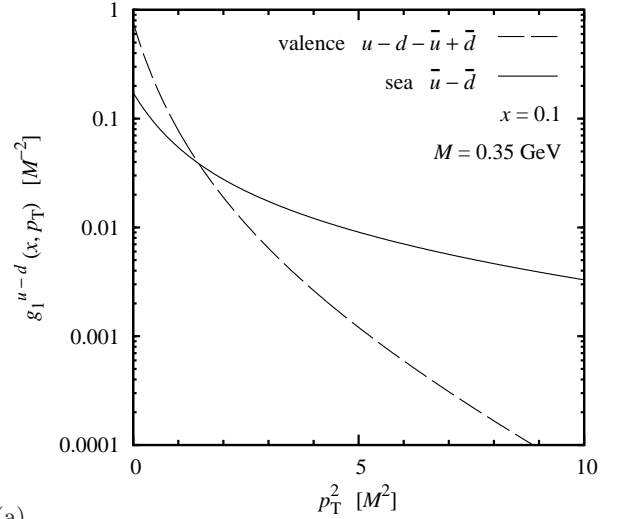}
\\[-4ex]
(a)
\\
\includegraphics[width=.47\textwidth]{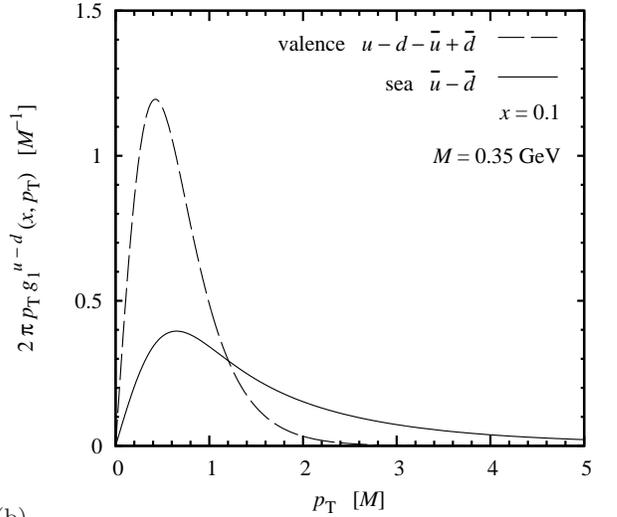}
\\[-4ex]
(b)
\end{tabular}
\caption[]{Transverse momentum distributions of flavor--nonsinglet 
polarized valence and sea quarks at $x = 0.1$. Panel (a) shows 
$g_1^{u - d - \bar u + \bar d}$ and $g_1^{\bar u - \bar d}$
as functions of $p_T^2$ on a logarithmic scale; panel (b) 
shows the radial distribution $2\pi p_T g_1^{u - d - \bar u + \bar d}$
and $2\pi p_T g_1^{\bar u - \bar d}$ on a linear scale, such that the area 
under the curves corresponds to the integral over $p_T$.
Dashed lines: Valence quark distribution $g_{1}^{u - d - \bar u + \bar d}$
(see Fig.~\ref{fig:g1_val}). Solid lines: Sea quark distribution 
$g_1^{\bar u - \bar d}$ (PV regularization). [Self--consistent soliton 
profile Eq.~(\ref{P_self}) with $M = 0.35\, \textrm{GeV}, 
M_N = 3.26 \, M$.]}
\label{fig:g1_val_sea}
\end{figure}
At transverse momenta $p_T^2 \gg M^2$ the convolution integral
Eq.~(\ref{g_pi_def}) can again be simplified by using
the collinear approximation, Eqs.~(\ref{neglect_kt}) and (\ref{neglect_k3}),
and becomes
\be
g_{1, {\rm grad}}^{\bar u - \bar d}(x, p_T) 
&=& \int_x^\infty \!\frac{dy}{y} \; 
g_{\rm cl} (y) \; g_{q\bar q}(x/y, p_T) . 
\hspace{2em}
\label{g_grad_lc_simple} 
\ee
Here $g_{\rm cl} (y)$ is the corresponding $k_T$--integrated 
light--cone momentum distribution of the classical chiral field,
cf.\ Eq.~(\ref{f_pi}),
\be
g_{\rm cl} (y) &\equiv& \int d^2 k_T \; g_{\rm cl} (y, \bm{k}_T) 
\nonumber \\[1ex]
&=& \frac{F_\pi^2 M_N^2 y}{3(2\pi)}
\int\!\frac{d^2k_T}{(2\pi)^2} 
\text{tr}_{\text{fl}} [\tau^3 \widetilde{U}_{\text{cl}} (\bm{k}) \,
\widetilde{U}_{\text{cl}} (\bm{k})^\dagger ] \hspace{3em}
\label{g_cl}
\\[1ex]
&& (k^3 = y M_N) ,
\nonumber
\ee
which can be evaluated using Eq.~(\ref{utilde_trace_tau3})
in Appendix~\ref{app:fourier}. The numerical distribution is
shown in Fig.~\ref{fig:fy}. One sees that at $y \rightarrow 0$ the
``polarized'' distribution is suppressed relative to the ``unpolarized''
one, as expected, and that at large values of $y \sim 1$ the distributions
approximately satisfy $3 g_{\rm cl}(y) \approx f_{\rm cl}(y)$, corresponding
to saturation of the large--$N_c$ inequality for the transverse momentum
densities (see below). The momentum distribution of the quark--antiquark
pair appearing in Eq.~(\ref{g_grad_lc_simple}) is the same as in 
the unpolarized case, cf.\ Eq.~(\ref{g_qqbar_equals_f_qqbar}), 
\beq
g_{q\bar q}(z, p_T) \;\; = \;\; f_{q\bar q}(z, p_T) ,
\label{g_qqbar_equals_f_qqbar_simple}
\eeq
where $z \equiv x/y$ and $f_{q\bar q}(z, p_T)$ is defined 
in Eq.~(\ref{f_qqbar_simple}) and explicitly given by 
Eq.~(\ref{f_qqbar_unreg}), up to modifications by the UV cutoff. 
As a result, the flavor--nonsinglet polarized distribution exhibits
a would--be power--like tail at large transverse momenta similar to the 
the unpolarized distribution,
\beq
g_1^{\bar u - \bar d}(x, p_T) \;\; \sim \;\; 
\frac{C_{g_1}^{\bar u - \bar d}(x)}{p_T^2} 
\hspace{2em} (p_T^2 \gg M^2). 
\eeq
Moreover, the coefficient of this tail is related to the UV divergence 
of the $p_T$--integrated distribution in the same way as in the 
unpolarized case, and is given by [cf.\ Eq.~(\ref{tail_nucleon_gradient})]
\beq
C_{g_1}^{\bar u - \bar d}(x) 
\;\; = \;\; g_{1, {\rm grad}}^{\bar u - \bar d}(x) .
\eeq
Thus, our earlier discussion of the UV regularization
and its effect on the transverse momentum distributions can be carried 
over directly to the polarized case.

For a numerical estimate of the flavor--nonsinglet polarized sea quark
distribution at all values of $p_T$ (including $p_T \sim M$)
we use the analogue of the ``interpolation formula'' Eq.~(\ref{interpolation}),
in which one adds the contribution from the discrete bound--state level,
Eq.~(\ref{gpt_val_res}), to the gradient expansion approximating the 
Dirac continuum contribution, Eq.~(\ref{g_grad_lc}), 
\beq
g_{1}^{\bar u - \bar d}(x, p_T) \;\; \approx \;\; 
g_{1, {\rm grad}}^{\bar u - \bar d}(x, p_T) \; + \; 
g_{1, {\rm lev}}^{\bar u - \bar d}(x, p_T) .
\label{interpolation_g1}
\eeq
The resulting polarized sea quark distribution is shown in 
Fig.~\ref{fig:g1_val_sea} and compared to the valence quark
distribution $g_{1}^{u - d - \bar u + \bar d}$ calculated in
Sec.~\ref{subsec:valence_polarized}. One sees that, as in the
flavor--singlet unpolarized case, the would--be power--like tail
strongly influences the numerical behavior of the flavor--nonsinglet 
polarized sea quark distribution at $p_T^2 > \textrm{few} \, M^2$ and
causes it to be qualitatively different from that of the valence quarks.
This again represents a direct effect of dynamical chiral symmetry 
breaking on the nucleon's partonic structure. Note that here the
effect occurs in a nonsinglet channel, in which the distributions
are likely to be much less affected by perturbative QCD evolution
than in the singlet case, making this nonperturbative effect 
even more striking (cf.\ discussion in Sec.~\ref{sec:summary}).

A more microscopic explanation for the similarity of the flavor--singlet
unpolarized and flavor--nonsinglet polarized distributions at $p_T^2 \gg M^2$
is provided in Sec.~\ref{sec:correlations}, where we show that the tails 
in the sea quark distributions are due to correlated quark--antiquark pairs 
in the nucleon's light--cone wave function. The correlated pairs appear 
in scalar--isoscalar ($\Sigma$) and pseudoscalar--isovector ($\Pi$) 
quantum numbers. The flavor--singlet unpolarized sea results from the 
overlap of like pairs ($\Sigma\Sigma, \Pi\Pi$), while the 
flavor--nonsinglet polarized one is due to the interference of
different types of pairs ($\Sigma\Pi, \Pi\Sigma$) in the initial
and final state. At $p_T^2 \gg M^2$ the wave functions of the $\Sigma$ and
$\Pi$ pairs become the same due to the restoration of chiral symmetry, 
Eq.~(\ref{restoration}), leading naturally to a relation between the two 
sea quark distributions. 
We note that this derivation of the flavor--nonsinglet polarized sea 
quark distribution gives a precise meaning to the notion of 
``$\Pi\Sigma$ interference,'' which was discussed in connection
with the meson cloud model of flavor asymmetries in 
Ref.~\cite{Dressler:1999zg}.

To conclude our discussion of the polarized sea quark distribution, we 
would like to see how the general large--$N_c$ inequality for
sea quark distributions, Eq.~(\ref{ineq_anti}), is realized in our model.
At $p_T^2 \gg M^2$ the distributions are given by the gradient expansion,
in the simplified form of Eqs.~(\ref{f_grad_lc_simple}) and 
(\ref{g_grad_lc_simple}), and we want to verify that
\beq
f_1^{\bar u+\bar d}(x,p_T) \; \pm \; 3 \, 
g_1^{\bar u-\bar d}(x,p_T) \;\; > \;\; 0
\label{inequality_sea}
\eeq
in this approximation. Because of 
Eq.~(\ref{g_qqbar_equals_f_qqbar_simple}) the differences 
on the left--hand side of Eq.~(\ref{inequality_sea})
can be written in the form
\be
\lefteqn{
f_1^{\bar u+\bar d}(x,p_T) \; \pm \; 3 \, g_1^{\bar u-\bar d}(x,p_T)} && 
\nonumber \\[2ex]
&=& \int_x^\infty \!\frac{dy}{y} \, \left[ f_{\rm cl} (y) 
\, \pm \, 3 g_{\rm cl} (y) \right] \; f_{q\bar q}(x/y, p_T) ,
\label{inequality_sea_grad}
\ee
where $f_{\bar q q}$ is explicitly positive; cf.\ the discussions
in Secs.~\ref{subsec:pair} and \ref{subsec:cutoff} and the 
representation of this function as light--cone wave function overlap 
derived in Sec.~\ref{subsec:restoration}, Eq.(\ref{f_qqbar_as_overlap}).
The difference of the momentum distributions of the classical field
in Eq.~(\ref{inequality_sea_grad}), in turn, is given by
\be
f_{\rm cl} (y) \; \pm \; 3 \, g_{\rm cl} (y) 
&=& \frac{F_\pi^2 M_N^2 y}{(2\pi)} \int\!\frac{d^2k_T}{(2\pi)^2} 
\nonumber
\\[1ex]
&\times& 
\text{tr}_{\text{fl}} [(1 \pm \tau^3) \widetilde{U}_{\text{cl}} (\bm{k}) \,
\widetilde{U}_{\text{cl}} (\bm{k})^\dagger ]
\hspace{2em}
\label{positivity_classical}
\\[2ex]
&& (k^3 = y M_N) ,
\nonumber 
\ee
which is explicitly positive, cf.\ Eq.~(\ref{utilde_trace_sumdiff}) in 
Appendix~\ref{app:fourier}. Thus we see that the
``restoration of chiral symmetry'' expressed in 
Eq.~(\ref{g_qqbar_equals_f_qqbar}) naturally guarantees that
the large--$N_c$ inequalities for the sea quark distributions
are satisfied at $p_T^2 \gg M^2$ in our scheme of approximations.

The results of the interpolation formulas Eq.~(\ref{interpolation}) and 
(\ref{interpolation_g1}) for the distributions at $p_T^2 \sim M^2$,
taken literally, would violate the inequality Eq.~(\ref{inequality_sea})
at $p_T^2 \lesssim 2\, M^2$. However, we noted in Sec.~\ref{subsec:numerical}
that in this region the unpolarized sea quark distribution cannot be 
estimated with any meaningful relative accuracy using this approximation.
We therefore cannot conclusively study the inequality at low $p_T$
using this approximation.
\section{Short--range correlations of partons}
\label{sec:correlations}
\subsection{Nucleon wave function at large momenta}
So far we studied the properties of valence and sea quarks in the 
nucleon by investigating their one--body momentum densities,
Eqs.~(\ref{fpt_def}) and (\ref{fpt_anti_def}). A more microscopic
understanding of our results can be obtained by considering the 
nucleon's partonic (or light--front) wave function in the chiral
quark--soliton model. Specifically, we want to show that in this model
sea quarks partly exist in correlated pairs of a size of the order of the 
cutoff scale $\Lambda^{-1} \ll R$, and that the ``tail'' in their 
transverse momentum density can directly be attributed to these
configurations. That is, the two--scale picture of the effective
dynamics resulting from chiral symmetry breaking implies the existence
of short--range quark--antiquark correlations in the partonic wave function.
This observation has far--reaching implications for our understanding 
of the partonic structure of the nucleon not only in the chiral
quark--soliton model but in QCD in general. It suggests an interesting 
analogy with short--range nucleon--nucleon correlations in nuclei, 
which give rise to high--momentum components of the nuclear spectral 
function governing single--particle knockout reactions and can be probed
directly in multiparticle correlation 
experiments \cite{Frankfurt:1981mk,Frankfurt:2008zv,Arrington:2011xs}. 
A detailed treatment of parton short--range correlations due to dynamical 
chiral symmetry breaking and their implications will be the subject of a 
subsequent publication. Here we want to discuss only the aspects relevant to 
understanding the behavior of the intrinsic transverse momentum distributions.

The light--front wave function of the nucleon in the chiral quark--soliton
model was derived and discussed in a general context 
in Refs.~\cite{Petrov:2002jr,Diakonov:2004as}. In this approach the 
many--body wave function of the fast--moving nucleon is constructed by
applying the creation operators of $N_c$ valence quarks and a coherent 
superposition of quark--antiquark pairs to the vacuum state of the 
effective chiral theory, i.e., the Dirac vacuum with baryon number zero. 
The construction can be carried out explicitly in the sense of a Fock 
state expansion, using the bound--state level occupied by $N_c$ quarks as 
a seed \cite{Lorce:2007as}. Here we are interested in the sea quark component
of the nucleon wave function at large transverse momenta, 
\beq
p_T^2 \;\; \gg \;\; M^2 .
\label{large_pt_wf}
\eeq
This component of the wave function can be calculated directly in a 
simple approximation, assuming the existence of a stable mean field.
It will be seen that this approximation is equivalent to the gradient
expansion for the one--body momentum densities of sea quarks described
in Sec.~\ref{sec:sea}.

The appearance of sea quarks in the light--cone wave function of the 
nucleon in the chiral quark--soliton model can be viewed as the creation 
of quark--antiquark pairs by the chiral field produced by the other 
constituents. In general this is a complicated process, which affects the
state of motion of the source particles by changing their longitudinal and 
transverse momenta. Also, the produced pair can re-interact with the 
chiral field and experience distortion of its wave function. The limit 
of large transverse momenta, Eq.~(\ref{large_pt_wf}), permits several 
important simplifications, which make it possible to describe this 
process in practice and calculate the sea quark component of the 
wave function in a controlled approximation.

First, the invariant mass difference in the pair creation process is 
dominated by the invariant mass of the pair, while the contribution
from the change of the state of motion of the source system can be 
neglected. Consider the creation of a pair with overall plus momentum 
fraction $y$ and transverse momentum $\bm{k}_T$ by a ``source'' with mass 
$\sim M_N$ (the precise coefficient does not matter) and 
initial transverse momentum zero. Let $z$ and $\bar z$ be the relative 
plus momentum fractions of the quark and antiquark in the pair, 
and $\bm{p}_T$ the transverse momentum of the quark 
(cf.\ Sec.~\ref{subsec:lightcone}). The change of invariant mass 
of the total system (source and quark--antiquark pair) in the process is
\be
\Delta s_{\rm tot} &=& \frac{M^2 + \bm{p}_T^2}{yz}
\; + \; \frac{M^2 + (\bm{k}_T - \bm{p}_T)^2}{y\bar z} 
\nonumber
\\
&+& \frac{M_N^2 + \bm{k}_T^2}{\bar y} - M_N^2
\nonumber
\\
&=& \frac{1}{y} \left[ \frac{M^2 + \bm{p}_T^2}{z}
\; + \; \frac{M^2 + (\bm{k}_T - \bm{p}_T)^2}{\bar z} 
\right.
\nonumber
\\
&+& \left. \frac{y^2 M_N^2 + y \bm{k}_T^2}{\bar y} \right] .
\label{delta_m2}
\ee
The longitudinal and transverse momentum transfer by the source
is of the order
\be
|\bm{k}_T| &\sim & M , \\
y M_N &\sim & M .
\ee
In the limit $p_T^2 \gg M^2$ we can neglect the last term in 
the parenthesis in Eq.~(\ref{delta_m2}) and drop $\bm{k}_T$ and $M$
in the other terms. The total invariant mass difference is
thus determined by the intrinsic invariant mass of the quark--antiquark
pair, $s$, cf.\ Eq.~(\ref{s_def}), and given by
\beq
\Delta s_{\rm tot} \; \approx \;\; \frac{p_T^2}{y z\bar z} 
\;\; = \;\; \frac{s}{y}.
\label{delta_m2_approx}
\eeq

Second, in the region $p_T^2 \gg M^2$ the interaction of the
quark--antiquark pair with the source is effectively given by 
the leading--order Born approximation in the nontrivial part 
of chiral field,
\beq
M [U^{\gamma_5}(x) - 1] ,
\eeq
describing the deviation from the vacuum. Higher--order interactions
come with higher powers of the constituent quark mass (as well as 
gradients of the chiral fields) and are suppressed by inverse powers 
of $p_T$. This means that at $p_T^2 \gg M^2$ the distortion of the
internal wave function of the quark--antiquark pair can be
neglected, and that it can be regarded as being in a plane--wave 
state after its creation.

Altogether, we see that a very simple picture of the sea quark component 
of the nucleon's partonic wave function emerges at $p_T^2 \gg M^2$.
The fast--moving nucleon creates color--singlet quark--antiquark pairs
through the interaction Hamiltonian
\beq
H_{\rm int}(t) \;\; = \;\; \int d^3 x \; \bar\psi(t, \bm{x}) \; M 
[U^{\gamma_5}(t, \bm{x}) - 1]_{\rm ret} \; \psi(t, \bm{x}) ,
\label{H_int}
\eeq
where $[U^{\gamma_5} - 1]_{\rm ret}$ is the nontrivial chiral field 
produced by the other constituents, in the sense of a retarded potential. 
The interaction is to be treated in first order, and the pair is 
produced in a plane--wave state. In this approximation the rest of
the nucleon producing the pair acts as a classical source
(henceforth called ``classical nucleon''); it transfers
longitudinal and transverse momentum as well as spin/isospin quantum numbers
to the pair but otherwise remains inert.

One may ask whether the approximations proposed here respect chiral 
invariance. As it stands, first--order Born approximation with the
Hamiltonian Eq.~(\ref{H_int}) does not result in chirally invariant
amplitudes. However, below we shall see that chiral invariance is
effectively restored in the momentum density of quarks and antiquarks 
at high $p_T$, as it should be.

%
%
\begin{figure}
\includegraphics[width=.34\textwidth]{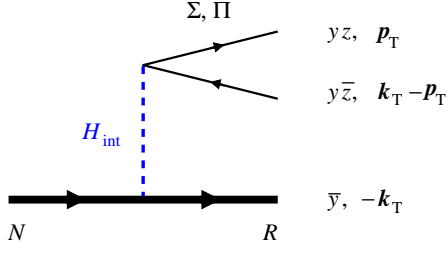} 
\caption{(Color online) 
Approximation to the nucleon's light--cone wave function at 
$p_T^2 \gg M^2$. The nucleon acts as a classical source, which produces a
color--singlet quark--antiquark pair through the retarded potential 
Eq.~(\ref{H_int}). The longitudinal momentum fractions and transverse 
momenta in the final state are indicated on the right 
($\bar y \equiv 1 - y$ etc.). The outgoing classical nucleon 
can be in a rotational state with $N$ or $\Delta$ quantum numbers.
The quark--antiquark pair can have scalar--isoscalar ($\Sigma$)
or pseudoscalar--isovector ($\Pi$) quantum numbers.}
\label{fig:pair}
\end{figure}
With the above approximations it is straightforward to calculate
the high--$p_T$ component of the nucleon's partonic wave function
(see Fig.~\ref{fig:pair}). 
Using ordinary time--ordered perturbation theory, we calculate the 
amplitude for a 
fast--moving nucleon at $t = -\infty$ to produce a quark--antiquark 
pair at time $t = 0$ via the interaction Eq.~(\ref{H_int}), 
assuming that the interaction is adiabatically switched on 
starting from $t = -\infty$. We obtain
\be
&& \langle q (\bm{p}_1, \sigma_1, f_1, c_1) \, 
\bar q (\bm{p}_2, \sigma_2, f_2, c_2), \, R(\bm{P}_R, I_R) \, | 
\nonumber
\\
&& \left[ (-i) \int_{-\infty}^0 \!\! dt \; H_{\rm int} (t) \right] \; 
| \, N (\bm{P}_N, I_N) \, \rangle
\nonumber \\
&=& (2\pi)^3 \; \delta^{(3)}(\bm{p}_1 + \bm{p}_2 + \bm{P}_R - \bm{P}_N) \;
\delta_{c_1 c_2}
\nonumber \\[1ex]
&\times & \frac{\bar u (\bm{p}_1, \sigma_1, f_1) \; \Gamma \; 
v(\bm{p}_2, \sigma_2, f_2)}{E_1 + E_2 + E_R - E_N} .
\label{topt}
\ee
Here the momentum and energy of the nucleon 
in the initial state are
\be
\bm{P}_N &=& (\bm{0}, \, P) ,
\\
E_N    &=& \sqrt{P^2 + M_N^2} ,
\ee
and the momenta and energies of the quark and antiquark are
\be
\bm{p}_1 &=& (\bm{p}_T, \; y z P) ,
\\[1ex]
\bm{p}_2 &=& (\bm{k}_T - \bm{p}_T, \; y \bar z P) ,
\\[1ex]
E_{1, 2} &=& \sqrt{|\bm{p}_{1,2}|^2 + M^2} ,
\ee
such that the total momentum of the pair is 
\beq
\bm{k} \;\; \equiv \;\; \bm{p}_1 + \bm{p}_2 \;\; = \;\; (\bm{k}_T, \; yP). 
\label{k_pair}
\eeq
The momentum of the recoiling classical nucleon state $R$ is fixed by 
the momentum--conserving delta function, and its energy is
\be
E_R &=& \sqrt{|\bm{P} - \bm{p}_1 - \bm{p}_2|^2 + M_N^2} .
\ee
The nucleon and quark momentum states are normalized according 
to the relativistic convention, 
\beq
\langle N(\bm{P}') | N(\bm{P}) \rangle = 2E_N \, (2\pi )^3 \;
\delta^{(3)}(\bm{P}' - \bm{P}), \hspace{1em} \text{etc.}
\label{normalization_nucleon}
\eeq
Eq.~(\ref{topt}) and all subsequent expressions are to be considered
in the limit $P \rightarrow \infty$, where the matrix element 
represents the nucleon wave function in the infinite--momentum frame,
in the approximation specified above.

In Eq.~(\ref{topt}), $\Gamma$ denotes the transition matrix element 
of the retarded potential between classical nucleon states,
\be
\Gamma &\equiv&
\langle R(\bm{P}_R, I_R) | \, [U^{\gamma_5}(0) - 1]_{\rm ret} \,
| N(\bm{P}_N, I_N) \rangle
\nonumber
\\[1ex]
&=& F_\pi^{-1} (\Sigma + i \gamma_5 \tau^a \Pi^a) ,
\label{retarded_fields}
\ee
where $\Sigma$ and $\Pi^a$ are the effective scalar and pseudoscalar
field of the transition. The spin--isospin quantum numbers of the
states are summarily denoted by $I_N \equiv (S_N, T_N, T_{N3}, S_{N3})$,
where $S_N = T_N$ is the spin/isospin and $T_{N3}$ and $S_{N3})$ 
their projections, and similarly for $I_R$. Note that the
recoiling classical nucleon can be in a different rotational state 
from the initial one; in the case of a transition induced by the
quark--antiquark operator considered here it can have 
$N$ or $\Delta$ quantum numbers (we explicitly evaluate the
matrix element below).
Furthermore, in Eq.~(\ref{topt}) $\sigma_{1, 2}$ and $f_{1, 2}$ are
the spin and flavor quantum numbers of the quark and antiquark states,
and $u$ and $v$ their plane--wave Dirac spinor and isospinor wave functions,
normalized as $\bar u u = - \bar v v = 2 M$. The color quantum 
numbers are denoted by $c_{1, 2}$, and the pair is in a 
color--singlet state. 

To evaluate Eq.~(\ref{topt}) further, we note that the energy denominator 
in the limit $P \rightarrow \infty$ is proportional to the invariant mass 
difference between the initial and final state, Eq.~(\ref{delta_m2}),
\beq
E_1 + E_2 + E_R - E_N \;\; \sim \;\; \frac{\Delta s_{\rm tot}}{2P} 
\;\; = \;\;
\;\; \frac{s}{2 y P} ,
\eeq
where we have used the approximation Eq.~(\ref{delta_m2_approx}) in
the last step. We see that we can write the right--hand side 
of Eq.~(\ref{topt}) as
\be
\lefteqn{\frac{\bar u (\bm{p}_1, \sigma_1, f_1) \; \Gamma \; 
v(\bm{p}_2, \sigma_2, f_2)}{E_1 + E_2 + E_R - E_N}}
\nonumber \\[1ex]
&=& 2yP \, \left[ \Sigma \; \psi_\Sigma(z, \bm{p}_T; 
\sigma_1, f_1; \sigma_2, f_2) 
\right.
\nonumber \\[1ex]
&& \left. \;\;\; + \;\; 
\Pi^a \; \psi_\Pi^a (z, \bm{p}_T; \sigma_1, f_1; \sigma_2, f_2 ) \right] .
\ee
Here $\psi_\Sigma$ and $\psi_\Pi^a$ are the infinite--momentum--frame 
wave functions of a quark--antiquark pair with scalar--isoscalar 
and pseudoscalar--isovector quantum numbers
in the effective chiral model,
\be
\psi_\Sigma 
&=& 
\frac{M}{F_\pi} \frac{\bar u(\bm{p}_1, \sigma_1) v(\bm{p}_2, \sigma_2)}{s} 
\; \delta_{f_1 f_2} ,
\label{psi_sigma_1}
\\
\psi_\Pi^a
&=& \frac{M}{F_\pi} 
\frac{\bar u(\bm{p}_1, \sigma_1) i \gamma_5 v(\bm{p}_2, \sigma_2)}{s} 
\; (\tau^a )_{f_1 f_2} . 
\label{psi_pi_1}
\ee
They are independent of the overall
longitudinal momentum and depend only on the quark momentum fraction $z$
and the transverse momentum $p_T$ (we neglect the transverse momentum
of the pair as a whole, $\bm{k}_T$, relative to the intrinsic 
transverse momentum $\bm{p}_T$), as well as on the spin and flavor
quantum numbers.
\subsection{Spin structure of pair correlations}
The explicit expressions for the plane--wave Dirac spinors 
of the quark and antiquark, with the spin projections $\sigma_1$ 
and $\sigma_2$ defined relative to the (fixed) $3$--axis, are
\be
u(\bm{p}_1, \sigma_1) &=& 
\left( \begin{array}{l} \displaystyle
\sqrt{E_1 + M} \; w(\sigma_1)
\\[2ex]
\displaystyle
\sqrt{E_1 - M} \; \frac{\bm{p}_1 \hat{\bm{\sigma}}}{|\bm{p}_1|} \; w(\sigma_1)
\end{array} \right) ,
\\[2ex]
v(\bm{p}_2, \sigma_2) &=& 
\left( \begin{array}{l} 
\displaystyle
- \sqrt{E_2 - M} \; \frac{\bm{p}_1 \hat{\bm{\sigma}}}{|\bm{p}_1|} 
\; \hat\sigma^2 \,
w(\sigma_2)  \\[2ex]
\displaystyle
- \sqrt{E_2 + M} \; \hat\sigma^2 \, w(\sigma_2)
\end{array} \right) , \hspace{2em}
\ee
where $\hat\sigma^i (i = 1, 2, 3)$ are the Pauli spin matrices (to avoid 
confusion with the spin quantum numbers we distinguish them by the hat)
and $w(\sigma_{1, 2})$ is a two--spinor with
\beq
w(\sigma_{1, 2} = 1/2) \; = \; 
\left(\begin{array}{c} 1 \\[.5ex] 0 \end{array}\right),
\hspace{.5em}
w(\sigma_{1, 2} = - 1/2) 
\; = \; \left(\begin{array}{c} 0 \\[.5ex] 1 \end{array}\right).
\eeq
Substituting these expressions and evaluating the bilinear forms in the 
limit $P \rightarrow \infty$, we obtain the spin structure of the pair
wave functions, Eqs.~(\ref{psi_sigma_1}) and (\ref{psi_pi_1}), as
\be
\psi_\Sigma &=& \frac{M}{F_\pi \sqrt{z\bar z}\, s} 
\; \delta_{f_1 f_2} 
\nonumber \\[.5ex]
&\times & 
[(\hat{\bm{\sigma}} \bm{p}_T)\hat\sigma^2 + (2z - 1) M \hat\sigma^3 
\hat\sigma^2]_{\sigma_1 \sigma_2} ,
\label{psi_sigma_2}
\\[1ex]
\psi_\Pi^a
&=& \frac{M}{F_\pi \sqrt{z\bar z}\, s} \; (\tau^a)_{f_1 f_2} 
\nonumber \\[.5ex]
&\times& i [-(\hat{\bm{\sigma}} \bm{p}_T) \hat\sigma^3 \hat\sigma^2 + M 
\hat\sigma^2]_{\sigma_1 \sigma_2} .
\label{psi_pi_2}
\ee
In the terms proportional to $\bm{p}_T$ the spin projections of the 
quark and antiquark are parallel (the matrix is diagonal); they
correspond to configurations with orbital angular momentum $L = 1$.
The presence of these configurations is a direct consequence of
the chirally--odd structure of the coupling of the quark--antiquark
pair to the chiral field ($1, \gamma_5$). Note that these terms 
have the same coefficient in both pairs, up to a phase factor.
In the terms proportional to $M$ in Eqs.~(\ref{psi_sigma_2})
and (\ref{psi_pi_2}) the spin projections of the quark and antiquark 
are antiparallel (the matrix is off--diagonal); in the case of 
$\psi_\Pi^a$ this term would correspond to the spin--flavor wave function 
of the pion in the nonrelativistic quark model. 
\subsection{Restoration of chiral symmetry}
\label{subsec:restoration}
An interesting simplification happens with the pair wave functions
of Eqs.~(\ref{psi_sigma_2}) and (\ref{psi_pi_2}) in the region 
$|\bm{p}_T| \gg M$, where we want to use them to in our approximation
to the nucleon wave function. In this region the first term in the spin wave 
function of Eqs.~(\ref{psi_sigma_2}) and (\ref{psi_pi_2}) dominates, 
and one finds
\be
\sum_{\sigma_1, \sigma_2} \sum_{f_1, f_2} |\psi_\Sigma|^2 &\approx &
\sum_{\sigma_1, \sigma_2} \sum_{f_1, f_2} |\psi_\Pi^a|^2 
\nonumber \\[1ex]
&\approx& \frac{4 M^2 p_T^2}{F_\pi^2 z\bar z \, s^2} 
\;\; \equiv \;\; |\psi_{\rm pair}|^2
\label{restoration}
\ee
(here $a = 1, 2, 3$ is fixed; no summation over $a$). Thus, chiral symmetry 
is effectively ``restored'' in the quark--antiquark pair wave function at 
high momenta. This finding is important in ensuring the chiral invariance
of the one--body density of sea quarks, as shown below. It also
explains the close connection between the flavor--singlet unpolarized
and the flavor--nonsinglet polarized sea quark distributions 
observed in Sec.~\ref{subsec:polarized_sea}.

Equation~(\ref{restoration}) also shows that the high--$p_T$ behavior
of the sea quark density is entirely governed by the $L = 1$ component
of the pair wave function induced by chiral symmetry breaking.
This fact has implications for our general understanding of chiral
symmetry breaking in a light--front wave function description of
the nucleon.
\subsection{Momentum distribution from pairs}
Let us now calculate the one--body momentum density of sea quarks
from the first--order ``wave function'' of Eq.~(\ref{topt}).
Generalizing Eqs.~(\ref{fpt_def}) and (\ref{fpt_anti_def}), the
flavor--singlet unpolarized antiquark density is obtained as
\be
&& \frac{P}{(2\pi)^3} \; \langle N (\bm{P}', I_N) | \;
\nonumber \\
&\times & \sum_{f = u, d} \sum_\sigma \;
b_{f\sigma}^\dagger (\bm{p}) b_{f\sigma}(\bm{p}) \;
| N (\bm{P}, I_N) \rangle
\nonumber \\
&=& 2 E_N (2\pi )^3 \delta^{(3)}(\bm{P}' - \bm{P}) \; 
f_1^{\bar u + \bar d}(x, \bm{p}_T)
\\[1ex]
&& \left[ \bm{p} = (\bm{p}_T, xP) \right] .
\nonumber
\ee
The delta function on the right--hand side appears because, in difference
from Eqs.~(\ref{fpt_def}) and (\ref{fpt_anti_def}), the center--of--mass
motion of the nucleon is now quantized and the states normalized according
to Eq.~(\ref{normalization_nucleon}). Substituting for the nucleon states 
the first--order quark--antiquark component of Eq.~(\ref{topt}), 
and resolving the constraints resulting from momentum conservation,
we get
\be
f_1^{\bar u + \bar d} (x, \bm{p}_T) &=& \frac{P}{(2\pi)^3} 
\int\!\frac{d^3 k}{(2\pi)^3} 
\; \frac{(2yP)^2 \, N_c A}{2 E_1 2 E_2 2 E_R 2 E_N} \hspace{2.5em}
\label{f_from_topt_1}
\\[1ex]
&=& \int  
dy \int d^2 k_T \; \frac{N_c A}{(2\pi)^6 4 y^2 z\bar z} ,
\label{f_from_topt_2}
\ee
where
\be
A &\equiv& \sum_{I_R} \sum_{f_1, f_2} \sum_{\sigma_1, \sigma_2} 
\nonumber \\
&\times& 
(\Sigma^\ast \psi_\Sigma^\ast \; + \; 
\Pi^{a\ast} \psi_\Pi^{a \ast})_{\sigma_1 f_1, \sigma_2 f_2}
\nonumber \\
&\times& 
(\Sigma \psi_\Sigma^{\phantom\ast} \; + \; 
\Pi^{b} \psi_\Pi^{b})_{\sigma_1 f_1, \sigma_2 f_2} .
\label{spin_sum}
\ee
In Eq.~(\ref{f_from_topt_1}) we have chosen the pair momentum $\bm{k}$ 
as integration variable, cf.\ Eq.~(\ref{k_pair}), so that $y = k^3/P$ 
and $z = x/y$ under the integral; in Eq.~(\ref{f_from_topt_2})
we have changed the integration variable from $k^3$ to $y$
and replaced the energies by their values in the 
$P \rightarrow \infty$ limit, with $E_R = (1 - y)P \approx P$
in our approximation. The factor $N_c$ results from the summation 
over the quark--antiquark colors.
The sum over spins and flavors in 
Eq.~(\ref{spin_sum}) is readily evaluated. Because of the 
different symmetry of the $\Pi$ and $\Sigma$ pair wave functions 
the cross terms vanish, and we have
\be
A &=& \sum_{I_R} \sum_{\sigma_1, \sigma_2} \sum_{f_1, f_2}  
\left( \Sigma^\ast \Sigma \; 
\psi_\Sigma^\ast \psi_\Sigma^{\phantom\ast} 
\, + \, \Pi^{a \ast} \Pi^b \; \psi_\Pi^{a\ast} \psi_\Pi^b \right)
\nonumber
\hspace{2.5em}
\\[1ex]
&\approx& \sum_{I_R} ( \Sigma^\ast \Sigma \; + \; \Pi^{a \ast} \Pi^a ) \, 
|\psi_{\rm pair}|^2  \hspace{1em} (p_T^2 \gg M^2).
\ee
In the last step we have taken into account that in the region 
$p_T^2 \gg M^2$ the pair wave functions become alike due to 
the ``restoration of chiral symmetry,'' cf.\ Eq.~(\ref{restoration}).
As a result, the transition $\Pi$ and $\Sigma$ fields effectively
appear in a chirally invariant combination, rendering our result
chirally invariant.

Combining everything, we can represent Eq.~(\ref{f_from_topt_2}) in 
the form of a convolution integral of momentum densities, as
obtained earlier from the gradient expansion, cf.\
Eq.~(\ref{f_grad_lc_simple}), 
\be
f_1^{\bar u + \bar d}(x, p_T) &=& \int_x^\infty \!\frac{dy}{y} \; 
f_{\rm cl} (y) \; f_{q\bar q}(x/y, p_T) , \hspace{2em}
\label{f_grad_topt} 
\ee
where now
\be
f_{\rm cl} (y) &=& \frac{y}{2 (2\pi)^3} \int d^2 \! k_T \;
\sum_{I_R} ( \Sigma^\ast \Sigma + \Pi^{a \ast} \Pi^a ), 
\nonumber \\
\label{f_pi_from_fields}
\\[1ex]
f_{q\bar q} (z, p_T) &=& \frac{N_c |\psi_{\rm pair}|^2}{(2\pi)^3 2 z \bar z}
\hspace{2em} (z = x/y).
\label{f_qqbar_as_overlap}
\ee
These formulas establish the desired connection with the gradient
expansion of the one--body densities in Sec.~\ref{sec:sea}.
The light--cone momentum density of the chiral field 
Eq.~(\ref{f_pi}) appears as
the square of the effective scalar and pseudoscalar fields associated
with the transition between the classical nucleon states in which 
the quark--antiquark pair was produced. The momentum distribution
of the quark--antiquark pair, Eq.~(\ref{f_qqbar_simple}), in turn, 
is given by the squared 
light--cone wave function of the pair, according to the standard
overlap formula (the particular factor $1/[(2\pi)^3 2 z\bar z]$ is a
consequence of our definition of the wave function).

The UV cutoff of the effective dynamics can be introduced into our 
treatment of pair correlated pairs in the nucleon's light--cone wave 
function in the same manner as discussed in Sec.~\ref{subsec:cutoff}.
The invariant mass cutoff Eq.~(\ref{f_qqbar_inv}) can naturally implemented
at the wave function level by multiplying the pair wave functions
with the cutoff function,
\be
\psi_\Sigma &\rightarrow & \psi_\Sigma \, F(s) ,
\\[.5ex] 
\psi_\Pi^a &\rightarrow & \psi_\Pi^a \, F(s) .
\ee
In fact, the physical significance of this regularization scheme
emerges here at the wave function level. Note that the restoration of 
chiral symmetry at $p_T^2 \gg M^2$ is preserved if the $\Sigma$-- and 
$\Pi$--type pairs are suppressed at large $s$ in the same manner.

It remains to show that the light--cone momentum density defined
by Eq.~(\ref{f_pi_from_fields}) coincides with the expression
Eq.~(\ref{f_pi}) obtained from the gradient expansion of the
one--body density. Here we can use the fact that the transition 
matrix element defined by Eq.~(\ref{retarded_fields}) is invariant 
under longitudinal boosts and behaves like a light--cone wave function.
The classical fields are (pseudo--) scalars, and the only effect of
a longitudinal boost is on their space--time dependence, while
the transition matrix element depends only on the four--momentum 
transfer between the initial and final classical nucleon state,
$k \equiv P_N - P_R$. More explicitly, the boost invariance of 
Eq.~(\ref{f_pi_from_fields}) can be demonstrated by considering
the case of small fields that are generated perturbatively 
from a pointlike classical source via a Lorentz--invariant retarded 
Green function. 

We can therefore evaluate Eq.~(\ref{f_pi_from_fields})
in the rest frame, where the initial and final baryon 3--momenta 
are
\be
\bm{P}_N &=& 0, \\
\bm{P}_R &=& (-\bm{k_T}, -yM_N),
\ee
and the 3--momentum transfer to the nucleon is 
\be
\bm{k} &=& \bm{P}_N - \bm{P}_R \;\; = \;\; (\bm{k_T}, yM_N) .
\ee
In the sense of the $1/N_c$ expansion its components are of the order
\beq
y M_N, \; |\bm{k}_T| \;\; = \;\; O(N_c^0) ,
\eeq
and the energy difference between the initial and final classical
nucleon states is
\be
E_R - M_N &=& \sqrt{|\bm{k}|^2 + M_N^2} - M_N
\\
&\approx&  \frac{|\bm{k}|^2}{2 M_N} \;\; = \;\; O(N_c^{-1}) .
\label{energy_difference_classical}
\ee
In this frame the classical nucleon effectively behaves as a 
nonrelativistic system. In particular, as a consequence of
Eq.~(\ref{energy_difference_classical}), the ``plus'' momentum
transfer between the states is in leading order of $1/N_c$ expansion
completely given by the 3--momentum transfer, and we obtain
\be
k^+ &=& P^+ - R^+ \;\; = \;\; M_N - E_R - R^3 
\nonumber \\[1ex]
&=& y M_N
\; + \; O(N_c^{-1}) 
\nonumber \\[1ex]
&=& y P^+ \; + \; O(N_c^{-1}) , 
\ee
as it should be, cf.\ the discussion in Sec.~\ref{subsec:restframe}.
In this frame the space--time dependent chiral field is given by
\beq
U^{\gamma_5}(t, \bm{x})_{\rm ret} 
\;\; = \;\; U^{\gamma_5}_{\rm cl}(\bm{x} - \bm{X}) ,
\eeq
where $\bm{X}$ is the center--of--mass coordinate of the classical
nucleon, and the matrix element between momentum eigenstates of the 
classical nucleon is calculated as \cite{Diakonov:1987ty}
\be
\lefteqn{
\langle \bm{P}_R| [ U^{\gamma_5}(0) - 1]_{\rm ret} | \bm{P}_N \rangle} &&
\nonumber \\[1ex]
&=& 2 M_N \int \! d^3 X  \; e^{i (\bm{P}_R - \bm{P}_N) \bm{X}}
\; [U^{\gamma_5}_{\rm cl}(\bm{x} - \bm{X}) - 1] \hspace{2em}
\nonumber
\\[1ex]
&=& 2 M_N \, U^{\gamma_5}_{\rm cl}(\bm{k})_{\; \bm{k} = (y M_N, \bm{k}_T)}  
\nonumber
\\[1ex]
&=& 2 M_N \, \left[ \frac{1 + \gamma_5}{2} U_{\rm cl}(\bm{k}) \, + \, 
\frac{1 - \gamma_5}{2} U_{\rm cl}(-\bm{k})^\dagger \right] ,
\hspace{2em}
\ee
cf.\ Eq.~(\ref{U_gamma_5_k}); the factor $2M_N$ results from the 
relativistic normalization of states, Eq.~(\ref{normalization_nucleon}).
Finally, the matrix element between rotational states of the classical 
nucleon is obtained by subjecting the classical field to an (iso--) 
spin rotation
\beq
U_{\rm cl} \;\; \rightarrow \;\; R U_{\rm cl} R^\dagger
\label{rotation_U}
\eeq
and calculating the transition matrix elements between the initial 
and recoiling rotational states, using 
\beq
\langle I_R | \ldots | I_N \rangle \;\; = \;\;
\int dR \; \phi_{I_R}(R)^\ast \; \ldots \; 
\phi_{I_N}(R) ,
\label{integral_rotations}
\eeq
where the integration is over the group measure
the rotational wave functions are given in terms of the
the Wigner D--functions \cite{Diakonov:1987ty}. In the case at hand
we actually do not need to calculate the integral 
Eq.~(\ref{integral_rotations}) for a given $I_R$, as the sum
over rotational quantum numbers in Eq.~(\ref{f_pi_from_fields})
produces a delta function that rigidly couples the rotations
in the matrix element and its complex conjugate,
\beq
\sum_{I_R} \phi_{I_R}(R') \; \phi_{I_R}(R)^\ast \;\; = \;\; \delta(R - R') .
\eeq
The rotational average thus reduces to a single integral as in the 
expectation value of the one--body density. Combining everything,
we finally obtain
\be
&& \sum_{I_R} ( \Sigma^\ast \Sigma + \Pi^{a \ast} \Pi^a )
\nonumber \\[1ex]
&=& \frac{F_\pi^2}{8} \, \sum_{I_R} 
\textrm{tr}\, [\Gamma^\dagger \Gamma]
\nonumber \\[1ex]
&=& 2 M_N^2 F_\pi^2 \, \textrm{tr}_{\rm fl} [ U_{\rm cl} (\bm{k})
U_{\rm cl} (\bm{k})^\dagger ] .
\ee
Inserting this result into Eq.~(\ref{f_pi_from_fields}) we obtain
precisely Eq.~(\ref{f_pi}). This completes the proof that the 
first--order nucleon wave function Eq.~(\ref{topt}) reproduces
the gradient expansion result for the one--body momentum density 
of sea quarks discussed in Sec.~\ref{sec:sea}.

The calculation of the sea quark density from correlated pairs in
the nucleon's light--cone wave function presented here can easily be 
extended to the flavor--nonsinglet polarized distribution.
One finds that this distribution originates from the interference
of $\Sigma$-- and $\Pi$--type pairs in the wave function of the
initial and final state; this structure was already discussed in 
connection with the meson cloud model in Ref.~\cite{Dressler:1999zg}.
The restoration of chiral symmetry in the light--cone wave function
of the pairs [cf.\ Eq.~(\ref{restoration}) for the probabilities]
naturally leads to the result that the $\Sigma\Pi$ and $\Pi\Sigma$
wave function overlap at $p_T^2 \gg M^2$ is of the same form
as the $\Sigma\Sigma$ and $\Pi\Pi$ ones producing the 
flavor--singlet unpolarized density. In the conventions of
Sec.~\ref{subsec:polarized_sea} this implies that
$g_{q\bar q}(z, p_T) \equiv f_{q\bar q}(z, p_T)$, 
Eq.~(\ref{g_qqbar_equals_f_qqbar_simple}), which was obtained
independently from the gradient expansion of the respective densities.
Furthermore, the pertinent transition fields $\Sigma$ and $\Pi^a$, 
now defined as matrix elements in which the initial/final nucleon
state has a definite spin projection 1/2 and isospin projection 1/2,
combine to produce the polarized momentum distribution Eq.~(\ref{g_pi_def}).
Thus, the existence of $\Sigma$ and $\Pi$ pair correlations and
the restoration of chiral symmetry at $p_T^2 \gg M^2$ naturally
explain the close connection between the flavor--singlet unpolarized
and the flavor--nonsinglet polarized sea in the chiral quark--soliton
model.
\section{Summary and discussion}
\label{sec:summary}
\subsection{Summary of model results}
In this article we have studied the effect of dynamical chiral 
symmetry breaking on the intrinsic transverse momentum distributions 
of partons, using the chiral quark--soliton model as an approximate 
description of the effective dynamics below the chiral symmetry--breaking 
scale. To conclude our investigation, we would like to summarize the 
findings from our model calculations, revisit the question of the 
embedding in QCD, and outline possible applications of our results to 
deep--inelastic processes.

The results of our study of transverse momentum distributions in
the chiral quark--soliton model can be summarized as follows:
\begin{itemize}
\item The constituent quark picture of the effective chiral dynamics 
implies a natural definition of the intrinsic transverse momentum 
distributions of partons, extending the established description of 
$p_T$--integrated parton densities in this approach. The model
describes the $p_T$--distributions of constituent quarks and
antiquarks, which are to be matched with QCD quarks, antiquarks 
and gluons at the chiral symmetry--breaking scale.
\item
The distribution of valence 
quarks (quarks minus antiquarks) is dominated
by momenta of the order of the inverse nucleon size, 
$p_T \sim M \sim R^{-1}$ and shows an approximate Gaussian shape.
The systematic differences between the unpolarized and polarized 
valence quark $p_T$ distributions attest to the relativistic character 
of the nucleon bound state.
\item
The distribution of sea quarks is qualitatively different and shows
a would--be power--like tail $\sim 1/p_T^2$ extending up to the 
UV cutoff. Its coefficient is determined by low--energy chiral dynamics 
and quasi model--independent. Such behavior is found in both the 
flavor--singlet unpolarized and the flavor--nonsinglet polarized 
distribution.
\item
The UV cutoff of the model influences the shape of the sea quark 
transverse momentum distribution at momenta $p_T^2 \sim \textrm{few} \, M^2$. 
Imposing rather general physical conditions on the regularization scheme 
(analyticity, charge conservation, large--distance behavior) stable 
numerical results for the sea quark distributions are obtained up to 
$p_T^2 \sim 10 \, M^2$. The sea quark $p_T$ distribution thus obtained
is qualitatively different from that of the valence quarks and
extends up to much larger values of $p_T$.
\item
The power--like tail in the sea quark distribution can be explained 
microscopically as the result of short--range correlations in the 
nucleon's light--cone wave function. The relevant configurations
are quark--antiquark pairs with transverse momenta $p_T^2 \gg M^2$
created by the classical chiral field. These pairs can have
scalar--isoscalar ($\Sigma$) and pseudoscalar--isovector ($\Pi$) 
quantum numbers, whose wave functions become identical at $p_T^2 \gg M^2$,
corresponding to an effective ``restoration of chiral symmetry''
at the cutoff scale.
\item 
The coordinate--space correlation function obtained as the Fourier--Bessel 
transform of the sea quark $p_T$ distribution is unambiguously defined in 
the model. At large transverse distances it decays exponentially with a 
characteristic range determined by the constituent quark mass and the 
spatial size of the mean field. At zero transverse distances it coincides 
with the $p_T$--integrated parton density. It is thus tightly constrained 
and largely independent of the regularization scheme.
\item 
The flavor--singlet unpolarized and the flavor nonsinglet polarized
sea quark $p_T$ distributions at $p_T^2 \gg M^2$ satisfy the large--$N_c$
inequalities for transverse momentum distributions, as a result of the
invariance properties of the classical chiral field and the ``restoration 
of chiral symmetry'' in quark--antiquark pairs with $p_T^2 \gg M^2$.
\end{itemize}

When interpreting the results of our study or using them for phenomenology, 
one should keep in mind that the specific numerical results for the valence 
and sea quark transverse momentum distributions (see 
Figs.~\ref{fig:f1_val_sea} and \ref{fig:g1_val_sea}) are model--dependent 
and should not be taken too literally. The numerical values of the sea 
quark transverse momentum distributions are strongly affected by the
UV cutoff, and while it is very encouraging that the physically motivated 
regularization schemes described here produce stable distributions, 
we cannot exclude that further refinement of the UV regularization
would change the numerical values. Furthermore, as explained in 
Sec.~\ref{sec:introduction} and discussed further in the following
subsections, the distributions refer to effective degrees of freedom 
defined by the model, and we presently cannot express them objectively
as matrix elements of a priori defined QCD operators in the nucleon.
Our main result is the \textit{qualitative difference} between the
valence and sea quark transverse momentum distributions. We expect
it to be model--independent because it is rooted in the basic
structure of the effective dynamics resulting from chiral symmetry
breaking and can at least qualitatively be explained in QCD 
proper (see Sec.~\ref{subsec:toward_QCD}). Any applications of
our results should therefore focus on this qualitative difference 
rather than the specific numerical values of the distributions
(see Sec.~\ref{sec:applications}).

%
%
\begin{figure*}
\begin{tabular}{ll}
\includegraphics[width=.45\textwidth]{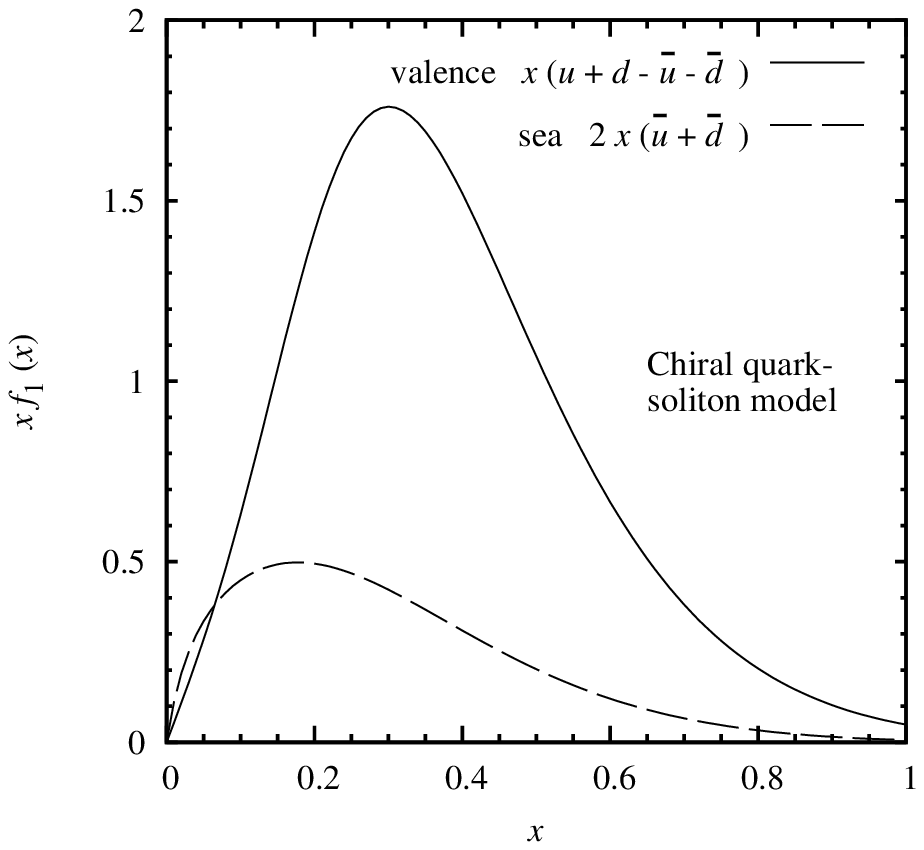}
& 
\includegraphics[width=.45\textwidth]{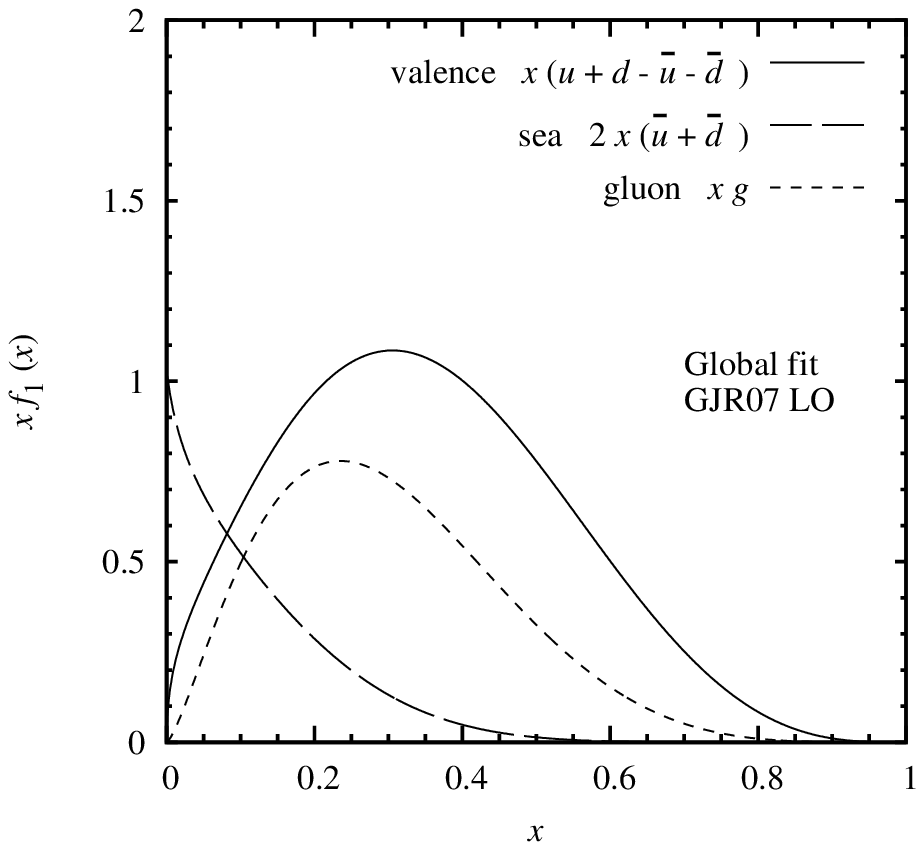}
\\[-2ex]
(a) & (b)
\end{tabular}
\caption[]{Unpolarized flavor--singlet parton momentum densities 
$x f_1^{u + d - \bar u - \bar d} (x)$ and $x f_1^{\bar u - \bar d} (x)$.
The sea quark momentum densities are plotted multiplied by a factor 2, 
with which they appear in the momentum sum rule. (a) Chiral 
quark--soliton model. Shown are the results obtained with the
numerical approximations Eqs.~(\ref{approx_val}) and (\ref{interpolation})
[self--consistent soliton profile Eq.~(\ref{P_self}),
$M = 0.35\, \textrm{GeV}, M_N = 3.26 \, M$; sea quark distribution 
with PV regularization]. The constituent quarks and antiquarks carry 
the entire light--cone momentum of the nucleon in the model. 
(b) QCD parton densities from the global fit of Ref.~\cite{Gluck:2007ck} 
(GJV07 LO, scale $\mu^2_{\rm LO} = 0.3\, \textrm{GeV}^2$). Gluons carry 
$\sim 30\%$ of the nucleon's light--cone momentum.}
\label{fig:fx}
\end{figure*}

The study of the $p_T$ distributions in the chiral quark--soliton model
presented here could be extended in various directions. One obviously 
should calculate also the $p_T$ distribution of the $1/N_c$--suppressed
combinations of the parton densities, namely the flavor--nonsinglet 
unpolarized densities, $f_1^{u - d}$ and $f_1^{\bar u - \bar d}$ and
the flavor--singlet polarized ones, $g_1^{u + d}$ 
and $g_1^{\bar u + \bar d}$. These distributions are proportional to
the angular velocity of the isorotational motion of the soliton, 
$\Omega \sim 1/N_c$ and the corresponding expressions can be derived 
in analogy to those of the leading distributions in 
Sec.~\ref{subsec:evaluation}, using the techniques described in 
Refs.~\cite{Diakonov:1996sr,Pobylitsa:1998tk}. An interesting question
is whether the $1/N_c$--suppressed sea quark distributions also exhibit
a power--like tail at $p_T^2 \gg M^2$, and to what extent 
this behavior would differ from that of the $1/N_c$--leading distributions. 
It is known that
the UV cutoff dependence of the flavor--nonsinglet unpolarized
distribution in the model is qualitatively different from that of the 
singlet distribution at parametrically small values 
$x \sim M/(\Lambda N_c)$ \cite{Pobylitsa:1998tk}, 
suggesting a qualitatively different behavior also of the $p_T$ 
distributions. 

Equally interesting is the $p_T$ dependence of the quark transversity 
distributions $h_1^q$ and $h_1^{\bar q}$. Earlier 
studies \cite{Pobylitsa:1996rs,Schweitzer:2001sr} revealed that the sea 
quark transversity distributions in the chiral quark--soliton model are
finite in the limit of large UV cutoff and numerically small, which
would indicate that there is no $\sim 1/p_T^2$ tail in the $p_T$
distribution. It would be especially interesting to explain the
different behavior of the transversity distributions at the wave 
function level, using the method developed in Sec.~\ref{sec:correlations}.

Another direction is the study of short--range parton correlations 
at the wave function level. In Sec.~\ref{sec:correlations} we 
calculate the nucleon wave function at $p_T^2 \gg M^2$ in the simplest 
possible approximation, which violates chiral invariance (although it 
is effectively restored at high $p_T$). It would be worth developing
a manifestly chirally invariant expansion scheme, which takes into
account the distortion of the pair wave function by the classical
field. This could be done using the induced vector field 
representation of the effective chiral dynamics, in which the 
original matrix field $U^{\gamma_5}$ is absorbed by a chiral rotation
of the massive fermion fields. Furthermore, one should further explore
the connection of our approximation with the Fock space expansion of
the large--$N_c$ baryon wave function of 
\cite{Petrov:2002jr,Diakonov:2004as,Lorce:2007as}.
\subsection{Matching with QCD}
\label{subsec:matching}
The approach to partonic structure presented here is based on the idea of 
an effective description of QCD below the chiral symmetry--breaking scale. 
The chiral quark--soliton model of the nucleon describes the transverse 
momentum distribution and correlations of constituent quarks and 
antiquarks --- effective degrees of freedom, which are to be matched with 
QCD quarks, antiquarks and gluons at the chiral symmetry--breaking scale. 
As explained in Sec.~\ref{sec:introduction}, this matching cannot be 
performed entirely on the basis of intrinsic properties of the effective 
chiral dynamics but requires additional information about its embedding 
in QCD, either in the form of a microscopic derivation of the effective 
model or of phenomenological input. A detailed treatment of this problem 
is beyond the scope of the present study. Still we would like to comment 
on several aspects and draw some conclusions based on the structure of 
the results obtained within the effective model.

To discuss the matching quantitatively it is instructive to compare the 
$p_T$--integrated parton densities in the chiral quark--soliton model, 
which were studied in several earlier 
works \cite{Diakonov:1996sr,Diakonov:1997vc,Weiss:1997rt},
with empirical parametrizations of the parton densities obtained
from global fits (see Fig.~\ref{fig:fx}). Fig.~\ref{fig:fx}a shows 
the unpolarized flavor--singlet momentum densities of valence and
sea quarks in the model, $x f_1^{u + d - \bar u - \bar d}(x)$
and $x f_1^{\bar u + \bar d}(x)$. For consistency we show here the results
obtained with the numerical approximations employed in the present 
study of $p_T$ distributions, Eqs.~(\ref{approx_val}) and 
(\ref{interpolation}); these approximations reproduce the exact numerical 
results for the $p_T$--integrated valence and sea quark 
densities \cite{Weiss:1997rt}
with a relative accuracy far better than $10\%$ and $20\%$, respectively,
in the region $x < 0.5$. Note that the momentum sum rule is satisfied 
within the model in leading order of the $1/N_c$ 
expansion \cite{Diakonov:1996sr}, which is another 
testimony to the consistency of the relativistic mean--field approximation.
The sum rule is preserved by the PV regularization used here;
see Ref.~\cite{Weiss:1997rt} for a detailed discussion. 
The distributions of constituent quarks and antiquarks obtained
in the model thus satisfy
\beq
\int dx \, x \, [ f_1^{u + d - \bar u - \bar d}(x) 
\; + \; 2 f_1^{\bar u + \bar d}(x)]_{\rm model}
\;\; = \;\; 1 .
\label{momentum_sr_model}
\eeq
In Fig.~\ref{fig:fx}a the sea quark distribution is plotted including 
the factor 2 with which it appears in Eq.~(\ref{momentum_sr_model}).
Figure~\ref{fig:fx}b shows the empirical parametrizations of the QCD
parton momentum densities obtained in a recent global fit of
DIS and other data \cite{Gluck:2007ck}. 
For clarity we show here the leading--order (LO) distributions,
which are independent of the renormalization scheme and can readily
be compared with the model distributions. The scale of these 
distributions is $\mu^2_{\rm LO} = 0.5 \, \textrm{GeV}^2$,
corresponding approximately to the chiral--symmetry breaking
scale $\rho^{-2} \approx 0.4 \, \textrm{GeV}^2$ identified
in the nonperturbative QCD approaches discussed in 
Sec.~\ref{sec:introduction}. 
In the QCD parton densities the momentum is shared between quarks, 
antiquarks and gluons, and the momentum sum rule reads
\beq
\int dx \, x \, [f_1^{u + d - \bar u - \bar d}(x) 
\; + \; 2 f_1^{\bar u + \bar d}(x) \; + \; g(x)]_{\rm QCD}
\;\; = \;\; 1 .
\label{momentum_sr_fit}
\eeq
According to the fit of Ref.~\cite{Gluck:2007ck} 
$\sim 30\%$ of the nucleon's momentum at the
low scale is carried by gluons. Comparing the model and the QCD parton
distributions in Fig.~\ref{fig:fx}a and b two features stand out. 
First, the total momentum carried by sea quarks is roughly the same;
however, the constituent sea quark distributions are noticeably
harder (stronger at larger $x$) than those of the QCD antiquarks. 
This shows that the matching of the model sea quark distribution with
QCD partons is generally nontrivial and may produce and evolution--like
effect. [Note that the model distributions should not be considered as 
genuine predictions in the parametrically small region 
$x \sim M^2 / (\Lambda^2 N_c)$, where they are sensitive to the details of 
the UV cutoff (see Sec.~\ref{sec:sea}).] Second, the momentum carried 
by the valence quarks is smaller in the case of the QCD distributions. 
This suggests that most of the QCD gluons should be ``generated'' from 
the model valence quark distributions in the matching process.

In the simplest approximation one can identify the constituent quarks 
and antiquarks of the effective model with QCD quarks and antiquarks at 
the chiral symmetry--breaking scale and set the gluon distribution in
the model to zero. For the $p_T$--integrated parton densities this 
approximation can be formally justified in the instanton model of the 
QCD vacuum, where it appears as the leading--order approximation in the 
expansion of the instanton packing fraction. Fig.~\ref{fig:fx} and
the above discussion show that this approximation is reasonable 
(at non--exceptional values of $x$) but of limited accuracy.
The limitations of this approximation should be kept in mind
when interpreting our results for the $p_T$ distributions.

An important lesson from the fit shown in Fig.~\ref{fig:fx}b is that 
the empirical gluon density at the low scale shows substantial 
strength at large values of $x > 0.2$ and has a shape comparable to
the valence quark distribution. Such behavior is difficult to
explain as the result of a composite structure of \textit{individual}
constituent quarks and antiquarks, which, like DGLAP evolution,
would produce gluons primarily at values of $x$ much
smaller than that of the quarks. It suggests that this component of the 
gluon density originates rather from \textit{correlations} between 
constituent quarks, which can leverage the sum of the $x$--values of the 
two correlated quarks for the produced gluons. 
This fact has not been appreciated
in most attempts to explain the empirical gluon density made so far.
Whether the relevant correlations are between quark--antiquark pairs in
the sea, as found in the present model based on effective chiral 
dynamics, or diquark--like correlations between valence quarks, 
is an important question which deserves further study.
\subsection{Toward parton correlations in QCD}
\label{subsec:toward_QCD}
The chiral quark--soliton model predicts short--range correlations 
between constituent quarks and antiquarks as a consequence of a 
basic property of the effective chiral dynamics: $\Lambda^2 \gg M^2$, 
or the parametric smallness of the dynamical quark mass compared to
the UV cutoff representing the chiral symmetry--breaking scale.
An important question is whether and how such parton correlations could 
be rigorously defined in QCD. The challenge lies in the fact that the 
relevant dynamical scale arises from chiral symmetry breaking in the 
QCD vacuum, which is not readily associated with partonic structure in 
a model--independent manner. Gribov's concept of the partonic wave 
function of a fast--moving hadron \cite{Gribov:1973jg} maintains the
connection between partons and vacuum structure and should in principle 
be appropriate for discussing parton correlations as proposed here; 
however, the concept was developed based on scalar field theory, and 
the extension to QCD presents many technical challenges (gauge invariance, 
UV divergences, renormalization). A more rigorous formulation of
parton short--range correlations in QCD may be possible with the 
new concept of multiparton distributions, which were introduced in 
the description of multiple hard scattering processes in high--energy
$pp$ collisions; their operator definition and renormalization properties
have been studied in several recent works \cite{Blok:2010ge,Diehl:2011yj}.

\begin{figure}
\includegraphics[width=.32\textwidth]{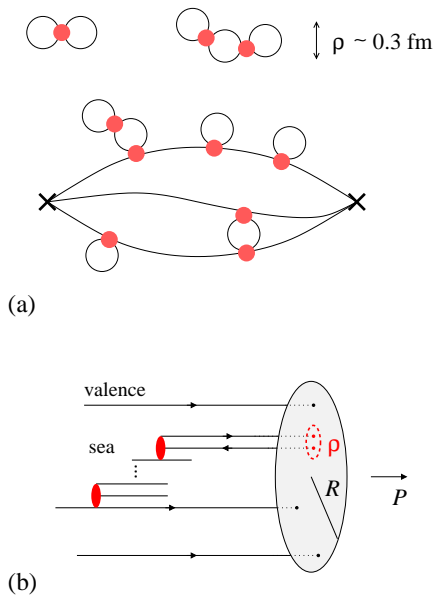} 
\caption{(Color online) Qualitative picture of the emergence of 
nonperturbative parton short--range correlations in QCD (details see text).
(a) Slow--moving nucleon. (b) Fast--moving nucleon (parton picture).}
\label{fig:corr}
\end{figure}

Short of a rigorous formulation, we can still develop a qualitative
picture of how parton short--range correlations emerge from dynamical
chiral symmetry breaking in QCD. In the usual equal--time formulation 
of relativistic dynamics the QCD vacuum is not empty, but populated by 
localized nonperturbative gluon fields. These fields create 
quark--antiquark pairs with a characteristic size 
$\rho \sim 0.3 \, \textrm{fm}$ [cf.\ Eq.~(\ref{m_0^2}) and following
discussion], which form the chiral condensate (see Fig.~\ref{fig:corr}a). 
Quarks propagating through this medium interact with the vacuum fields 
and effectively acquire a dynamical mass, which determines much of 
hadronic structure. This phenomenon has been 
investigated extensively in slow--moving hadronic states 
$(P \lesssim \rho^{-1})$, whose properties can be studied
using Euclidean (imaginary--time) correlation functions. A partonic description
appears when considering hadrons which move with a momentum much larger than 
the scale of the vacuum fluctuations, $P \gg \rho^{-1}$. Following 
Gribov \cite{Gribov:1973jg},
in this limit one can separate the quanta carrying a finite fraction of the 
hadron momentum from those ``left behind'' in the vacuum, and the hadron
becomes a closed system in the quantum--mechanical sense, amenable to
a wave function description. When we imagine approaching the regime of 
large momentum $P \gg \rho^{-1}$ gradually, it is clear that some of the 
quark--antiquark pairs in the vacuum will be ``dragged along'' and become 
the sea quarks in the nucleon's partonic wave function
(see Fig.~\ref{fig:corr}b). These pairs of 
course inherit the typical size $\rho$ with which they exist in the vacuum
and thus induce nonperturbative short--range correlations in the nucleon's 
partonic wave function. In this argument we implicitly assume that 
the nonperturbative wave function is defined ``at a scale of the 
order $\rho^{-2}$,'' and that configurations with transverse momenta
$p_T^2 \gtrsim \rho^{-2}$ will be built up by perturbative QCD radiation.
While leaving aside many important questions (UV divergences, 
renormalization) this simple picture qualitatively explains how parton 
short--range correlations emerge from chiral symmetry breaking 
in QCD \cite{Dorokhov:1993fc}.
%
%

The picture described here implies that the small--size $q\bar q$ pairs 
in the nucleon wave function are generally accompanied by strong gauge 
fields with transverse momenta of the order $\rho^{-1}$. These gauge 
fields need not necessarily project onto physical gluon states in the 
limit $P \rightarrow \infty$ (e.g., they can correspond to unphysical 
polarization states), but can exert forces on the quark and antiquark 
corresponding to higher--twist effects. A transverse momentum--dependent 
hard scattering process involving sea quarks with $p_T \sim \rho^{-1}$
thus generally takes place in the presence of a strong gluon field.
One should therefore expect sizable corrections to the impulse
approximation, in which one takes into account the $p_T \sim \rho^{-1}$ 
of the initial quark/antiquark but not the equally strong final--state 
interaction with the small--size gluon field. For this reason we cannot 
use our calculated $p_T$ distributions directly to make numerical predictions 
for the transverse momentum distributions in hard processes.
Our conclusions regarding semi--inclusive DIS presented below assume
only on the existence of the short--distance scale and do not rely
on the impulse approximation. 

Modeling the effect of the nonperturbative gluon fields at the chiral 
symmetry--breaking scale in hard scattering processes will be essential
for putting the proposed picture of short--range correlations in QCD on 
a more quantitative footing. The instanton vacuum model has proved to be
a valuable tool for estimating the effect of such fields in inclusive 
DIS, where one can apply the local operator product
expansion to identify the scaling (leading--twist) and power--suppressed
(higher--twist) parts of the cross 
section \cite{Balla:1997hf,Dressler:1999zi,Weiss:2002qq}. 
Whether the instanton 
vacuum model could be adapted to estimate also final--state interactions 
in semi--inclusive processes is an interesting question for further study.

Another interesting question concerns the color structure of parton
short--range correlations in QCD. In the effective model used in the 
present study the chiral symmetry--breaking color fields are integrated 
out, and the resulting dynamics generates short--range quark--antiquark
correlations only in color--singlet states ($\Sigma, \Pi$). The parameter 
$N_c$ in the model really plays the role of a degeneracy or weight
factor and no longer refers to actual color interactions. In a more
microscopic approach such as the instanton vacuum, short--range
correlations could in principle appear also between quark--antiquark 
pairs in color--octet states, if the color is compensated by physical
gluon fields. Such pair correlations would have different properties
as initial conditions for QCD radiation starting from the scale $\rho^{-2}$,
and would influence the final state in semi--inclusive processes in a
different way, compared to color--singlet correlations. Investigating the 
possibility of such color--octet correlations in the instanton vacuum 
model would be an interesting problem for further study.

In sum, our arguments suggest that the short--distance scale associated 
with dynamical chiral symmetry breaking in the QCD vacuum leaves a
characteristic imprint on the nucleon's partonic structure. It implies
a definite pattern of intrinsic transverse momentum distributions,
in which the distribution of valence quarks is concentrated at $p_T^2$ 
of the order of the inverse hadronic size $R^{-2}$, while that of sea 
quarks extends up to the chiral symmetry--breaking scale 
$\rho^{-2} \gg R^{-2}$. At the level of the partonic wave function, 
the sea quarks show
short--range pair correlations that reflect their creation by 
nonperturbative gluon fields of transverse size $\rho$.
These conclusions are generic and rely only on the existence of the
nonperturbative short--distance scale in the QCD vacuum; they do not 
depend on the particular implementation of this scenario in the 
effective dynamical model employed here.
\section{Applications to deep--inelastic processes}
\label{sec:applications}
Our findings about the intrinsic transverse momentum distributions
of partons have many implications for DIS
experiments with identified particles in the final state. 
A quantitative description of the transverse momentum spectra in 
particle production requires further assumptions about perturbative
QCD radiation and final--state interactions and detailed modeling of 
the fragmentation process, and is beyond the scope of the present article. 
In the following we only want to point out several obvious consequences 
and applications of our results that can be stated model--independently 
and merit further detailed study.
\subsection{Perturbative QCD radiation}
When applying our results to deep--inelastic processes with identified
particles it is important to take into account the effects of perturbative 
QCD radiation. The picture of intrinsic transverse momentum 
distributions and parton 
short--range correlations in the nucleon wave function described in 
Sec.~\ref{sec:summary} applies at the chiral symmetry--breaking scale 
$\rho^{-2}$. Generally, in a hard process perturbative QCD radiation 
builds up configurations with invariant masses (or virtualities) of the 
order of the hard scale, $Q^2$. In inclusive DIS
this radiation is described by DGLAP evolution and well understood.
In processes where one measures the transverse momenta of particles
in the final state the relevant radiation processes are generally much 
more complex; for semi--inclusive DIS this problem was recently studied 
in Ref.~\cite{Aybat:2011zv}. A key question is whether the nonperturbative 
short--distance scale implied by chiral symmetry breaking, 
$\rho^{-1} \approx 0.6\, \textrm{GeV}$, acts as an infrared cutoff 
for perturbative radiation in such processes, or whether such
radiation (possibly with Sudakov suppression) is still relevant 
at lower scales. Experimental evidence on this point is ambiguous.
Data on jet structure in $e^+e^-$ annihilation can be explained by 
allowing perturbative radiation to substantially lower scales. 
At the same time, there are hints from exclusive processes that 
low--virtuality radiation is suppressed, e.g.\ in the surprisingly similar 
magnitude and energy dependence of transverse and longitudinal 
cross sections in $\rho^0$ meson production \cite{Aaron:2009xp}.
What these observations imply for semi--inclusive particle production 
is an interesting question which merits further study.
\subsection{Semi--inclusive measurements}
Semi--inclusive DIS with single identified hadrons in the current
fragmentation region, $\gamma^\ast + N \rightarrow h + X$, is a standard
tool for separating the different charge and flavor components of the 
nucleon's parton densities, using the fact that the 
fragmentation process is sensitive to the charge and flavor of the
struck quark. In these measurements one integrates over the transverse
momentum $P_{T, h}$ of the identified hadron $h$ and aims to extract the 
cross section as a function of the fraction $z_h$ of the virtual photon 
energy in the target rest frame carried away by the hadron.
Generally, the observed transverse momentum $P_{T, h}$ is compounded
from the intrinsic transverse momentum of the struck parton, QCD 
final--state interactions and perturbative radiation, and the 
transverse momentum incurred during the soft fragmentation process. 
Our findings about the intrinsic transverse momentum distributions imply 
that hadrons produced in scattering from antiquarks generally have a much 
broader $P_{T, h}$ distribution than those produced from quarks, if
the transverse momenta incurred from final--state interactions and
fragmentation are comparable in both cases. One example is the
production of $K^+$ (valence quark content $u\bar s$) and 
$K^- (\bar u s)$, where our picture predicts a broader $P_{T, h}$ 
distribution for the latter, assuming the production is dominated 
by scattering from $u$ and $\bar u$ quarks in the proton.
In experiments with incomplete coverage in $P_{T, K\pm}$ this can result
in a modification of the observed numbers of $K^+$ and $K^-$ that is
not related to the $u$ and $\bar u$ number densities in the target 
and must be corrected in the charge/flavor separation.

Another instance where this effect plays a role are measurements
where relations between fragmentation functions are used to isolate
certain combinations of parton densities
\cite{Frankfurt:1989wq,Christova:2000nz}. For example, the 
cross sections for semi--inclusive $\pi^+$ and $\pi^-$ production,
integrated over the pion transverse momentum $P_{T, \pi}$, is
up to an overall kinematic factor given by
\be
\sigma^{\pi^\pm} &\propto& e_u^2 \; [f_1^u(x) \, D_1^{u/\pi^\pm}(z_\pi)
+ f_1^{\bar u}(x) \, D_1^{\bar u/\pi^\pm}(z_\pi)]
\nonumber \\
&+& e_d^2 \; [f_1^d(x) \, D_1^{d/\pi^\pm}(z_\pi) + f_1^{\bar d}(x) 
\, D_1^{\bar d/ \pi^\pm}(z_\pi)] , 
\hspace{2em}
\label{pi_integrated}
\ee
where $e_{u, d}$ are the quark charges and 
$D_1^{u/\pi^+}(z_\pi)$, etc., the fragmentation 
functions describing the inclusive probability of a quark/antiquark 
to produce a $\pi^\pm$ carrying fraction $z_\pi$ of its longitudinal momentum, 
integrated over the soft transverse momenta. Taking the difference of 
$\pi^+$ and $\pi^-$ cross sections, and using the relations
between the fragmentation functions following from charge conjugation
invariance,
\be
D_1^{u/\pi^\pm}(z_\pi) &=& D_1^{\bar u/\pi^\mp}(z_\pi),
\nonumber \\
D_1^{d/\pi^\pm}(z_\pi) &=& D_1^{\bar d/\pi^\mp}(z_\pi),
\label{fragmentation_charge}
\ee
and from isospin symmetry,
\be
D_1^{u/\pi^\pm}(z_\pi) &=& D_1^{d/\pi^\mp}(z_\pi) ,
\nonumber
\\
D_1^{\bar u/\pi^\pm}(z_\pi) &=& D_1^{\bar d/\pi^\mp}(z_\pi) ,
\label{fragmentation_isospin}
\ee
one obtains
\be
\sigma^{\pi^+} - \sigma^{\pi^-} &\propto& 
(e_u^2 - e_d^2) [f_1^u(x) - f_1^{\bar u}(x) - f_1^d(x) + f_1^{\bar d}(x)] 
\nonumber \\
&\times& [D_1^{u/\pi^+}(z_\pi) - D_1^{u/\pi^-}(z_\pi)] .
\ee
A measurement of this cross section difference thus provides direct
access to the valence quark densities in the target; the sea quark
densities drop out because of the relations between the fragmentation
functions. However, this reasoning requires modification if the effects 
of transverse momenta
are taken into account \textit{and} the experiment covers only part of 
the relevant $P_{T, \pi}$--distribution. To illustrate the point, we may 
use the simple parton model with intrinsic transverse momenta in the 
parton density and the fragmentation function, where it is assumed
that all transverse momentum integrals converge because of some
intrinsic soft scale (our conclusion is model--independent and holds also 
in the presence of QCD radiation and final--state interactions). 
In this model the transverse momentum distribution of the identified 
hadron is given by the convolution of the intrinsic transverse momentum 
distribution of quarks in the target and that incurred in the 
fragmentation process,
\be
\sigma^{\pi^\pm}(P_{T, \pi}) &\propto& 
\int \! d^2 p_T \int \! d^2 K_T \; \delta^{(2)}(\bm{P}_{T, \pi}
- z_\pi\bm{p}_T - \bm{K}_T) 
\nonumber \\
&\times & [ e_u^2 \, f_1^u(x, p_T) \, D_1^{u/\pi^\pm}(z_\pi, K_T)
\; + \; \ldots ] ,
\label{pi_diff_int}
\ee
where the ellipsis denotes the corresponding other terms appearing
in Eq.~(\ref{pi_integrated}). Here $D_1^{u/\pi^+}(z_\pi, K_T)$ etc.\ denote 
the transverse momentum dependent fragmentation function, which 
satisfy
\beq
\int \! d^2 K_T \; D_1^{u/\pi^+}(z_\pi, K_T) \;\; = \;\; D_1^{u/\pi^+}(z_\pi) ,
\hspace{2em} \textrm{etc.}
\eeq
The $K_T$--dependent fragmentation functions obviously obey the same
charge conjugation and isospin symmetry relations as the $K_T$--integrated 
ones, Eqs.~(\ref{fragmentation_charge}) and (\ref{fragmentation_isospin}).
As a result, one obtains a formula analogous to Eq.~(\ref{pi_diff_int}) 
for the difference of cross sections measured at the same $P_{T, \pi}$:
\be
\lefteqn{[\sigma^{\pi^+} - \sigma^{\pi^-}](P_{T, \pi})} && 
\nonumber
\\[.5ex]
&\propto& \int \! d^2 p_T \int \! d^2 K_T \;
\delta^{(2)}(\bm{P}_{T, \pi} - z_\pi\bm{p}_T - \bm{K}_T) 
\nonumber 
\\[.5ex]
&\times& (e_u^2 - e_d^2) \left[ f_1^u(x, p_T) - f_1^{\bar u}(x, p_T)
\phantom{f^{\bar d}}
\right.
\nonumber
\\[.5ex]
&& \left. - \; f_1^d(x, p_T) + f_1^{\bar d}(x, p_T) \right]
\nonumber 
\\[.5ex]
&\times& [ D_1^{u/\pi^+}(z_\pi, K_T) - D_1^{u/\pi^-}(z_\pi, K_T)] 
\label{pi_diff_pt}
\ee
The cross section difference is proportional to the difference of 
$p_T$--dependent quark and antiquark distributions. If one integrated 
over the pion transverse momentum $\bm{P}_{T, \pi}$ 
\textit{without restriction,} the delta function in Eq.~(\ref{pi_diff_pt}) 
would disappear, and the unrestricted integrals over $p_T$ and $K_T$ 
would reduce the transverse momentum--dependent parton distribution and 
fragmentation functions to the integrated functions of Eq.~(\ref{pi_diff_int}).
However, one can no longer extract the \textit{$p_T$--integrated}
valence quark density from Eq.~(\ref{pi_diff_pt}) from measurements 
with incomplete $P_{T, \pi}$ coverage if quarks and antiquarks have 
different intrinsic $p_T$ distributions, as implied by our arguments based 
on the QCD vacuum structure. To see this, let us write the $p_T$--dependent
quark and antiquark distributions in the simple parton model in the form
\be
f_1^u(x, p_T) &=& f_1^u(x) \, F_1^u (x, p_T),
\\
f_1^{\bar u}(x, p_T) &=& f_1^{\bar u}(x) \, F_1^{\bar u} (x, p_T),
\ee
and similarly for the $d$ quarks and antiquarks. Here the 
functions $F_1^{a}$ describe the normalized $p_T$ 
profile of the quarks and antiquarks at a given $x$,
\beq
\int \! d^2 p_T \; F_1^{a} (x, p_T) \;\; = \;\; 1 ,
\hspace{2em}  (a = u, \bar u, d, \bar d).
\label{profile_normalization}
\eeq
We can then write the differences of $p_T$--dependent distributions
appearing in Eq.~(\ref{pi_diff_pt}) in the form
\be
\lefteqn{f_1^u(x, p_T) - f_1^{\bar u}(x, p_T)} && 
\nonumber \\
&=& [f_1^u(x) - f_1^{\bar u}(x)] \, F_1^u(x, p_T)
\nonumber \\
&+& f_1^{\bar u}(x) \, [F_1^u(x, p_T) - F_1^{\bar u}(x, p_T)] ,
\hspace{1em} \textrm{etc.}
\ee
If one integrated over $p_T$ without restriction, as would correspond 
to integration of Eq.~(\ref{pi_diff_pt}) over all $P_{T, \pi}$, the 
difference of profile functions on the last line would integrate to
zero because of the normalization conditions 
Eq.~(\ref{profile_normalization}), and the measured cross section
difference would be proportional to the integrated valence quark
density alone. However, without integration 
over all $p_T$ the cross section difference contains an admixture 
of the antiquark distribution because generally 
$F_1^{\bar u}(x, p_T) \neq F_1^u(x, p_T)$. Consequently, one cannot
extract the $p_T$--integrated valence quark density from measurements 
of Eq.~(\ref{pi_diff_pt}) over an incomplete range of $P_{T, \pi}$
without additional assumptions about the intrinsic transverse momentum
dependence. This conclusion is general and not limited to the simple
parton model used to illustrate the point. It also affects the use of
the pion charge asymmetry (either as an absolute cross section
difference or as a ratio of cross sections) in the flavor separation
of polarized parton densities with proton and nuclear targets.
In practice, there might still be considerable experimental
advantages in using observables such as Eq.~(\ref{pi_diff_pt})
for the study of $p_T$--integrated parton densities; however, 
this requires detailed study based on the measured or modeled
transverse momentum distributions.

Generally, our findings underscore the importance of accurate measurements 
of the basic transverse momentum distributions of hadrons ($\pi, K$) in 
unpolarized semi--inclusive DIS, up to $P_{T, h} \sim 1\, \textrm{GeV}$, 
and differentially in $x, z$ and $Q^2$. Simple comparisons between 
the $P_T$ distribution of different particles (such as $K^+$ and $K^-$) 
can serve as model--independent tests of the predicted pattern of 
intrinsic transverse momentum distributions. Measurements with a
deuteron target would be particularly useful, as they directly access
the flavor--singlet (isoscalar) combination of the parton distributions
and would allow one to search for signals of a difference between the 
transverse momentum distributions $f_1^{u + d}$ and $f_1^{\bar u + \bar d}$
without the complications of flavor separation. Such measurements provide 
essential information about the mechanism of particle production in 
semi--inclusive DIS and should be done before one studies more subtle 
observables such as spin asymmetries. 
\subsection{Correlation measurements}
\label{subsec:correlation}
\begin{figure}
\includegraphics[width=.36\textwidth]{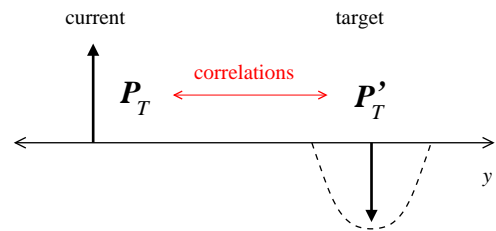} 
\caption{(Color online) 
Measurements of particle production in the central and target 
fragmentation regions in correlation with an identified hadron in the 
current fragmentation region. $y$ denotes the rapidity.}
\label{fig:rapidity}
\end{figure}

Much more information can be obtained from semi--inclusive experiments
that measure particle production in the central and target fragmentation
regions in correlation with an identified hadron in the current 
fragmentation region (see Fig.~\ref{fig:rapidity}). Such measurements can 
answer the question of what ``balances'' the $P_{T, h}$ of hadrons observed
in the current fragmentation region, which is the key for unraveling the
production mechanism in semi--inclusive DIS. As emphasized earlier,
the $P_{T, h}$ of a hadron observed in the current fragmentation region can 
come from  the parton intrinsic transverse momentum in the 
target, final--state interactions and QCD radiation in the hard process,
or the soft fragmentation process. Single--inclusive measurements
alone cannot discriminate between these different sources.
Correlation measurements offer additional observables that can test
at least the relative importance of the various mechanisms.
A fully quantitative theory of such measurements is a 
complex problem and beyond the scope of the present article. 
Here we only wish to outline in what kinematic region such experiments 
could be performed such that they have a simple interpretation in terms
of nonperturbative nucleon structure.

In DIS experiments with measurements of multiparticle final states
it is convenient to describe the produced particles in terms of 
their rapidity
\beq
y_h \;\; \equiv \;\; \frac{1}{2} 
\ln \frac{E_h + P_{\parallel, h}}{E_h - P_{\parallel, h}} ,
\eeq
where $E_h = (P_{\parallel, h}^2 + P_{T, h}^2 + m_h^2)^{1/2}$ is the hadron 
energy and $P_{\parallel, h}$ the longitudinal momentum, defined as
the component in the direction of the virtual photon momentum. 
The rapidity changes by a constant
under Lorentz boosts along the longitudinal direction, so that rapidity
differences are frame--independent. The Lorentz--invariant rapidity 
interval over which particles with a given transverse momentum are 
distributed is
\beq
Y \;\; \equiv \;\; y_{h, {\rm max}} - y_{h, {\rm min}}
\;\; \approx \;\; \ln [W^2 / (P_{T, h}^2 + m_h^2)] ,
\label{rapidity_interval}
\eeq
where $W$ is the $\gamma^\ast N$ center--of--mass (CM) energy
and it is assumed that $W^2 \gg P_{T, h}^2 + m_h^2$.
Studies of hadronic final states in DIS at $W \sim \textrm{few GeV}$ 
show that the target and current fragmentation regions 
occupy at least one unit of rapidity. To cleanly 
separate the two regions, rapidity intervals of $Y \sim 4$ are needed. 
Our picture based on QCD vacuum structure suggests that one look for
correlations between pions with $P_{T, \pi}^2 \sim 0.5 \, \textrm{GeV}^2$ 
in the current fragmentation region and hadrons with similar 
transverse momenta in the target fragmentation region. 
From Eq.~(\ref{rapidity_interval}) we see that this requires squared 
CM energies around $W^2 \approx 30 \, \textrm{GeV}^2$. 
At these energies sea quarks with $x = 0.05 \, (0.1)$ could be probed 
at scales $Q^2 \approx x W^2 = 1.5 \, (3.0) \, \textrm{GeV}^2$,
where perturbative QCD radiation still plays a minor role in 
inclusive particle production. The kinematic region described here
thus represents a ``window'' where one can expect to see nonperturbative 
correlations between sea quarks reflected in the hadronic final state. 
%
%

By choosing a leading hadron in the current fragmentation region 
with $z \gtrsim 0.5$ ($z$ is the fraction of the virtual photon's 
laboratory energy
carried by the produced hadron) one can minimize also the effect
of transverse momentum broadening through fragmentation and
approximately infer the intrinsic $p_T$ of the struck (anti)quark
in the target. By measuring particle production in the central
and target fragmentation regions one can then establish how this
transverse momentum is balanced in the final state.
If the observed $P_{T, h}$ of the forward hadron indeed originated from 
the nonperturbative intrinsic transverse momentum of quarks in the nucleon,
the picture of pair correlations described in Sec.~\ref{sec:correlations}
implies that the $P_{T, h}$ should be balanced by the other (anti)quark in 
the pair, which materializes in the target fragmentation region,
leading to long--range correlations in rapidity between the current 
and target fragmentation regions \cite{foot_corr}. If, however, 
the main source of the observed $P_{T, h}$ were the fragmentation process, 
one expects short--range correlations within the current 
fragmentation region.

If $Q^2$ is increased from the values of $\sim\textrm{few GeV}^2$ discussed
above, QCD radiation should play an increasing role. In inclusive particle 
production this is described by DGLAP evolution and well understood
theoretically and experimentally. In correlation measurements of the kind 
described here, these emissions should diminish the correlations 
expected from soft interactions by providing additional possibilities 
for the balancing of the observed $P_{T, h}$ in the current fragmentation 
region.

Correlations between the current and target fragmentation region could
be studied by measuring either mesons or baryons, or a collection of
hadrons, in the target fragmentation region. The role of baryons in
balancing the $P_{T, h}$ of the current jet is likely more important
in processes where a valence quark in the target is removed,
potentially providing a handle to separate the intrinsic transverse
momenta of valence and sea quarks.

Correlation measurements of the kind described here would be feasible 
in the kinematic region covered by the CERN COMPASS experiment
(squared lepton--nucleon CM energy $s \approx 300 \, \textrm{GeV}^2$)
\cite{COMPASS}
or with a medium--energy Electron--Ion Collider \cite{Accardi:2011mz},
if suitable detection capabilities for forward particles are provided.
The merit of such correlation measurements in a smaller rapidity 
interval with the Jefferson Lab 12 GeV Upgrade 
($s \approx 20\, \textrm{GeV}^2$), especially with the CLAS12 detector, 
deserves further study \cite{JLab12}.
\subsection{Multiparton processes}
More direct tests of the idea of parton short--range correlations
may become possible with the concept of multiparton distributions,
whose proper formulation in QCD is a subject of present 
work \cite{Blok:2010ge,Diehl:2011yj}. Such distributions are required in the
description of $pp$ collisions with multiple hard processes,
where correlations manifest themselves as an enhancement of the
multiple process rate compared to the uncorrelated expectation.
The data on three--jet plus photon production in $\bar pp$ collisions 
from the Tevatron CDF \cite{Abe:1997bp} and D0 \cite{Abazov:2009gc} 
experiments show a significant enhancement compared to the uncorrelated
expectation \cite{Frankfurt:2003td}, suggesting the presence of 
substantial correlations in the partonic wave function.
Whether these correlations are due to perturbative QCD radiation
or the nonperturbative vacuum structure discussed here is a
challenging question which requires further study.
\subsection{Exclusive meson production}
The nonperturbative short--range correlations in the nucleon's
partonic wave function discussed here (see Fig.~\ref{fig:corr})
have potential implications beyond semi--in\-clusive processes, 
for example for hard exclusive meson 
production $\gamma^\ast N \rightarrow M + N$. In $\pi^+$ production
at $Q^2 \sim \textrm{few GeV}^2$ and $x \gtrsim 0.1$, a substantial
longitudinal cross section should result from the ``knockout''
of a correlated $q\bar q$ pair in the nucleon with appropriate 
quantum numbers, which then forms a $\pi^+$ by way of soft
nonperturbative interactions. This mechanism is theoretically
subleading compared to the hard scattering mechanism in the
limit $Q^2 \rightarrow \infty$ (in which the equivalent correlations 
are generated by perturbative gluon exchange) but may be practically
dominant at all realistic values of $Q^2$. It may also play a role 
in exclusive $\rho^0$ production in the region $W \sim$ 2--4 GeV
\cite{Morrow:2008ek}, which shows an energy dependence consistent with 
the $t$--channel exchange of a $q\bar q$ pair with scalar 
quantum numbers \cite{Guidal:2007cw}. The theoretical formulation
of such knockout processes in exclusive production is the subject
of on-going work. Generally, exclusive processes offer many possibilities 
for probing correlations in the partonic wave function due to QCD
vacuum structure, including their $x$--distribution and quantum numbers.
\section*{Acknowledgments}
In this study we greatly benefited from numerous discussions with D.~Diakonov,
V.~Petrov, P.~Pobylitsa, and M.~Polyakov during earlier joint work on the 
chiral quark--soliton model. We are grateful also to C.~Lorc\'e for 
enlightening discussions of the properties of the nucleon light--cone 
wave function. 

M.~S.\ acknowledges the hospitality of Jefferson Lab during the work 
on this study. This work is supported by the U.S. DOE under Grant 
No.~DE-FGO2-93ER40771. Notice: Authored by Jefferson Science Associates, 
LLC under U.S.\ DOE Contract No.~DE-AC05-06OR23177. The U.S.\ Government 
retains a non--exclusive, paid--up, irrevocable, world--wide license to 
publish or reproduce this manuscript for U.S.\ Government purposes.
\appendix
\section{Parametrization of soliton profile}
\label{app:profile}
\begin{figure}
\includegraphics[width=.45\textwidth]{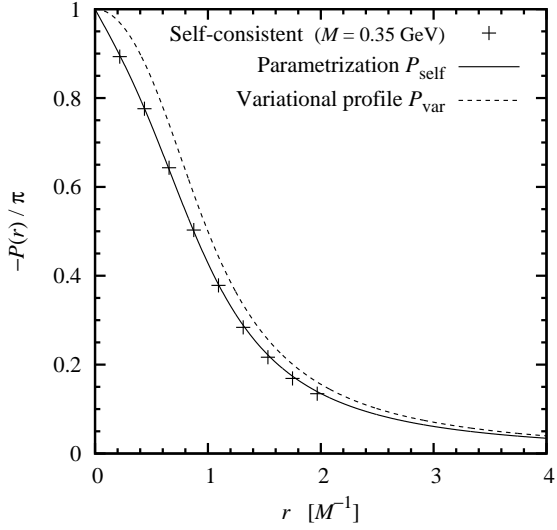}
\caption[]{Soliton profile $P(r)$ as function of the radius $r$.
Points: Self--consistent profile obtained by minimizing the classical
energy (PV regularization, $M = 0.35 \, \textrm{GeV}, M_\pi = 0$) 
\cite{Weiss:1997rt}.
Solid line: Parametrization $P_{\rm self}$, 
Eqs.~(\ref{P_self})--(\ref{P_self_end}).
Dashed line: Variational profile $P_{\rm var}$, Eq.~(\ref{P_var}),
for $M_\pi = 0$.}
\label{fig:prof}
\end{figure}

The profile function $P(r)$ of the classical chiral field, 
Eq.~(\ref{hedgehog}) is determined by minimizing the classical
energy of the soliton, Eq.~(\ref{E}). Numerical minimization
was performed with various types of cutoff of the Dirac sea 
contribution and result in very similar values of the classical
energy; see Ref.~\cite{Christov:1995vm} for a review.
For a Pauli--Villars (PV) cutoff, described in detail in Sec.~\ref{sec:sea},
the self--consistent profile was determined in Ref.~\cite{Weiss:1997rt}.
The result obtained with a dynamical quark mass $M = 0.35\, \textrm{GeV}$,
a PV cutoff with $M_{\rm PV}^2 /M^2 = 2.52$, cf.\ Eq.~(\ref{mpv}),
and the chiral limit $M_\pi = 0$, is shown in Fig.~\ref{fig:prof}. 
The figure shows the mesh points in the variable $r$ at which the profile 
was determined by numerical minimization; at larger distances the profile 
effectively exhibits an asymptotic behavior as 
\beq
P(r) \; \sim \; -A/r^2 .
\label{P_large_r}
\eeq
With these parameters the classical energy of the soliton, i.e., the 
nucleon mass in leading order of the $1/N_c$ expansion, 
Eq.~(\ref{M_N_as_minimum}), was obtained as \cite{Weiss:1997rt}
\beq
M_N \;\; \approx \;\; 3.26 \, M .
\eeq
The nucleon's isovector axial coupling, obtained by calculating the
matrix element of the axial current operator as a sum over quark
single--particle levels, is $g_A=1.03$ with these parameters
\cite{Schweitzer:thesis}. In the chiral limit the isovector axial 
coupling is proportional to the coefficient of the leading $r^{-2}$ 
asymptotic behavior of the soliton profile at large distances, 
Eq.~(\ref{P_large_r}),
\beq
g_A \;\; = \;\; \frac{8\pi}{3} \, F_\pi^2 A ,
\label{g_A_tail}
\eeq
and can alternatively be determined in this way. The equivalence of the 
two ways of calculating $g_A$ for the self--consistent profile of 
Ref.~\cite{Weiss:1997rt} was verified in Ref.~\cite{Schweitzer:thesis}.
%
%

For the purposes of our study we need an analytic parametrization of 
the soliton profile which has the correct limiting behavior at small 
and large distances and reproduces the numerical self--consistent
profile with reasonable accuracy. It can be constructed in the form
\beq
P_{\rm self}(r) \;\; = \;\; -2 \, \arctan \, T(r) ,
\label{P_self}
\eeq
where the function $T(r)$ satisfies the following two conditions:
(a) $T(r) \propto r^{-1}$ at $r \rightarrow 0$, to ensure linear behavior
of the profile near the center; (b) $r^2 T(r) = A + O(r^{-4})$ 
for $r \rightarrow \infty$, to guarantee the correct $r^{-2}$
asymptotic behavior of the profile and the absence of subleading $r^{-4}$ 
terms, which was noted in Ref.~\cite{Diakonov:1987ty} in the
context of a long--distance expansion of the equations of motion. 
A simple choice which fulfills these conditions is
\beq
T(r) \;\; = \;\; \frac{r_0^2}{r^2} \, \tanh (Br) .
\label{P_self_T}
\eeq
The parameter $r_0^2 = A/2$ determines the large--distance behavior
of the profile; we fix it by requiring that the parametrization 
reproduce the ``exact'' numerical value of 
$g_A = 1.03$ \cite{Schweitzer:thesis} via Eq.~(\ref{g_A_tail}), which gives
\beq
r_0^2 \;\; = \;\; 0.87 \, M^{-2}.
\eeq
The parameter $B$ is then fixed by fitting the behavior of the profile
at small distances, which results in
\beq
B \;\; = \;\; 1.6 \, M .
\label{P_self_end}
\eeq
The simple parametrization Eqs.~(\ref{P_self})--(\ref{P_self_end})
provides an excellent fit to the numerical self--consistent profile
over all distances (see Fig.~\ref{fig:prof}). 

Also shown in the figure is the variational profile of 
Ref.~\cite{Diakonov:1987ty},
\be
P_{\rm var} (r) &=&
-2\, \arctan \frac{r_0^2}{r^2} ,
\label{P_var}
\\[1ex]
r_0^2 &=& 1.0 \, M^{-2} ,
\ee
which was used extensively in earlier calculations of partonic 
structure \cite{Diakonov:1996sr,Diakonov:1997vc}. This profile
corresponds to the limit $B \rightarrow \infty$ of our more general
parametrization, Eqs.~(\ref{P_self}) and (\ref{P_self_T}),
and has an unphysical quadratic behavior near the
center, which, however, is unimportant for the quantities studied
here (cf.\ the discussion in Sec.~\ref{subsec:momentum_classical}). 
With the replacement
\beq
\frac{r_0^2}{r^2} \;\; \rightarrow \;\; \frac{r_0^2}{r^2} \;
(1 + M_\pi r ) \exp (-M_\pi r)
\eeq
Eq.~(\ref{P_var}) also allows one to study the effect of a finite pion mass.
\section{Bound--state level wave function}
\label{app:level}
In this appendix we give the explicit form of the bound--state level 
wave function and its Fourier transform, as used in the calculation of
the level contribution to the transverse momentum distributions of 
quarks and antiquarks in Sec.~\ref{sec:valence}. Most of the relevant
expressions were given in Appendix~B of Ref.~\cite{Diakonov:1996sr},
and we include them here for reference only (see also 
Footnote~\cite{foot_gamma5} regarding the convention for gamma matrices).

In the standard representation of the gamma matrices, the bound--state 
solution of the Dirac equation in the nucleon rest frame, where the
classical chiral field is given by Eq.~(\ref{hedgehog}),
can be written in the form
\be
\Phi_{\rm lev} (\bm{x}) &=& \frac{1}{\sqrt{4\pi}}
\left(
\begin{array}{r} h (r) \\[1ex] 
\displaystyle
-i \frac{(\bm{x}\bm{\sigma})}{r} \, j(r) 
\end{array} \right) | 0 \rangle ,
\label{level_spinor}
\ee
where $r \equiv |\bm{x}|$ and $| 0 \rangle$ 
is the spin--isospin wave function where spin and isospin
are coupled to zero total, 
\be
(\sigma_i + \tau_i ) | 0 \rangle &=& 0, \hspace{2em} \langle 0 | 0 \rangle
\;\; = \;\; 1 .
\ee
The functions $h$ and $j$ are solutions of the radial equation
\be
\lefteqn{
\left(\begin{array}{cc}
 M \cos P(r) & 
{\displaystyle -\frac{\partial}{\partial r} - \frac{2}{r} - M \sin P(r)}\\
{\displaystyle \frac{\partial}{\partial r} - M \sin P(r)} & 
- M \cos P(r)
\end{array}\right) } && 
\nonumber \\
&\times& 
\left(\begin{array}{c}
h(r) \\[1ex] j(r) 
\end{array}\right) 
\;\; = \;\; E_{\rm lev}
\left(\begin{array}{c}
h(r) \\[1ex] j(r) 
\end{array}\right) . \hspace{3em}
\label{level_equation}
\ee
The level wave function Eq.(\ref{level_spinor}) is normalized to
\be
\int d^3 x \; \Phi_{\rm lev}^\dagger (\bm{x})
\Phi_{\rm lev} (\bm{x}) &=& 1,
\ee
corresponding to
\be
\int_0^\infty dr \, r^2 \left[ h^2 (r) + j^2 (r) \right] &=& 1.
\ee
\par
The wave function in momentum representation, defined 
according to Eq.(\ref{phi_momentum_def}),
\be
\Phi_{\rm lev} (\bm{p}) &=& \int d^3 x \; e^{-i \bm{p}\bm{x}} \;
\Phi_{\rm lev} (\bm{x}) ,
\ee
can be written in the form
\be
\Phi_{\rm lev} (\bm{p}) &=& 
\frac{\sqrt{2} \pi}{p}
\left(
\begin{array}{r} h (p) \\[1ex] 
- {\displaystyle \frac{(\bm{p} \bm{\sigma})}{p}} \, j(p) 
\end{array} \right) | 0 \rangle ,
\label{level_spinor_momentum}
\ee
where $p \equiv |\bm{p}|$. The radial functions in the 
momentum representation here are given by
\be
h(p) &=& \int\limits_{0}^{\infty} dr \, r^2\, h(r) R_{p0}(r),
\nonumber
\\
j(p) &=& \int\limits_{0}^{\infty} dr \, r^2\, j(r) R_{p1}(r),
\label{level_h_j_momentum}
\ee
where $R_{pl}$ are the free radial wave functions of the 
continuous spectrum, defined as
\be
R_{pl}(r) &=&  
\sqrt{\frac{2}{\pi}} p \, j_l (pr)
\; = \; 
\sqrt{\frac{2}{\pi}} 
\left( -\frac{1}{p} \frac{d}{dr} \right)^l \frac{\sin pr}{r} ,
\hspace{2em}
\label{R_pl_def}
\ee
where $j_l (l = 0, 1, \ldots)$ 
denote the spherical Bessel functions, and normalized according to 
\be
\int_0^\infty dr \, r^2 \, R_{pl}(r) R_{p'l}(r) &=& \delta (p - p') .
\ee
The normalization condition for the momentum representation of the
bound--state wave function, Eq.(\ref{level_spinor_momentum}), is
\be
\int \frac{d^3 p}{(2 \pi )^3} \; \Phi_{\rm lev} (\bm{p})^\dagger 
\Phi_{\rm lev} (\bm{p}) &=& 1,
\ee
corresponding to
\be
\int_0^\infty dp \left[ h^2(p) + j^2(p) \right] &=& 1.
\ee

When computing matrix elements between bound--state level wave functions,
one can take advantage of the properties of the spin--isospin 
singlet state. Specifically,
\be
\ldots \tau^i | 0 \rangle &=& \ldots (-\sigma^i) | 0 \rangle ,
\\[1ex]
\langle 0 | \tau^i \ldots &=& \langle 0 | (-\sigma^i) \ldots ,
\\[1ex]
\langle 0 | \sigma^i | 0 \rangle
&=& \langle 0 | \tau^i | 0 \rangle \;\; = \;\; 0.
\ee
Using these identities one can convert matrix elements of products of 
$\sigma$ and $\tau$ matrices into equivalent expressions involving 
only one kind of matrix, which can then be evaluated using standard
techniques.
\section{Fourier transform of soliton field}
\label{app:fourier}
When evaluating the sea quark distribution using the gradient expansion
we need explicit expressions of the 3--dimensional Fourier transform of 
the static classical chiral field in the nucleon rest frame. In coordinate 
space the field is given by Eq.~(\ref{hedgehog}) and can be
expanded as
\beq
U_{\text{cl}} (\bm{x}) 
\;\; = \;\; \cos P(r) + \frac{i (\bm{x}\bm{\tau})}{r} \sin P(r) ,
\eeq
where $r \equiv |\bm{x}|$. 
The Fourier transform Eq.~(\ref{U_fourier_def}) is of the
form
\be
\tilde{U}_{\text{cl}} (\bm{k})
&\equiv& \int \! d^3 x\; e^{-i \bm{k}\bm{x}} \;
[U_{\text{cl}} (\bm{x}) - 1]
\nonumber
\\[1ex]
&=&
4\pi \left[ s(k) + \frac{(\bm{k}\bm{\tau})}{k} p(k) \right] ,
\ee
where $k \equiv |\bm{k}|$. The scalar functions $s(k)$ and $p(k)$ are
determined as
\be
s (k) &=& \int_0^\infty dr \, r^2 \; j_0 (kr) \;
\left[ \cos P(r) - 1 \right] ,
\\
p (k) &=& \int_0^\infty dr \, r^2 \; j_1(kr)\; \sin P(r) ,
\ee
where $j_0$ and $j_1$ are the spherical Bessel functions, cf.\ 
Eq.~(\ref{R_pl_def}). The traces in Eq.~(\ref{f_pi}) and Eq.~(\ref{g_cl}) 
are then obtained as
\be
\textrm{tr}_{\rm fl} [\widetilde U_{\text{cl}} (\bm{k}) 
\widetilde U_{\text{cl}} (\bm{k})^\dagger ]
&=& 32 \pi^2 [ s^2 (k) \, + \, p^2(k) ] ,
\label{utilde_trace}
\\[1ex]
\textrm{tr}_{\rm fl} [\tau^3 \widetilde U_{\text{cl}} (\bm{k}) 
\widetilde U_{\text{cl}} (\bm{k})^\dagger ]
&=& 64 \pi^2 \, s(k) \, p(k) \frac{k^3}{k} .
\label{utilde_trace_tau3}
\ee
In proving the large--$N_c$ inequalities for the sea quark distributions,
cf.\ Eq.~(\ref{inequality_sea_grad}), it is important to note that
the sum and difference of the forms in Eqs.~(\ref{utilde_trace})
and (\ref{utilde_trace_tau3}) is non--negative. Namely,
\be
\lefteqn{\textrm{tr}_{\rm fl} [(1 \pm \tau^3) 
\widetilde U_{\text{cl}} (\bm{k}) 
\widetilde U_{\text{cl}} (\bm{k})^\dagger ]} &&
\nonumber
\\[1ex]
&=& 32 \pi^2 \left[ s^2(k) \, + \, p^2 (k) \, \pm \, 2 s(k) \, p(k) \, 
\frac{k^3}{k} \right]
\nonumber
\\
&\geq & 32 \pi^2 \left[ |s(k)|^2 \, + \, |p(k)|^2 \, - \, 2 |s(k)| \, |p(k)| 
\, \frac{|k^3|}{k} \right]
\nonumber
\\
&\geq& 32 \pi^2 \left( |s(k)| - |p(k)| \right)^2 ,
\label{utilde_trace_sumdiff}
\ee
where the last inequality follows from $|k^3| \leq k$.

\begin{thebibliography}{99}
\bibitem{foot_m0} The quantity $m_0^2$ is conventionally defined as 
the ratio $m_0^2 = \langle\bar\psi \sigma^{\alpha\beta} 
F_{\alpha\beta} \psi\rangle / \langle\bar\psi \psi\rangle$.
The dimension--5 operator in the numerator is equivalent to 
$2\bar\psi \nabla^2 \psi$ by virtue of the Heisenberg equations of motion.
\bibitem{foot:width} A smaller value of the polarized $\langle p_T^2 \rangle$
compared to the unpolarized one is also obtained in phenomenological models
based on Gaussian distributions. For a Gaussian distribution 
$\langle p_T^2 \rangle$ is proportional to the width, and a larger width 
for the polarized distribution would lead to a violation of positivity at
large $p_T$. While certainly not rigorous, this argument makes plausible
why for near--Gaussian shapes the polarized $\langle p_T^2 \rangle$
should be smaller than the unpolarized one.
\bibitem{foot_corr} 
How the nonperturbative
$q\bar q$ correlations in the nucleon wave function manifest themselves 
in the hadronic final state may be affected by the possibility that 
the nonperturbative gluon field inducing the correlation could 
materialize in the final state. As discussed above, fits to DIS data 
show that at the chiral symmetry--breaking scale about $\sim 30\%$ 
of the nucleon's momentum is carried by gluons. This suggests that 
at least part of the nonperturbative gluon
fields in the fast--moving nucleon project on physical gluon states 
that can carry longitudinal momentum away into the final state.
\end{thebibliography}
\end{document}